\documentclass[10pt,twocolumn]{article}

\usepackage{lipsum} 		
\usepackage{blindtext} 		

\usepackage[superscript, biblabel]{cite}  	

\makeatletter
\renewcommand\@biblabel[1]{#1.}
\makeatother

\usepackage{etoolbox}
\patchcmd{\thebibliography}{\section*{\refname}}{}{}{}

\usepackage[margin=1.0in,hmarginratio=1:1,top=32mm,columnsep=15pt]{geometry}

\usepackage{color}				
\definecolor{dark-gray}{gray}{0.1}	
\usepackage{times}				
\usepackage{microtype} 			
\usepackage[english]{babel} 		

\usepackage[hang, font={footnotesize}, labelfont={bf,up}, textfont={sf,up}]{caption} 
\usepackage{booktabs} 	
\usepackage{mwe}  		

\usepackage{enumitem} 		
\setlist[itemize]{noitemsep} 	

\usepackage{abstract} 
\AtBeginDocument{}  		

\usepackage{titlesec} 			
\renewcommand\thesection{\Roman{section}.} 		  		
\renewcommand\thesubsection{\thesection\Alph{subsection}.} 	
\renewcommand\thesubsubsection{\thesubsection\arabic{subsubsection}.} 
\titleformat{\section}[block]{\normalfont\sffamily\bfseries}{\thesection}{1em}{\MakeUppercase}{} 	
\titleformat{\subsection}[block]{\normalfont\sffamily\bfseries}{\thesubsection}{1em}{}{}  
\titleformat{\subsubsection}[block]{\normalfont\sffamily\bfseries}{\thesubsubsection}{1em}{}{}  
\titlespacing*{\section}{0.0em}{1em}{0.25em}		
\titlespacing*{\subsection}{0.0em}{1em}{0.25em}	
\usepackage{indentfirst}						

\usepackage{tocloft}
\usepackage{fancyhdr} 

\fancypagestyle{plain}{
\fancyhf{}
\lhead{\color{dark-gray}\textit{}} 
\rhead{ {\it Submitted to ANS/NT (2021). LA-UR-20-30028 DRAFT} }
}

\usepackage{titling} 		

\usepackage{hyperref} 	
\hypersetup{
	backref=true,       
    	pagebackref=true,               
    	hyperindex=true,                
    	colorlinks=true,                
    	breaklinks=true,                
    	urlcolor= black,                
    	linkcolor=black,                
    	bookmarks=true,                 
    	bookmarksopen=false,
    	filecolor=black,
    	citecolor=blue
}

\pretitle{\begin{center}\large\bfseries} 	
\posttitle{\end{center}} 				
\title{\vspace{-0.3in} \sffamily{Nuclear Science for the Manhattan Project   \\
\& Comparison to Today's ENDF Data} }	
\author{%
\normalsize Mark B. Chadwick\thanks{corresponding author: mbchadwick@lanl.gov}\\[-0.5ex] 
\normalsize Los Alamos National Laboratory \\[-0.5ex] 
\normalsize Los Alamos, NM 87545
}
\date{ } 
\renewcommand{\maketitlehookd}{%
\begin{abstract}
\vspace*{-0.5in}
\noindent {\normalfont\textbf{Abstract}\textemdash}
\noindent 
Nuclear physics advances in the US and Britain, from 1939-1945, are described. The Manhattan Project's work led to an explosion in our knowledge of nuclear science. A conference in April 1943 at Los Alamos provided a simple formula used to compute critical masses, and laid out the 
research program needed to determine the key nuclear constants. In short order, four university accelerators were disassembled and reassembled at Los Alamos, and methods were 
established to make measurements on extremely small samples owing to the initial lack of availability of enriched $^{235}$U and plutonium.
I trace the program that measured fission cross sections, fission emitted neutron multiplicities and their energy spectra, and transport cross sections, comparing the measurements with our best understanding today as embodied in the 
Evaluated Nuclear Data File ENDF/B-VIII.0. The large nuclear data uncertainties at the beginning of the project, which often exceeded 25-50\%, were reduced by 1945 often to less than 5-10\%. $^{235}$U and $^{239}$Pu fission cross section assessments in the fast MeV range were reduced with more accurate measurements, and the neutron multiplicity ${\overline{\nu}}$ increased. By a lucky coincidence of canceling errors, the initial critical mass estimates were close to the final estimated masses. Some images from historical documents from our Los Alamos archives are shown. Many of the original measurements from these early years have not previously been widely available. Through this work, these data have now been archived in the international experimental nuclear reaction data library (EXFOR) in  a collaboration with the IAEA  and Brookhaven National Laboratory.
\begin{center}
\par
\noindent 
\vspace*{-0.05in}
\end{center}
\end{abstract}
}

\begin{document}

\maketitle	


\tableofcontents

\section{Introduction}

\begin{figure}[htbp]
\begin{center}
\includegraphics[width=2.in]{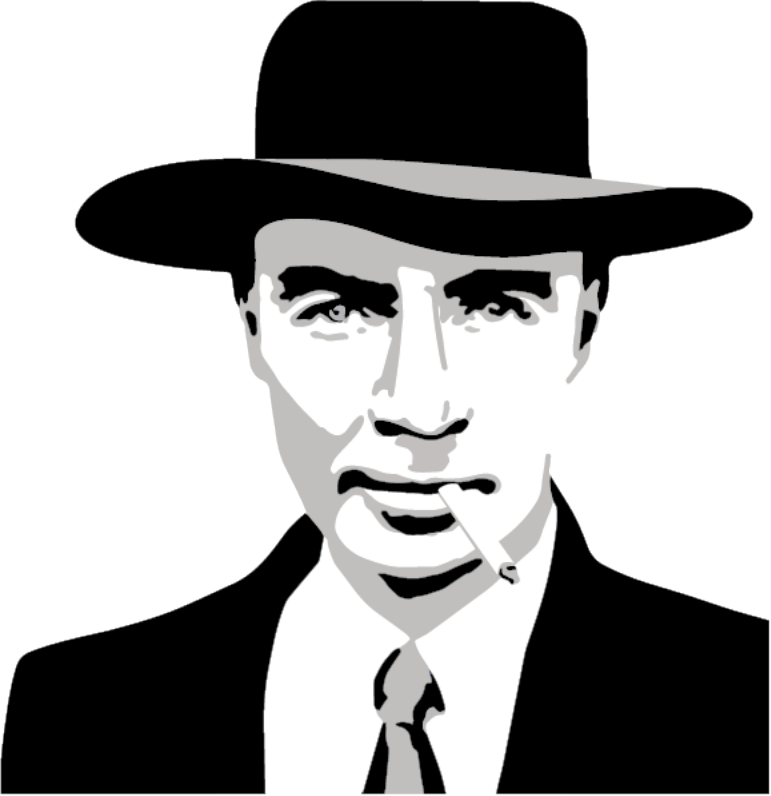}
\caption{J.R. Oppenheimer. Credit: Sarah Tasseff, LANL.}
\label{fig:oppy-face}
\end{center}
\end{figure}

In 1939 the world of physics was shaken by Bohr and Wheeler’s remarkable Physical Review paper, “The Mechanisms of Nuclear Fission”\cite{Bohr:1939}, which provided physics insights and a mathematical understanding of the fission process, which had only recently been discovered. It was inspired by Meitner and Frisch’s “liquid-drop" fission model. That same paper reported on the first fast fission uranium cross section measurements from Princeton (Ladenburg, Kanner, Barschall and van Voorhis) and from the Carnegie Institution of Washington DC (Tuve). Merle Tuve's measurement was particularly important since it was for neutron energies below 1 MeV, which Bohr and Wheeler had predicted would be below the $^{238}$U fission threshold 
 and therefore could be ascribed to the $^{235}$U isotope. The importance of these measurements was such that James Chadwick, in an interview three decades later \cite{Chadwick:1969}, still remembered  the page number on which the measurements were published (p.\,444)!


Also, in 1939 Rudolf Peierls in Birmingham had realized that a fast unmoderated neutron chain reaction was possible in the $^{235}$U isotope and derived a formula for its critical mass\cite{Close:2019}. He did not compute the actual critical mass values in his 1939 Proceeding of the Cambridge Philosophical Society paper\cite{Peierls:1939} (that came some months later, with Frisch), though he pointed out in a note added in proof how Bohr and Wheeler’s paper supported the arguments that he made.

Otto Frisch had come to Birmingham in the summer of 1939 and began to work with Peierls on whether a “super bomb” could be constructed with the $^{235}$U isotope supporting a chain reaction. At that time, both Frisch and Peierls 
were classed as enemy aliens; they were allowed to work on nuclear physics as it was thought to be less sensitive compared to radar.
\cite{Peierls:1990?} Using Peierls’ formula, together with estimates for the cross sections and the fission neutron multiplicity, Frisch and Peierls calculated\cite{Frisch:1940}  a $^{235}$U critical mass that was surprisingly (and erroneously) small, 0.6 kg (the correct value we now know is 46 kg for a sphere of pure $^{235}$U). The mistake they made was for the $^{235}$U fission cross section in the fast energy region, which they took as 10\,b (versus the best value today of 1.2b), influenced it seems by a measurement in Paris by Goldstein for the  total $^{nat}$U (not fission) cross section of 10-11\,b in the fast energy range.\cite{Bernstein:2011}  Nevertheless, their 1940 Frisch-Peieirls memorandum\cite{Frisch:1940} was tremendously insightful and was taken seriously by the British government, influencing the creation of the 
MAUD\cite{MAUD:1941} committee\footnote{The author, MBC, is pleased to write on MAUD, given that  its chairman G.P. Thomson was MBC's “academic grandfather” (P.E. Hodgson was Thomson’s PhD student);  he knew the instigator, Peierls, at Oxford in the 1980s when Peierls held the Wykham Professor of Physics chair at New College and MBC was a lowly Scholar; and the MAUD report’s author James Chadwick shares a common ancestor with MBC (sometime after LUCA).} 
  to determine the feasibility of developing an atomic bomb. (The name MAUD had its origins in a mistaken interpretation of a reference to Bohr's housekeeper, 
  Maud Ray, in a telegram from Bohr to Frisch).
  MAUD was chaired by G.P. Thomson and included various British scientific luminaries including James Chadwick, Britain's leading nuclear physicist. Two Americans also joined some of  the committee meetings, Bainbridge from Harvard and Lauritsen from 
 Caltech\cite{Moore:2020c}, as did two leading French scientists, Halban and Kowarski.

Chadwick realized that the Frisch-Peierls fission cross section estimate of 10\,b was much bigger than Tuve’s result, and consequently that their calculated critical mass was surely too small \cite{Chadwick:1969,Gowing:196?}. Frisch and Peierls seemed to be unaware of Tuve’s measurement, which is strange since Bohr and Wheeler's Physical Review paper was known to them. Tuve’s data at 0.6 MeV and 1 MeV were on natural uranium, reporting 3\,mb and 12\,mb respectively, and it was already known from Bohr and Wheeler that $^{238}$U has a fission threshold of about one MeV,\footnote{The $^{238}$U(n,f) is 
only 1\,mb at 0.6 MeV and 14\,mb at 1 MeV, but rises to 538\,mb at 2 MeV.}
 allowing an inference of the $^{235}$U cross section at 0.6 MeV by multiplying by its (inverse) isotopic abundance: 3\,mb$\times$129 = ~0.4\,b. But Chadwick also correctly guessed that Tuve’s value might be too low because it is so much smaller than the geometric cross section, which is about 1.7\,b, and he remembered \cite{Chadwick:1969} Bohr’s expectation that fission cross sections should not exceed the geometrical cross section (true in the fast region, not true at low energies where quantum mechanics comes in). Chadwick therefore quickly established a program of research in 1940/1941 at Liverpool’s cyclotron laboratory, with Frisch, to measure these cross sections.

Chadwick played a leading role in writing the MAUD report in 1941, a remarkable document for its clarity and prescience (originally secret of course and still hard to find - it is reproduced in Margaret Gowing’s book\cite{Gowing:196?}). It concluded in the likelihood of creating an atomic weapon assuming the successful isotope separation of $^{235}$U: “The committee said that the scheme for a uranium bomb is practicable and likely to lead to decisive results in the war.” Together with Harvard President James Conant, the US had sent two senior scientists, Urey and Pegram, to England in October 1941 to learn about the uranium work there; Chadwick told them that “if pure $^{235}$U could be made available, there was a 99\% chance of being able to produce an explosive reaction” (Gowing\cite{Gowing:196?}, p.117-119). The parallel US National Academy of Sciences (NAS) studies\cite{NAS:1941,Reed:2007}  chaired by Arthur Compton came to the same conclusion,``This seems to be as sure as any untried prediction based upon theory and experiment can be''. Conant later wrote\cite{Rhodes:19?} that the US government’s growing seriousness regarding the potential for an atomic bomb was influenced by both physicists in England who “had concluded that the construction of a bomb made out of uranium 235 was entirely feasible” as well as E.O. Lawrence’s 1941 proposal to use the newly-discovered plutonium. The NAS and MAUD reports helped launch what was to become the Manhattan Engineering District (MED) in August 1942 under General Groves \cite{MED:1946}, with the opening of the Los Alamos laboratory effort, code-named “Project Y”, in March 1943.

\begin{figure*}[t]
\begin{center}
\includegraphics[angle=-90,width=\textwidth]{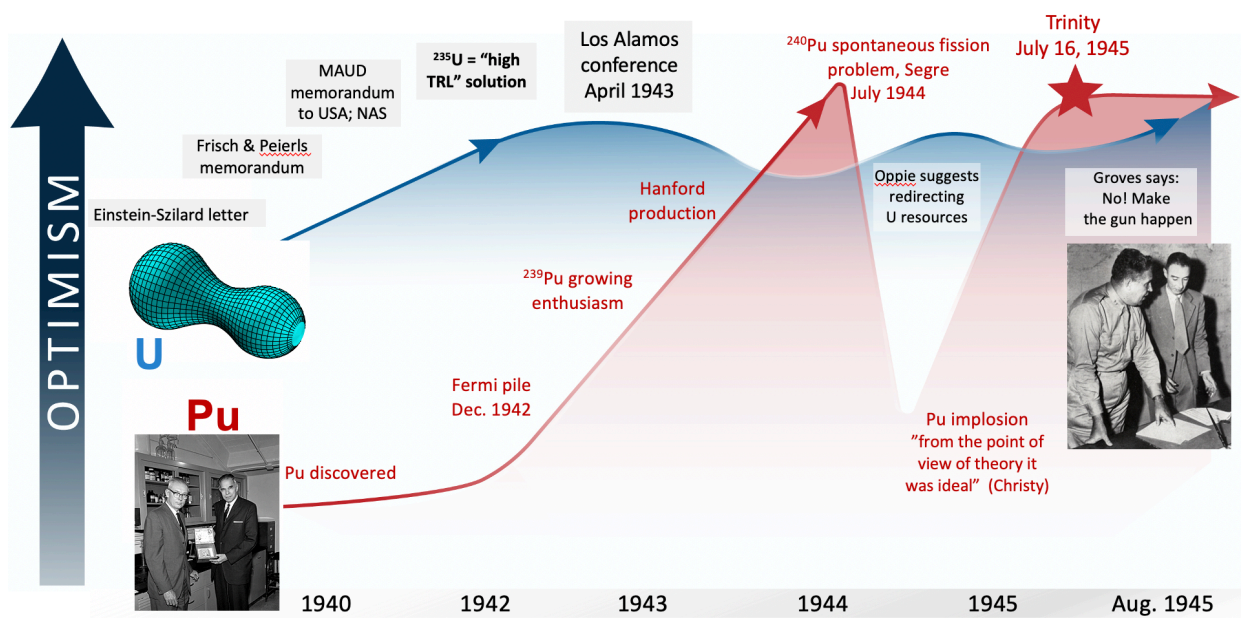}
\vspace{-2.cm}
\caption{Illustration of the dual approaches to building an atomic bomb during the Manhattan Project. The vertical axis denotes the level of optimism  and maturity 
for  each approach. ``TRL" denotes Technology Readiness Level, see text. Taken from Chadwick\cite{Chadwick:2020}. Credit: Peter Moller for the 3D fissioning uranium image, which comes from his
realistic nuclear theory calculation.}
\label{fig:dualpath}
\end{center}
\end{figure*} 


\begin{figure*}
\begin{center}
\vspace{-2. cm}
\includegraphics[angle=-90,width=\textwidth]{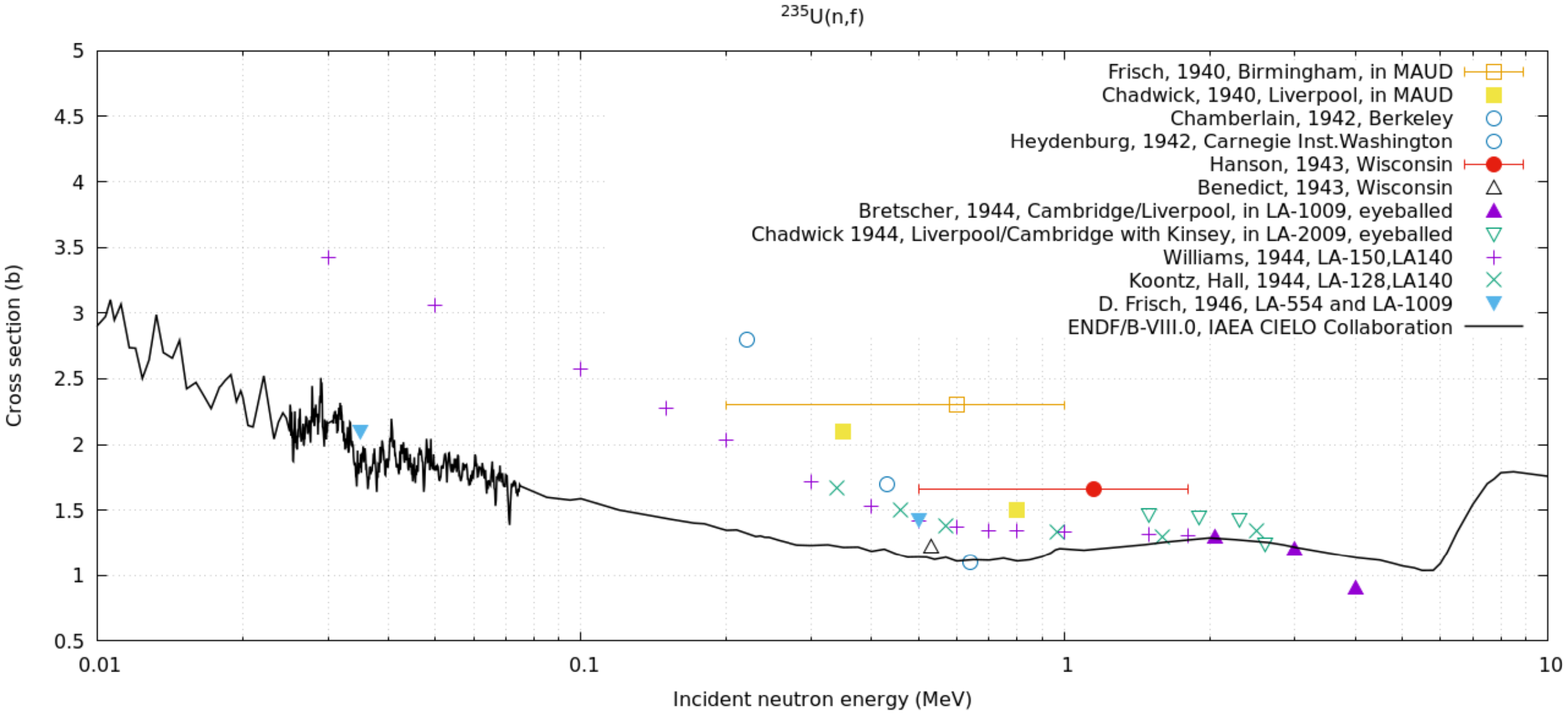}
\vspace{-2.55cm}
\caption{$^{235}$U fission cross section from 10 keV to 10 MeV; the data shown are discussed in some detail in this paper and the curve is the modern 
ENDF evaluation. The earlier measurements had systematic errors resulting in over-estimations of the cross section. Frisch and Peierls' 1940 first guess of 
10\,b can't be seen on this scale.}
\label{fig:u5fiss-fast}
\vspace{-0.5cm}
\end{center}
\end{figure*}

Robert Oppenheimer (Fig.~\ref{fig:oppy-face}) and General Leslie Groves intentionally pursued dual $^{235}$U and $^{239}$Pu pathways to develop the bomb, see Fig.~\ref{fig:dualpath}. The uranium gun bomb approach was what we would call today the “high technology readiness level” (TRL) path, where the physics was becoming well established and had been discussed since 1940, and the principal challenge was to produce adequate quantities of enriched $^{235}$U at Oak Ridge’s MED Y-12 facility. (So, in today’s parlance, it might be called high-TRL but lower “Material Readiness Level”, MRL). In contrast, the plutonium path was much lower-TRL and MRL. Plutonium had just been discovered in Berkeley in December 1940 by Seaborg, McMillan, Segr{\`e}, Kennedy and Wahl, and even by 1942 only a ~2 $\mu$g ``macroscopic'' quantity had been produced as purified PuO$_2$ at Chicago's Metallurgical Laboratory\cite{Norman:2015} (``Met Lab") -- Fermi’s successful reactor experiments at Chicago in December 1942 identified a future path to make plutonium in kg quantities. Even by 1943, much of plutonium’s material, metallurgical, chemical, and nuclear properties could only be guessed at.

The high potential value of plutonium for a bomb was independently perceived  around 1939-1940 by scientists in the US, Britain, and Germany, based on on the understanding of fission embodied in Bohr and Wheeler’s 1939 paper, probably even before plutonium was first created in 1940. Early insights from the US and Germany have been discussed by Bernstein\cite{Bernstein:1996}, including by Louis Turner (Princeton) in 1940 and  by  von Weizs{\"a}cker in July 1940 on neptunium-239 ``Eka-Rhenium'' produced from n+$^{238}$U capture (the plutonium-239 insight in Germany came one year later, in 1941, by Houtermans).
Below, British insights around the same time will be described. At that time it was typical to refer to plutonium as “94”; more exotically, Bretscher in Cambridge  referred to it as “the body 94X239”. Scientists quickly appreciated its two main advantages: it would likely have even more favorable fission properties compared to $^{235}$U; and it could be bred and chemically separated from the abundant $^{238}$U isotope in a reactor, as opposed to the challenging isotope-separation route for the minor $^{235}$U isotope. Histories of this era give credit to Bretscher, Feather, and Rotblat, for their first 1940 considerations of the  utility of plutonium for a weapon (Ref.~\cite{Brown:?}, p. 208). Surprisingly, this important insight is described by many sources, including in Margaret Gowing’s classic book\cite{Gowing:196?}, but without a reference to a Bretscher primary source. I was able to track down such a source in Egon Bretscher’s papers held by Churchill College Cambridge’s Archives.
 Bretscher did indeed write with great perception and intelligence. In Report II, December 19, 1940,  Bretscher and Feather \cite{Bretscher:1940D37} wrote (page 1) ``The large cross section of element 94 to be expected is of the greatest importance in the following respects: (a) it permits (after isolation of 94) the production of a super-explosive mass. The critical radius would seem to be considerably smaller than in the case with U(235) ...''.  See the end of Sec.~IV.A for more on this.

 Bretscher made valuable contributions at Los Alamos. He was a co-author on the important Nuclear Physics Handbook evaluations, LA-140 (1944) and LA-140A (1945), with Weisskopf, Inglis and Davis (discussed further, below) and he led work to measure the DD and DT cross sections for Teller’s Super thermonuclear studies, following earlier work at Purdue University. 
 
 Figure~\ref{fig:u5fiss-fast} illustrates the measurements shown in these Handbooks that will be discussed below in detail in this paper. It is evident that many of these 1940s  measurements over-estimated the true value as embodied in our best understanding today (the solid curve, for 
ENDF/B-VIII.0); reasons for this are discussed later in this paper. The challenges of assessing ``unrecognized sources of uncertainties" (USU) was the subject
of a recent useful paper by Capote {\it et al.} \cite{Capote:2020}. It also focused on $^{235}$U fission cross sections over the years, extending the data in 
Fig.~\ref{fig:u5fiss-fast} from 1945 to the 1980s in a figure that shows differences to a 1970 evaluation by Poenitz \cite{Poenitz:1970}.

\begin{figure}[ht]
\centering \begin{minipage}[b]{0.45\linewidth}
\includegraphics[width=1.5in]{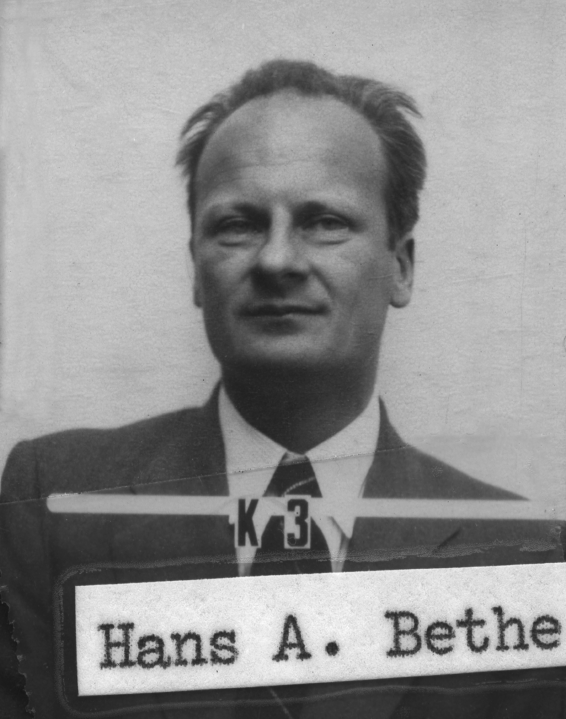}
\end{minipage}
\quad \begin{minipage}[b]{0.45\linewidth}
  \includegraphics[width=1.5in]{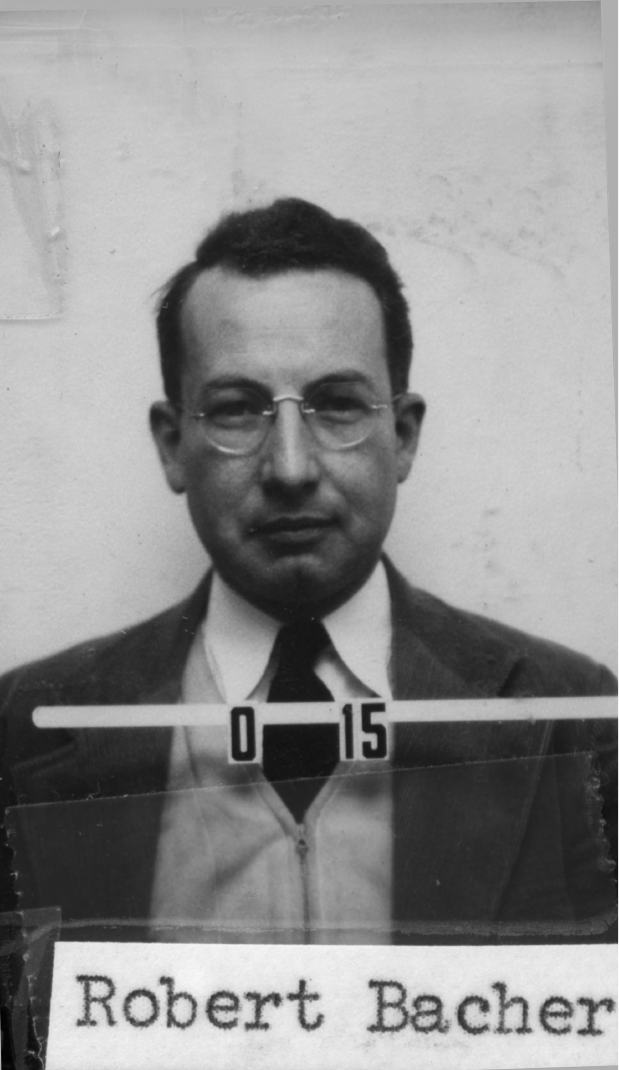}
\end{minipage}
 \caption{Hans Bethe and Robert Bacher lab badges. They led the Theoretical Division and the Physics Division, respectively.} 
 \label{fig:BetheBacher}
\end{figure}

\begin{figure*}[htbp]
\begin{center}
  \includegraphics[width=6.5in]{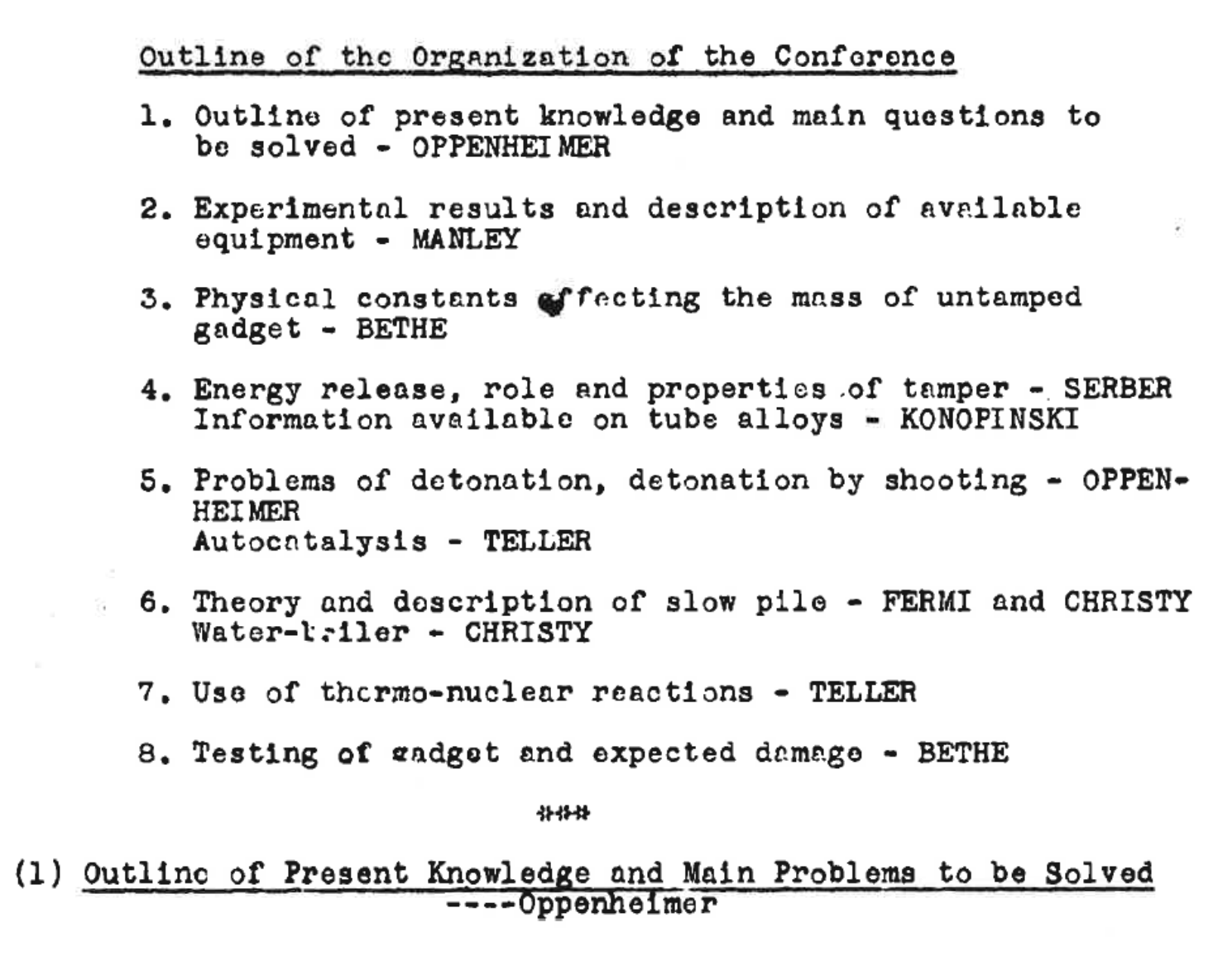}
\caption{The agenda for the first Los Alamos conference, April 1943, from LA-2.}
\label{fig:agenda}
\end{center}
\end{figure*}

The main purpose of the nuclear research program at Los Alamos during the Manhattan Project was to establish an understanding of the fundamental nuclear constants needed to determine critical masses, both bare and tamped, for the bomb designs, and to understand the breadth of nuclear physics to ensure 
there were no surprises or ``show-stoppers''.
 A two-pronged approach was taken, one being to measure the fundamental ``differential'' cross sections and neutron energy and scattering angle spectra, the other being to measure integral quantities. This paper focuses on the former approach, while other papers in this issue by Hutchinson, Kimpland, Myers {\it et al.}\cite{Hutchinson:2020a,Hutchinson:2020b,Hutchinson:2020c} and Ref.~\cite{Hutchinson:2021d}  focus on the latter approach: integral critical assembly measurements. But it should be remembered that for most of the Manhattan Project, substantial amounts of enriched $^{235}$U and plutonium were simply not available in Los Alamos – they only started to arrive in late 1944 and 1945.  Metal ``25''\footnote{Los Alamos shorthand 
 used the last number of the isotope's Z followed by the last number of its A value, so $^{235}$U is ``25'', and is also sometimes written u5; $^{239}$Pu is ``49'' and also pu9, {\it etc.}} spheres started to become available in October 1944, but the larger spheres were not available until April 1945 as described by Hutchinson\cite{Hutchinson:2020a}; plutonium ``49''  spheres were only available after February/March1945. Thus, the integral “fast” critical assembly experiments could only begin in 1945 (some in late 1944). Most of the design work for the uranium gun bomb and for the plutonium implosion bomb was based on the fundamental cross section understandings made between 1943 and late 1944, as described in this paper.

This present paper traces the U.S. and British nuclear science developments in the period leading up to the Manhattan Project, and the subsequent advances made by the U.S., British, and Canadian scientists and engineers at Los Alamos through 1945. Some of the history of early nuclear physics measurements (at universities and at Los Alamos) has been told by the Manhattan Project’s second Physics Division Leader, Robert Wilson\cite{Wilson:1947}, in 1947 (LA-1009), and more recently in the 1993 excellent and comprehensive book {\it Critical Assembly}\cite{Hoddeson:1993} by Los Alamos’ historians. 
The many measurements made at Los Alamos during the Manhattan Project are also documented in the ``LA-'' Los Alamos reports in the National 
Security Research Center (NSRC) archives; many of these are also available online as unclassified reports from the Los Alamos Library. A most remarkable set of documents is the Theoretical Division's monthly progress reports, put together by Hans Bethe with his group leaders (Fig.~\ref{fig:BetheBacher}). Their brilliance, and their diligent scholarship,
\footnote{Allan Carlson of NIST, who knew Hugh Richards at U. Wisconsin (who had been at Los Alamos), recollects Richards' story \cite{Richards:1993}
 of Bethe dictating a paper to his secretary (Hugh's wife). He walked back and forth dictating in perfect English with no mistakes or changes - as if he were reading a manuscript.}
 helps readers reconstruct the evolution of Project Y's work in theory, experiment and simulation.
 I will describe how the evolving understanding of fission cross sections and fission neutron multiplicities led to drastic swings in estimates of the critical masses for $^{235}$U and $^{239}$Pu, with corresponding mood swings of depression and elation among the Los Alamos scientists as the required material production quantities went up and down. The data that were measured are compared against our modern best understanding of the cross sections in our Evaluated Nuclear Data File ENDF/B-VIII.0 database\cite{Brown:2019} and in the Plutonium Handbook\cite{Chadwick:2019}. I will also show how uncertainties in cross sections, which were initially large 
({\it e.g.} Oppenheimer suggested an uncertainty of 0.5\,b on a $^{235}$U fission cross section he estimated as 2\,b in a 1942 letter to Peierls\cite{Oppenheimer:1942}), were substantially reduced to typically $<$ 5-10\% by the end of the project in 1945. Furthermore, some of the data described here have never before been openly available, and these data and related documentation are now being made available to the broader nuclear science community, via the IAEA’s Nuclear Data Section and Brookhaven National Laboratory (BNL) who have put them into the EXFOR file.

This paper does not attempt to provide a comprehensive review of all Manhattan Project nuclear science, and only covers the nuclear data used to compute fast bare critical masses.  Therefore many fascinating topics  are not covered here or are only covered briefly, including: tamped critical masses 
(see this issue, Ref.~\cite{Hutchinson:2020a});
fission delayed-neutron measurements; spontaneous 
fission properties; (alpha,n) reactions that meant impurities would be a concern and plutonium purification methods had to be developed
(see this issue, Ref.~\cite{Martz:2020}); lower energy neutron resonance and thermal fission measurements; fission product radiochemical measurements (see this issue, Ref.~\cite{Hanson:2020}),
and neutron scattering measurements on non-actinides.

\begin{figure}[ht]
\centering \begin{minipage}[b]{0.45\linewidth}
\includegraphics[width=1.5in]{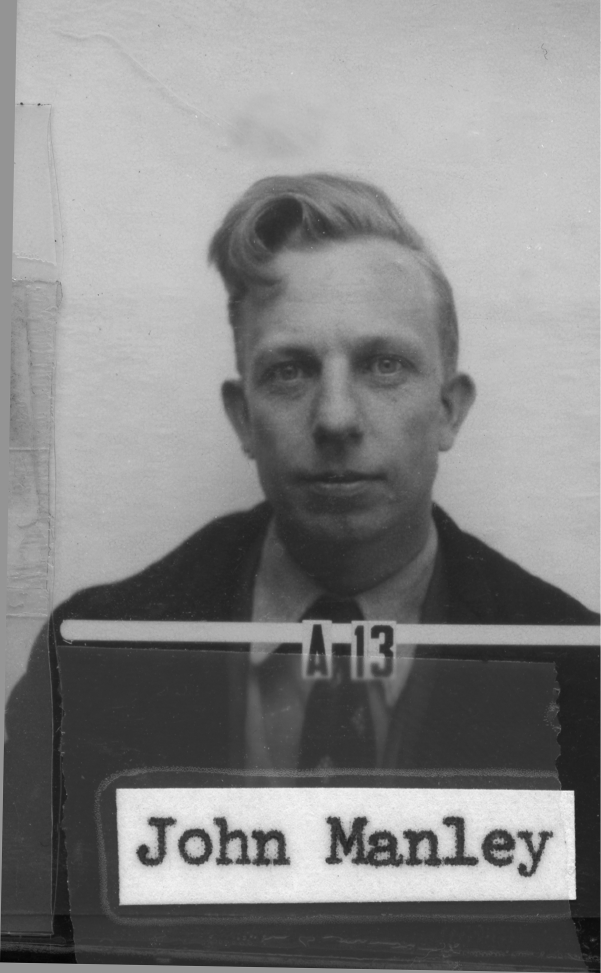}
\end{minipage}
\quad \begin{minipage}[b]{0.45\linewidth}
  \includegraphics[width=1.5in]{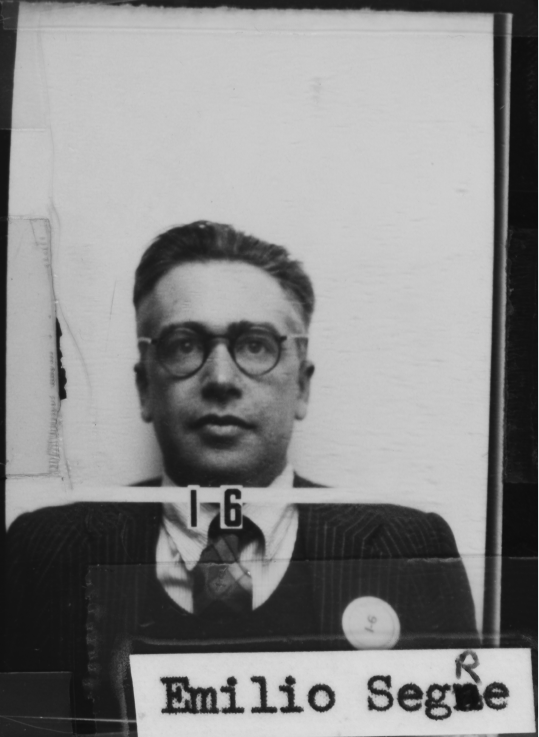}
\end{minipage}
 \caption{John Manley and Emilio Segr{\`e} lab badges.} 
 \label{fig:ManleySegre}
\end{figure}

\begin{figure*}[htbp]
\begin{center}
\includegraphics[width=5.25in]{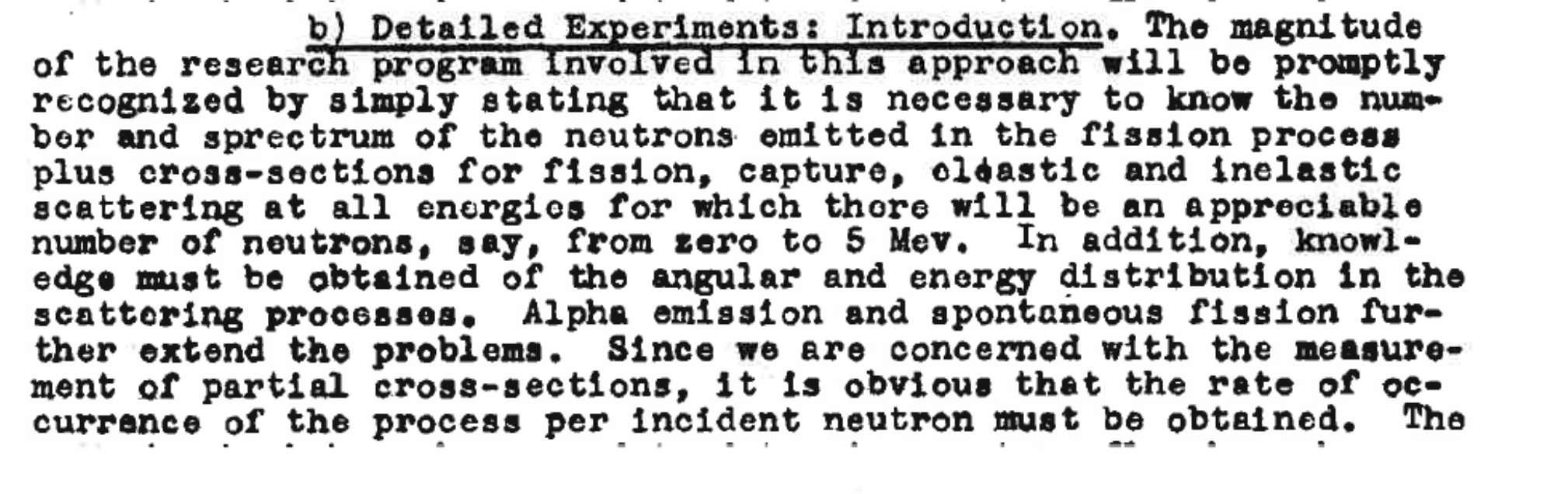}
\caption{Manley's summary of the nuclear science program needed for the Manhattan Project's success, from LA-2, March 1943.}
\label{fig:manley-agenda}
\end{center}
\end{figure*}

In April 1943, at the beginning of Los Alamos’ Project Y work, a conference \cite{Oppenheimer:1943,McMillan:1943} was held in which the current state of understanding was summarized, see 
Fig.~\ref{fig:agenda}. Oppenheimer’s opening paper started with the fundamental nuclear data – neutron cross sections, energy spectra, and so on – that are needed to compute a fast critical mass of $^{235}$U and $^{239}$Pu. He showed a critical mass analytic formula that was in use at the time to compute critical masses. This perspective defined the nuclear science and technology research program, worked out in detail by Bethe, Manley and Bacher (Figs.~\ref{fig:BetheBacher}, \ref{fig:ManleySegre}) that was put in place for the next ~30 months, see Fig.~\ref{fig:manley-agenda}. In this paper, I will follow this same approach. I will first describe the critical mass formula that was used between 1943 and 1945, and show how calculations of critical masses changed as the Los Alamos scientists established increasingly-accurate experimental results.

  \section{Critical mass calculations}

Soon after Frisch and Peierls' secret March 1940 memorandum\cite{Frisch:1940} that predicted the $^{235}$U critical mass to be 0.6 kg, scientists in Britain and the US realized it was
likely a substantial under-prediction. The subsequent fission cross section measurements at Birmingham, and then Liverpool - even though they used natural uranium targets - were able to infer the $^{235}$U fission cross section more accurately, as we describe later. This led the British MAUD Committee\cite{MAUD:1941}, in the summer of 1941, to provide an updated assessment  of the critical mass  in the range 9 kg (``most likely") to 43 kg (``pessimistic''). As we will see, this range proved to be reasonable with MAUD's pessimistic value being close to
the true value of 46 kg for pure $^{235}$U (the critical mass would be higher still for highly-enriched uranium versus pure $^{235}$U).

\begin{figure*}[htbp]
\begin{center}
\includegraphics[width=6.25in]{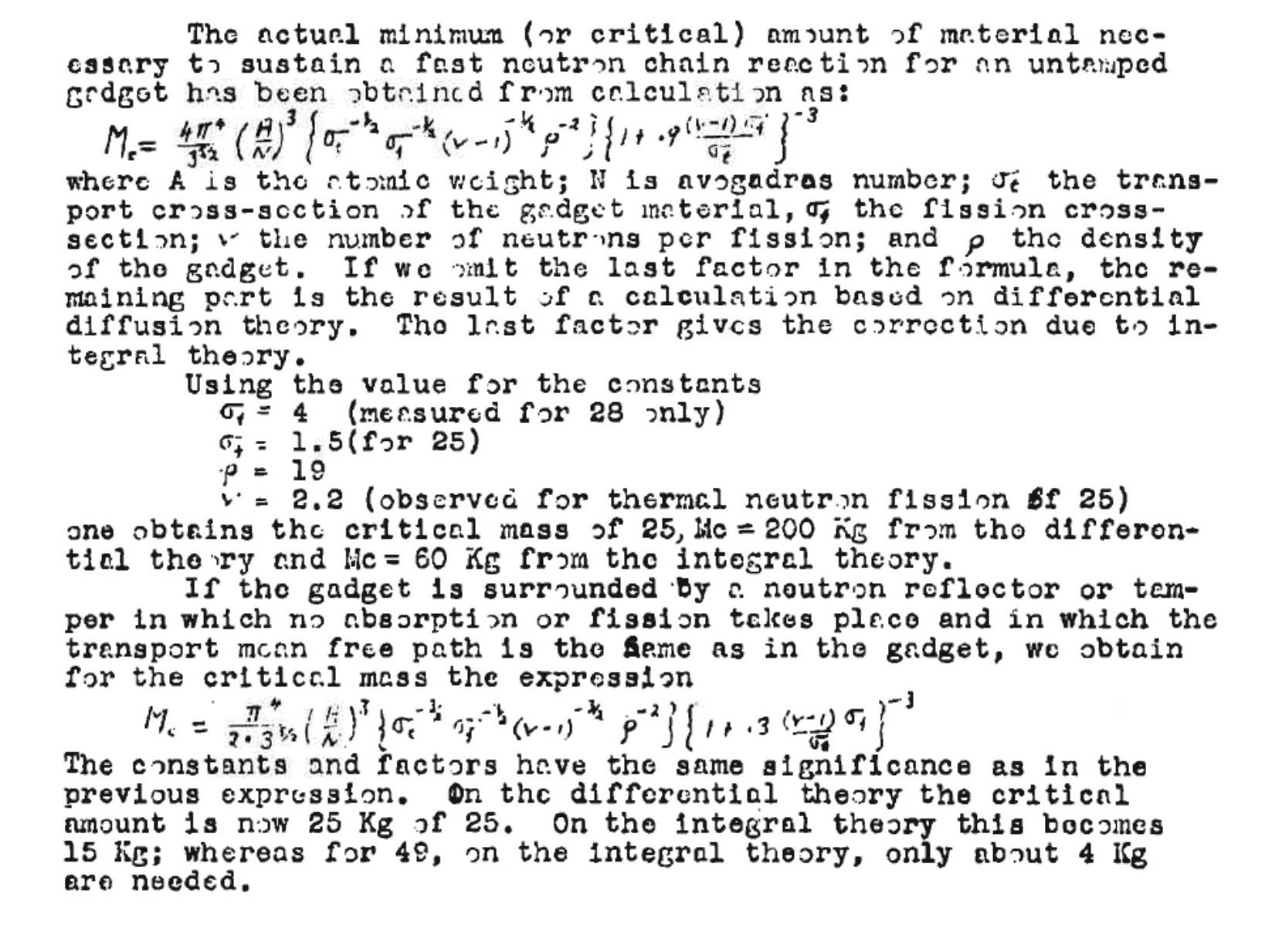}
\caption{Oppenheimer's summary of critical mass calculations from Los Alamos' first conference, April 1943, LA-2. Because the image is of poor quality, 
the first equation appears to have typos, with the parameters $\sigma_t,\sigma_f, ({\overline{\nu}} -1)$ raised  to the power $-1/2$. A careful study though shows that they are correctly raised to the power $-3/2$, as in Eq.~(\ref{eq:m-crit}).}
\label{fig:oppy-cm}
\end{center}
\end{figure*}

In the US in November 1941, Compton's National Academy of Sciences (NAS) committee did an assessment and estimated the $^{235}$U critical mass to be in the range 
2--100 kg\cite{Reed:2007,NAS:1941}. Subsequently, at Compton's request, in February 1942 Breit and Oppenheimer were asked to study the problem. They came to a result of 5 kg, a value
not far from the MAUD Committee's ``most likely'' value of 9 kg (Ref.\cite{Hoddeson:1993}, p.27). By the beginning of Project Y's work at Los Alamos, in March 1943, more work had been done; other scientists such as Serber had been engaged, and at that stage the estimate of the bare $^{235}$U critical mass was about 60 kg (see Oppenheimer's summary in Fig.~\ref{fig:oppy-cm}, as well as Serber's Primer \cite{Serber:1992} Los Alamos report LA-1 describing how the use of more exact diffusion theory reduces the calculated critical mass estimate from 200 kg to 60 kg). Also, Richard Tolman wrote\cite{Tolman:1943} a useful ``Memorandum on Los Alamos Project as of March 1943'', summarizing many aspects of the current state of knowledge of nuclear and material science,  where he assessed that the bare critical mass for 
$^{235}$U is 30$\pm$15\,kg, while that for $^{239}$Pu is 10\,kg (5--20\,kg range), estimates that would prove to be rather good.

 It was always realized that 
the greatest challenge for the Manhattan Project was the production of adequate amounts of fissionable nuclear materials. In comparison, the design of the 
bomb was somewhat easier, certainly for the uranium gun bomb (Little Boy), although once it was realized that this same gun approach for plutonium (Thin Man) would not work owing
to the high spontaneous fission rate of the contaminant $^{240}$Pu, the design of an implosion plutonium bomb (Fat Man) would prove to have substantial technical challenges, as described by
Chadwick and Chadwick\cite{Chadwick:2020a}. For this reason, an accurate theoretical  prediction of the critical masses was one of the highest priorities at Los Alamos. Because it was realized that substantial amounts of  $^{235}$U and $^{239}$Pu would not be sent to Los Alamos for a year or two (enabling a direct integral measurements of the critical mass), Project Y focused on theoretical predictions of bare and tamped (reflected) critical masses and in determining the underlying nuclear cross section data needed for these 
predictions.

Beginning with Peierls'  1939 work to derive the critical mass of an unmoderated fast critical assembly, simple analytic equations were used to calculate the fast (unmoderated) critical mass for $^{235}$U and $^{239}$Pu. The historical evolution of such formulae will not be in the scope of this paper; instead, I present the formula used throughout the Manhattan Project. The rest of the paper follows the Los Alamos scientists approach and describes the research advances that they made to determine the fundamental nuclear constants that go into this formula. 

\begin{figure*}[htbp]
\begin{center}
\includegraphics[width=6.25in]{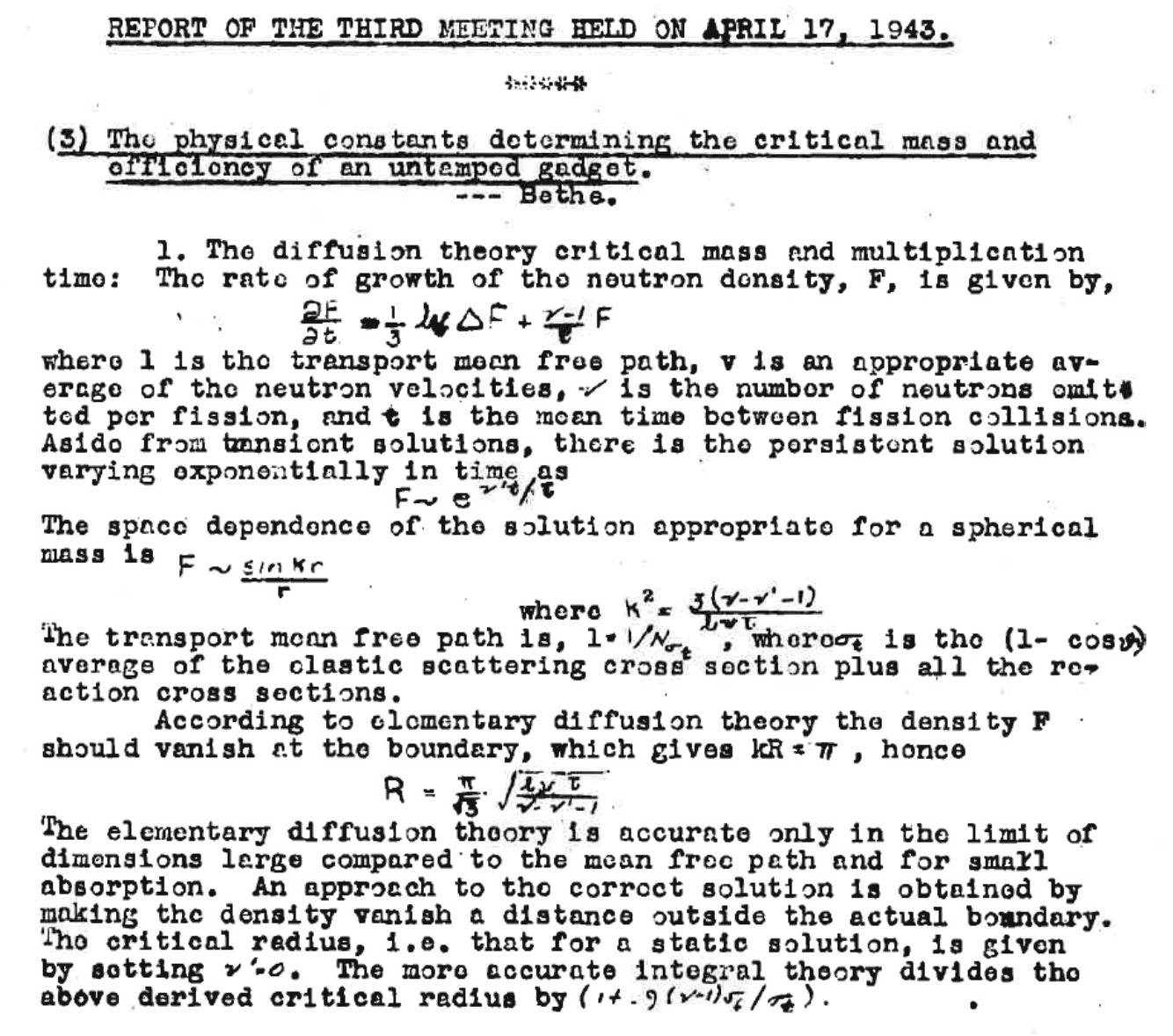}
\caption{Bethe's summary of critical mass calculations from Los Alamos' first conference, April 1943, LA-2\cite{Oppenheimer:1943}.}
\label{fig:bethe-cm}
\end{center}
\end{figure*}

The bare critical mass $M_C$ (see  Fig.~\ref{fig:oppy-cm}) was taken as:

$$
M_C={{4\pi^4}\over{3^{5/2}}}\left({A\over{N_A}}\right)^3 \left[\sigma_T^{-1/2}\sigma_F^{-1/2}( {\overline{\nu}} -1)^{-1/2}\right]^3\left(1\over{\rho}\right)^2\\
$$
\begin{equation}
 ~~~~~~~~~~~~~~\times \left(1+0.9 ({\overline{\nu}} -1)\sigma_F/\sigma_T\right)^{-3}
 \label{eq:m-crit}
\end{equation}

where $\rho$ is the density in $g/cm^3$, $\sigma_F$ is the fission cross section and ${\overline{\nu}}$ the prompt fission neutron multiplicity, and $\sigma_T$ is the transport cross section representing the total cross section for all processes except those scatterings leading to neutrons continuing in the forward direction, and 
$N_A$ is Avogadro's constant. The nuclear data values are adopted for just one energy group, relevant for fast energies near an MeV. The lower part of this equation contains a correction factor 
derived from ``more accurate integral theory", see the bottom of Bethe's notes in Fig.~\ref{fig:bethe-cm}. An image showing Oppenheimer's 
summary of the state of affairs with regard to critical masses is given in Fig.~\ref{fig:oppy-cm}, which is from his opening paper
 in LA-2\cite{Oppenheimer:1943} at the ``Los Alamos Conference, April 15, 1943", the first meeting at Project Y, which must have been one of the 
 most stimulating physics conferences ever held!

Bethe's summary\cite{Oppenheimer:1943} from LA-2 is shown in Fig.~\ref{fig:bethe-cm}, where he outlines the neutron diffusion theory that is used to calculate the critical radius,

\begin{equation}
\label{eq:r-crit}
r_C={\pi\over{\sqrt{3}}}\sqrt{ 1\over{( {\overline{\nu}} -1) .{n\sigma_T}.{n\sigma_F})} }\times f
\end{equation}
where $n$ is the atom number density, $\sigma_F$ is the fission cross section and ${\overline{\nu}}$ the prompt fission neutron multiplicity, and $\sigma_T$ is the transport cross section representing the total cross section for all processes except those scatterings leading to neutrons continuing in the forward direction.
$f$ is a correction factor derived from ``more accurate integral theory" (see Bethe's summary, Fig.~\ref{fig:bethe-cm}), 

\begin{equation}
\label{eq:f-crit}
f={1\over{(1+0.9 ({\overline{\nu}} -1)\sigma_F/\sigma_T)}}
\end{equation}

This critical radius, when used for a solid sphere, provides the critical mass shown in Eq.~(\ref{eq:m-crit}) using $M_C={4/3\pi}r_C^3\rho$. 
Given its simplicity the formula is remarkably accurate. This can be seen by comparing  the critical mass calculated with this formula with that obtained using our most sophisticated Monte Carlo neutron transport code today, MCNP6$^{\tiny\textregistered}$\footnote{MCNP$^{\tiny\textregistered}$ and Monte Carlo N-Particle$^{\tiny\textregistered}$ are registered trademarks owned by Triad National Security, LLC, manager and operator of Los Alamos National Laboratory. Any third party use of such registered marks should be properly attributed to Triad National Security, LLC, including the use of the ``$^{\tiny\textregistered}$'' trademark symbol, as appropriate. Any questions regarding licensing, proper use, and/or proper attribution of Triad National Security, LLC marks should be directed to trademarks@lanl.gov.} with ENDF/B-VIII.0. For the formula, ENDF/B-VIII.0 parameters are taken for a typical average neutron energy, 1.5 MeV. \footnote{Jen Alwin calculates the average energy causing fission in a critical sphere of $^{235}$U to be 1.48 MeV. For Jezebel, a sphere of plutonium, the average neutron energy causing fission is 1.88 MeV \cite{Chadwick:2010}} This comparison is shown in the last row of Table~\ref{table:u5critmass} where the exact MCNP6$^{\tiny\textregistered}$ result is 46.36 kg but the formula~Eq.\ref{eq:m-crit} gives 55 kg. The critical mass formula over-prediction is  approximately 20\%, but the critical radius over-prediction is just 6\%, rather impressive for so simple an equation that has just one energy group, {\it i.e.} it considers  relevant nuclear constants at just one average neutron energy (and as seen in Fig.~\ref{fig:u5fiss-fast}, the $^{235}$U fission cross section is fairly constant in the fast range, unlike the $^{238}$U cross section). By late 1944, such simple formulae were starting to be replaced by advances in neutronics simulations that 
were enabled by the first IBM punched card accounting 
machines\cite{Archer:2020a,Archer:2020b,Lewis:2020}
 to include multi-group treatments, inelastic scattering effects, and polynomial representations of angular scattering effects.

Derivations of these formulae are not provided here (see Serber's Primer\cite{Serber:1992} pp. 25-27 for a useful discussion); instead I will give the reader a feel for the physical basis for the critical radius Eq.~\ref{eq:r-crit}. One might expect that the critical radius would be proportional to the mean free path before a fission collision, $l_F^{MFP}$ and inversely proportional to the neutron production produced in that collision, 
(${\overline{\nu}}$-1), that is, proportional to  $l_F^{MFP}$/(${\overline{\nu}}$-1)=1/(${\overline{\nu}}$-1).$n.\sigma_F$, $n$ being the atom number density. But we might also expect the critical radius to be proportional to the mean free path for all collisions since this includes scattering  the neutrons and limits their ability to stream out and escape,
i.e. $l_T^{MFP}$=1/($n.\sigma_T$). In fact, neutron diffusion theory finds that the critical radius $r_C$ is proportional to both these quantities as their geometric mean (the  
square-root of their product), with a pre-factor of $\pi/\sqrt{3}$, which is Eq.~(\ref{eq:r-crit}), and the result Bethe shows in Fig.~\ref{fig:bethe-cm} (noting that $v\tau$ in Bethe's equation, the product of the neutron velocity and the time between fission collisions, is  $v\tau$=1/$n\sigma_F$). As a point of reference, for a critical sphere of $^{235}$U with mass 46\,kg, the critical radius is 8.8\,cm, whereas the fission mean free path is $l_F^{MFP}$=17\,cm, and the transport mean free path is  $l_T^{MFP}$= 4\,cm.

\begin{figure}[htbp]
\begin{center}
\includegraphics[width=3.25in]{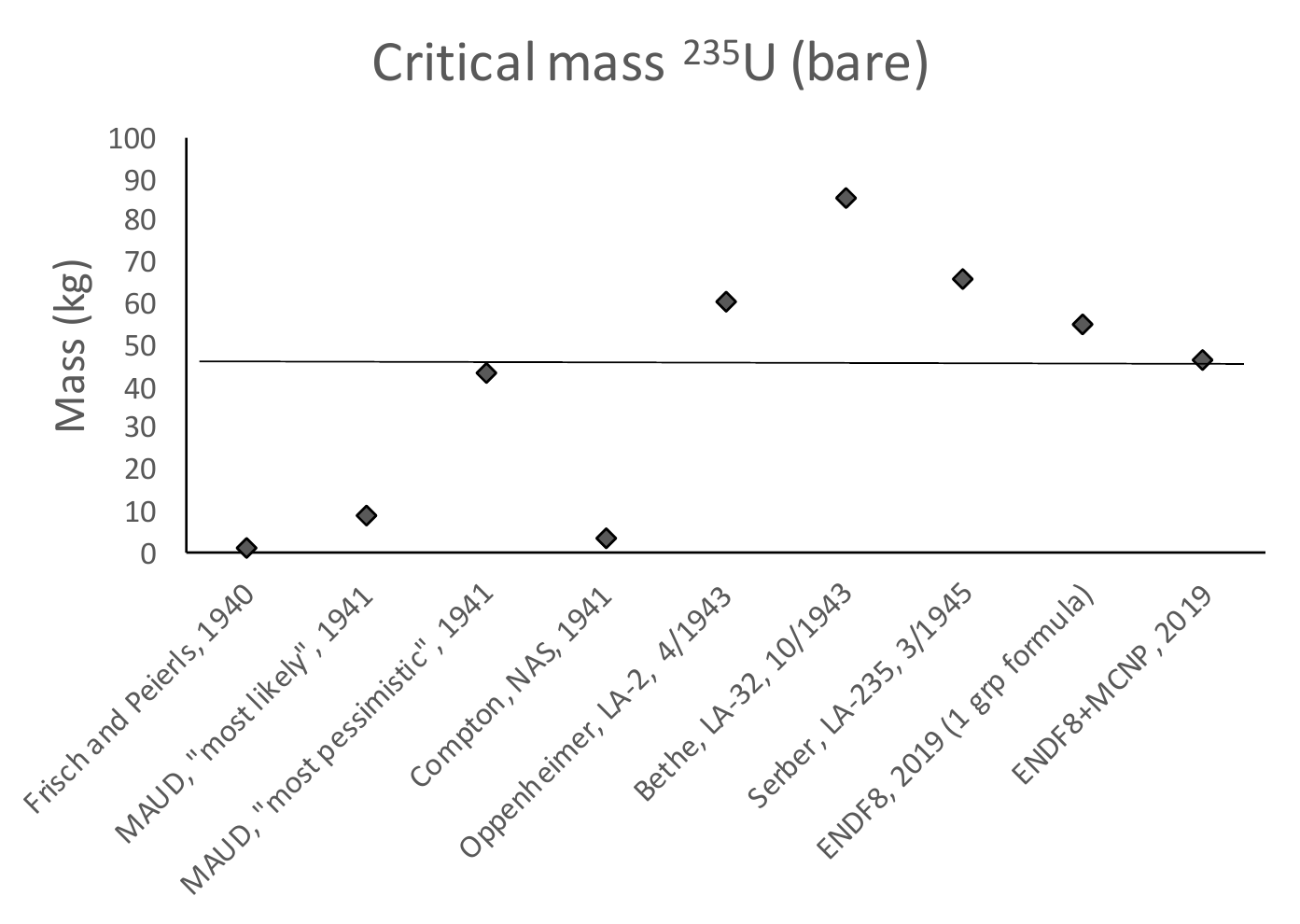}
\includegraphics[width=3.25in]{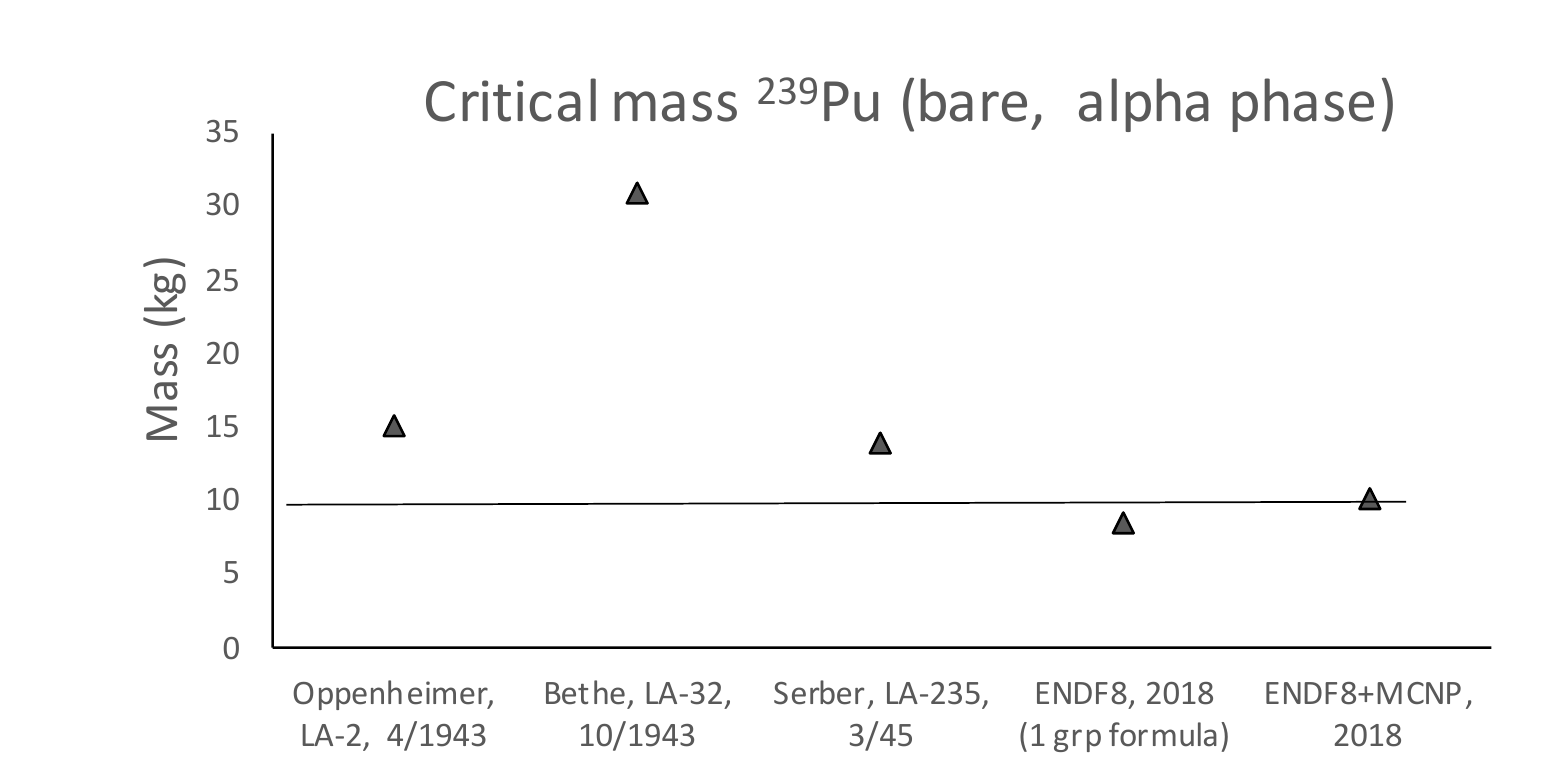}
\caption{Bare critical mass assessments changing with time, for $^{235}$U (upper panel) and $^{239}$Pu (lower panel).   The horizontal lines 
represent our best values today, from MCNP6$^{\tiny\textregistered}$  simulations using ENDF/B-VIII.0. Later in Fig.~\ref{fig:u5critmassUQ} the 
$^{235}$U figure is repeated on a logarithmic scale, with uncertainties added.}
\label{fig:critmass}
\end{center}
\end{figure}

Using this formula, critical masses are calculated using different assumptions regarding the nuclear constants, in Table~\ref{table:u5critmass} for $^{235}$U and 
Table~\ref{table:pu9critmass} for $^{239}$Pu, assuming idealized spheres of the single isotope. Figure~\ref{fig:critmass} shows these bare critical masses. The tables show the constants used, and the last column provides the 
critical mass  value calculated and, in parenthesis, the value reported in the original reference. (Note that in all rows except the last, the values in the last column 
are from the analytic equation; in the last row for ENDF/B-VIII.0, the reported value in parenthesis comes instead from a full high-fidelity MCNP$^{\tiny\textregistered}$ simulation.)
 In some cases there is very good agreement between the value I calculate from Eq.~\ref{eq:m-crit}  and that reported by the original authors,
indicating, for example, that Bethe in  LA-32 or the British in MAUD (1941) faithfully computed what they intended to compute. In other cases, for example Oppenheimer's result in Table~\ref{table:u5critmass} (60 kg in LA-2, see Figure~\ref{fig:oppy-cm})  there is a lack of consistency, for the nuclear parameters Oppenheimer lists in Figure~\ref{fig:oppy-cm} result in 77 kg for $^{235}$U, not the 60 kg he quotes. Perhaps an example of Oppenheimer's sloppiness when it came to math errors?\footnote{For all of Oppenheimer's brilliance as a scientist and a leader, he had a reputation for making sloppy math errors. Serber 
said\cite{Serber:1992} {\it ``When Dirac published -- in the 
Proceedings of the Royal Society, say -- it was all elegantly written, all the formulas carefully composed, everything just right. Oppenheimer's stuff would 
come out in a little letter in the  Physical Review, and some part might be off by a factor of $\pi$ or something like that. The little things might not be quite
right. But as far as the essentials went, Oppenheimer did some of the important things first."; In the 1941 NAS report\cite{NAS:1941}, Compton wrote ``Oppenheimer (probably through neglect of some factor considered by the others -- details of his calculation have not been submitted) finds efficiencies about ten times larger''! }}

\begin{table*}
\centering
\caption{$^{235}$U bare critical mass calculated with the listed nuclear data ``constants" using Eq.(1). In the last column, the critical mass in parenthesis is the one reported in the original reference. The last row
shows our best values today (ENDF/B-VIII.0) at 1.5 MeV neutron energy, the critical mass in parenthesis here coming from an MCNP6$^{\tiny\textregistered}$ calculation.}
\begin{tabular}{lllllll}
\hline
Author, & $\sigma_F$ &  {$\bar{\nu}$} & $\sigma_T$  & $\rho$ & Crit. Mass \\
 Date     & b & & b& g/cm$^3$ &  $^{235}$U (kg)   \\
\hline
Frisch, '40 & 10.& 2.3& 10.&  15. &  0.44(0.6) \\
Peierls\cite{Frisch:1940}  & & & & & & \\ \hline
MAUD'41 & 2. & 3. & 5.  & 19.6 & 8.5(9.) \\
likely\cite{MAUD:1941}  & & & & & & \\ \hline
MAUD'41 & 1.5 & 2.5 & 3.5  & 19.6 & 44.5(43.) \\
pessimistic\cite{MAUD:1941} & & & & & & \\ \hline
Compton case {\it a}& 3. & 3. & 9.  & 18.6 & 2.6(3.4) \\
'41, NAS\cite{Reed:2007,NAS:1941} & & & & & & \\ \hline
Oppenh. & 1.5 & 2.2 & 4.  & 19. & 77.(60.)\\
'43, LA-2\cite{Oppenheimer:1943}  & & & & & & \\ \hline
Bethe & 1.6 & 2. & 5.  & 19. & 85.(83.)\\
'43, LA-32\cite{Bethe:1943} & & & & & & \\ \hline
Serber\cite{Serber:1945},'45 & 1.34 & 2.4 & 4.7  & 18.7 & 66. (--)\\
LA-235 & & & & & & \\ \hline
ENDF\cite{Brown:2019} & 1.24 & 2.57 & 5.0  & 18.9 & 55.(46.)\\
2018 & & & & & & \\ \hline
\hline
\end{tabular}
\label{table:u5critmass}
\end{table*}

\begin{table*}
\centering
\caption{$^{239}$Pu bare alpha-phase critical mass calculated with the listed nuclear data ``constants" using Eq.(1). In the last column, the critical mass in parenthesis is the one reported in the original reference.  The last row
shows our best values today (ENDF/B-VIII.0) at 1.5 MeV neutron energy, the critical mass in parenthesis here coming from an MCNP6$^{\tiny\textregistered}$ calculation.}
\begin{tabular}{lllllll}
\hline
Author, & $\sigma_F$ &  {$\bar{\nu}$} & $\sigma_T$  & $\rho$ & Crit. Mass \\
 Date     & b & & b& g/cm$^3$ &  $^{235}$U (kg)   \\
\hline
Oppenh.\cite{Oppenheimer:1943}  & 3. & 2.2 & 4.  & 19. & 13.(15.)\\
LA-2, '43 & & & & & & \\ \hline
Bethe\cite{Bethe:1943} & 2.6 & 2. & 5.  & 19. & 29.(31.)\\
LA-32, '43 & & & & & & \\ \hline
Serber\cite{Serber:1945} & 1.85 & 2.8 & 4.8  & 19.4 & 14.(--) \\
LA-235, '45 & & & & & & \\ \hline
ENDF\cite{Brown:2019}  & 1.93 & 3.09 & 5.29 & 19.6 & 8.6(10.2)\\
2018 & & & & & & \\ \hline
\end{tabular}
\label{table:pu9critmass}
\end{table*}

For brevity this paper describes just bare critical mass comparisons, but the focus at Los Alamos in 1943 quickly became tamped (reflected) critical masses 
(as described in Hutchinson's paper\cite{Hutchinson:2020a} in this issue). 
Tamped critical assemblies typically require a little less than half the amount of $^{235}$U or $^{239}$Pu compared to the bare assemblies, and therefore 
they use the available precious special nuclear material more efficiently. Indeed, this is the reason that 
Tables~\ref{table:u5critmass}, \ref{table:pu9critmass} (far-right column) do not show a reported critical mass from Serber and Rarika's 1945 paper, LA--235 \cite{Serber:1945} -- the focus then was entirely on tamped assemblies. 
A formula for the tamped mass is also shown in Oppenheimer's paper in Fig.~\ref{fig:oppy-cm}. 
In that Figure, we see that in April 1943 Oppenheimer estimated a tamped critical mass for $^{239}$Pu of 4 kg; today with our best simulation tools, 
MCNP6$^{\tiny\textregistered}$ with ENDF/B-VIII.0 data, we find the same result, 4.0 kg of $\alpha$-phase plutonium,\footnote{
Pure plutonium is in the alpha phase at room temperature; this is distinguished from its delta-phase when alloyed with gallium\cite{Martz:2020}, which has a lower density and larger critical mass, {\it e.g.} as used in the Jezebel critical assembly.}
for an infinite $^{238}$U tamper. As was the case for $^{235}$U, we will show that again, 
this agreement was a lucky outcome resulting from canceling errors between fission $\sigma_F$  and ${\bar{\nu}}$ assessments.

The rest of this paper is devoted to discussing the developing  understanding of the nuclear cross sections used in Tables~\ref{table:u5critmass}, \ref{table:pu9critmass}, based on the research programs 
in Britain and the USA through 1945.

\section{Evaluations of Cross Section Data}
   The concept of periodically documenting our best understanding of nuclear cross sections began at Los Alamos with Bethe\cite{Bethe:1943b}, setting the stage for a series of documents with the name ``Los Alamos Handbook of Nuclear Physics''. After the war, this scholarly approach was continued 
   at Los Alamos ({\it e.g.} Hansen and Roach\cite{Hansen:1961}),  and was adopted by other laboratories, 
   {\it e.g.} Goldsmith at Brookhaven Laboratory\cite{Goldsmith:1947}, leading to Brookhaven's well-known BNL-325 documents 
   \cite{NSR1952HUZZ,NSR1955HUZY}
   on neutron cross sections. It was later extended by
   the international nuclear science communities who created the Evaluated Nuclear Data Files (ENDF) databases\cite{Holden:2001} in the US, JEFF\cite{Plompen:2020} in Europe, BROND in Russia, JENDL in Japan, and CENDL in China. The ENDF databases have been documented in detail in a series of publications in Elsevier's Nuclear Data Sheets\cite{Chadwick:2006,Chadwick:2011,Brown:2019}.
   
   The term ``evaluated data'' describes the process by which subject matter experts have reviewed available information from measurements, together with physics insights from theory and modeling, to determine “best values” that are recommended. Theory and modeling often play a role in extending the measured data to regions of energy or scattering angle that have not been measured. Such evaluated data can then be used in neutron transport codes for applied nuclear technology calculations.

\begin{table}
\caption{Evaluations of cross sections 1941-1945. Handbook refers to issues of the {\it Los Alamos Handbook of Nuclear Physics. The ``LA-" references denote work done at Los Alamos.}}
\centering
\begin{tabular}{lll}
\hline
Author & Date & Reference \\
\hline
Chadwick, &  & \\ 
Thomson & 1941 & MAUD, UK\\ \hline
Oppenheimer,  &  & \\
Bethe &  1943 & LA-2\\ \hline
Bethe, &  &   \\
Christy & 1943 & LA-11 Handbook \\ \hline
Bretscher, Davis, &            &  \\
Inglis, Weisskopf & 9/1944 & LA-140 Handboook \\ \hline
Inglis, Weisskopf & 3/1945 & LA-140A Handboook \\ \hline
Serber & 3/1945 & LA-235\\ \hline
Wilson & 1947 & LA-1009\\
& & Nuclear Physics\\
\hline
\end{tabular}
\label{table:evaluations}
\end{table}

  Table~\ref{table:evaluations} lists the main review documents written during the project, especially the three versions of the Los Alamos Handbook of Nuclear Physics. The earlier MAUD reference is included because it documented what was known in 1941, much of which proved to be remarkably accurate and played an important role in influencing subsequent efforts. 
  Beyond these evaluated data compilations,
there were many calculational papers written during the Manhattan Project, by Bethe,
Christy, Peierls, Serber, and others, that summarized the data they were
using to calculate critical masses and yields. 
  Table~\ref{table:evaluations}  highlights just one of those, the LA-235 article by Serber \cite{Serber:1945}, which provides a useful summary of what was being used in the Theoretical Division towards the end of the project.
  
   Robert Wilson’s comprehensive summing-up of what was learned at Los Alamos is the last entry, being written in 1947. This ``Nuclear Physics''  LA-1009 
   review\cite{Wilson:1947}  is quite remarkable in the breadth of its scope, and in describing what had been accomplished during the project under intense time deadlines.

\begin{table}
\caption{Measurements newly added to EXFOR through this work, via the IAEA Nuclear Data Section and BNL. In the references, ``CF'' reports stands for Central File Number; these reports are now held at ORNL/OSTI.}
\centering
\begin{tabular}{lll}
\hline
Author & Lab & Reaction \\
\hline
Tuve\cite{Bohr:1939} & Carnegie & $^{235,8}$U(n,f) \\ \hline
Chadwick\cite{MAUD:1941} & Liverpool & $^{235,8}$U(n,f) \\ \hline
Frisch\cite{MAUD:1941} & Birmingham & $^{235,8}$U(n,f) \\ \hline
{Chamberlain,\cite{Chamberlain:1942} }& Berkeley & $^{235,8}$U(n,f) \\
Kennedy, Segr{\`e}& Berkeley &  \\ \hline
{Heydenburg\cite{Heydenburg:1942}} & Carnegie & $^{235,8}$U(n,f) \\ \hline
{Hanson\cite{Hanson:1943}} & Wisconsin &  $^{235,8}$U(n,f) \\ \hline
{Benedict\cite{Benedict:1943}} & Wisconsin &  $^{235,8}$U(n,f) \\ \hline
{Williams\cite{Williams:1943}} & Wisconsin &  $^1$H(n,p)\\ \hline
{Bloch\cite{McMillan:1943}} & Stanford &  $^{235}$U(n,f) PFNS \\ \hline
{Bretscher\cite{Bretscher:1944}} & Cambridge &  $^{235}$U(n,f) \\ \hline
{Chadwick,\cite{Chadwick:1944}} & Cambridge &  $^{235}$U(n,f) \\ 
 Kinsey& /Liverpool &   \\ \hline
 {Wiegland,Segr{\`e}\cite{Wiegland:1943}} & Los Alamos & $^{239}$Pu/$^{235}$U(n,f) \\ \hline
 {Koontz, Hall\cite{Koontz:1944}} & Los Alamos &  $^{235}$U(n,f) \\ \hline
 {Williams\cite{Williams:1944}} & Los Alamos &  $^{239}$Pu/$^{235}$U(n,f) \\ \hline
 {Snyder,Williams\cite{Snyder:1944}} & Los Alamos & $^{239}$Pu/$^{235}$U(n,{$\overline{\nu}$)} \\ \hline
{Wilson\cite{Wilson:1944} }& Los Alamos & $^{239}$Pu/$^{235}$U(n,{$\overline{\nu}$)}\\ \hline
\end{tabular}
\label{table:exfor}
\end{table}

The present work  is making a number of historically-important measurements available to the broader nuclear science and technology community, via the internationally-maintained EXFOR database, through collaboration with the IAEA’s Nuclear Data Section and BNL. These publications are listed in Table~\ref{table:exfor}.

Below, we summarize the early work in Britain and the US at Berkeley, Carnegie (Washington DC), Wisconsin, Chicago, and Rice before the Manhattan Project began, and what was subsequently accomplished at Los Alamos by 1945. We compare the advances made with our best understanding of these data today, as embodied in our latest ENDF/B-VIII.0 evaluated cross section database.

\section{Fission Cross Sections}

In the 1930s and early 1940s, fast cross sections had been measured in three neutron energy ranges, requiring different sources. Photoneutrons from 0.1 -- 1 MeV  were used from radioactive 
gamma sources ({\it e.g.} Y-Be with radioactive $^{88}$Y made in a cyclotron from the (p,n) reaction on Sr; and a Ra-Be photoneutron source\cite{Wattenburg:1947}) allowing $^{235}$U fission to be measured without interference from $^{238}$U; between 2 -- 3 MeV, D$(d,n)$ neutrons were made from low-voltage accelerators; and around 4 MeV, broad-energy source neutrons were
made from Ra-Be sources. For the higher ($>$2 MeV) source neutron energies, fission on natural uranium targets was dominated by $^{238}$U fission. These methods, and the subsequent development of monoenergetic sources ({\it e.g.} via Li$(p,n)$ reactions with 
accurately-controlled accelerator voltages) were reviewed by Goldsmith\cite{Goldsmith:1947} in 1947.

\subsection{Early Work in Britain}
The MAUD committee's deliberations, including their study of the Frisch-Peierls memorandum from March 1940, led to urgent efforts to measure 
the fission cross section. Before moving to Liverpool in August, Frisch and Titterton set up fission measurements on natural uranium using a Ra--Be photoneutron source at Birmingham. This was restricted to obtaining $^{235}$U fission data for neutron energies 0.2-1 MeV that were below the $^{238}$U fission threshold. They found a $^{235}$U fission cross section of 2.3\,b that showed that the previous 10\,b estimate was too high 
(Ref.\cite{Peierls:1990?} p.131)\cite{Peierls:1945}. Afterwards in 1945 Chadwick wrote that it was “The first reasonable measurement, but still very rough” in handwritten marginal notes in Peierls' 1945 summary which is being made available by Moore in this Issue\cite{Peierls:1945} . 
From Bretscher's papers at Churchill College Cambridge, it is evident that in late 1940 he also made a measurement using the  high tension ``H.T. set''  (high voltage) accelerator at Cambridge\cite{Bretscher:1940D37b}, again with a natural uranium source, using neutrons 
from the d+C reaction. By combining his data with the aforementioned  Princeton measurements by Ladenburg {\it et al.}, he was able to conclude that for neutrons produced near 1.8 MeV energy, the fission cross section of $^{235}$U had to be less than 2.8\,b. (We now know this is indeed the case; its value is about half of this). Bretscher concluded by saying  ``The only way to get a better value for this cross section seems now to enrich U in U(235).''\cite{Bretscher:1940D37b}

In the Autumn of 1940, Chadwick, Frisch and collaborators  (Ref.~\cite{Brown:?}, p. 203) used the Liverpool cyclotron with unenriched natural uranium targets with a Li(p,n) source reaction to measure the cross section below 1 MeV (to a factor of 2 or better). Peierls stated that they showed ``Frisch’s previous figure as somewhat high and also that the cross section decreased with increasing energy'' \cite{Peierls:1945}. These Liverpool and Birmingham values were reproduced in the MAUD report, which stated: “The Liverpool measurements give values for the fission cross section of $^{235}$U varying from 2.1E-24 cm2 for neutrons of about 0.35 MeV to 1.5 E-24cm2 for neutrons of about 0.8 MeV.” (versus our best values today of 1.2\,b and 1.1\,b, respectively). Then MAUD says “Another measurement of this quantity has been obtained by Dr. Frisch using a mixed beam of neutrons comprising energies from 0.2 MeV to nearly 1 MeV. The value obtained was 2.3E-24cm2” (this was the  aforementioned 1940 Birmingham measurement).

A uranium sample enriched in $^{235}$U to 15\% was sent from Lawrence’s Berkeley laboratory to Liverpool in December 1942, and enabled Bretscher to determine the $^{235}$U fission cross section above the $^{238}$U fission threshold (see below).  Brown says (Ref.~\cite{Brown:?}, p. 24) of Lawrence, “it says much about his opinion of Chadwick and the Liverpool department that he was prepared to part with the precious sample”. From the British perspective, the sample was sent just in time, for Roosevelt was convinced that a new policy be put in place to limit information flow between countries if that information could not be used to win the war. Conant’s January 1942 memorandum on areas that would be impacted included fast neutron reactions (Ref.~\cite{Brown:?}, p. 235). There were also discussions on sending a plutonium sample to Britain, but the risk of its loss in transit from enemy attack was considered to be too high.

Later in 1943, following the August 19, 1943 Quebec agreement between ``The U.K. and the U.S.A. in the Matter of Tube Alloys”, the British learned for the first time about the Los Alamos site in New Mexico where there were already hundreds of scientists working under Oppenheimer. Approximately 25 British scientists would move to Los Alamos beginning in December 1943, and consequently the British nuclear science effort was then curtailed. Chadwick led the British Mission in Los Alamos, although in practice spent most of his time in Washington, helping ensure the smooth-running of the collaboration’s administrative aspects. Peierls deputized for Chadwick and had a highly impactful role at Los Alamos as Bethe’s Theoretical Division’s T-1 group leader for implosion physics (at Los Alamos he was known as much for his hydrodynamics expertise as for his nuclear physics\cite{Chadwick:2020a}). Frisch became the leader of the Gadget Division's criticality group, G-1, and made numerous nuclear physics contributions at Los Alamos (for example, the Dragon experiment described in this issue\cite{Hutchinson:2020b}), and Peierls tells the nice story that Frisch was equally renowned for his talents on the piano, and at one musical evening in Los Alamos the remark was overheard: “This guy is wasting his time doing physics!” (Ref.\cite{Peierls:1990?}, p.133).

Other notable measurements made in England during the war are two fast $^{235}$U fission measurements made in 1944, by Bretscher\cite{Bretscher:1944}, and by Chadwick and Kinsey\cite{Chadwick:1944}, reported in Koontz’s August 1944 Los Alamos report LA-128 and Robert Wilson’s 1947 Nuclear Physics review LA-1009\cite{Wilson:1947}. These are shown below in Fig.~\ref{fig:u5fiss.1.5}  and Table~\ref{table:u5fiss-1.5} and are seen to be fairly accurate. They were made using the enriched $^{235}$U target sent from the USA, before Bretscher and Chadwick came to Los Alamos in 1944. This same LA-128 (p. 19) mentions Chadwick-Kinsey data used a three-counter method to measure the flux.

 In this paper's Introduction, Bretscher, Feather, and Rotblat's early1940 insights into the high potential of $^{239}$Pu were noted. 
 Here I  transcribe Bretscher's progress report from December 1940:\cite{Bretscher:1940D37}
 
 {\it
``Report on Work Carried out in Cambridge, September - December, 1940'', from Churchill College Cambridge's Archive Center who hold 
papers of The Royal Commission on Historical Manuscripts, Egon Bretscher, CBE, Tube Alloys, D.37 (1940): 
Two distinct lines of research have been followed, (i) by Halban’s and Kowarski’s team on the conditions necessary for the realization of a divergent chain of fissions produced by slow neutrons, and (ii) by Bretscher and his collaborators, in the Cavendish H.T. Laboratory, on problems bearing upon the realisation of a divergent chain with fast (or medium fast) neutrons. It was assumed at the outset that only (ii) but immediate relevance to the problem of producing super-explosives, but, as the following Reports show, it now appears likely that the success which has already been achieved in (i) (one method of producing a divergent chain has been shown to the practicable – see Report 1) will enable us to approach (ii) with the greatest chance of success, also. Briefly, this comes from the following considerations. Early work was based on the expectation that separation of the rare isotope $^{235}$U$_{92}$ was the necessarily preliminary to the practical realisation of the divergent-chain super-explosive. It was considered likely that the fission properties of $^{235}$U$_{92}$ would be satisfactory in this respect – but a difficult practical problem was encountered in showing conclusively that this was the case. (Part of Report 2 deals with an attempt in this direction). On the other hand, general considerations indicate that the body $^{239}$X$_{94}$ would probably be even more satisfactory as regards fission properties than $^{235}$U$_{92}$ and the realisation of the slow neutron chain process makes possible the production of this body, even more readily than $^{235}$U$_{92}$ may be produced by isotope separation. For preliminary work on the fission properties of $^{239}$X$^{94}$, it seems that sufficient material may be obtained by irradiation of large quantities of uranium on the Cavendish H.T. set. $^{239}$X$_{94}$ is left after two successive $\beta$ disintegrations of $^{239}$U$_{92}$, formed from the abundant uranium isotope $^{238}$U$_{92}$ by slow neutron capture. The advantage of depending upon a reaction of the abundant isotope (99.3\%) does not require elaboration.} 
	
	Note that Report II is by Bretscher (page 1), Report I by Halban and Kowarski (page 9), both in document  D.37.
 James Chadwick pointed out that it was discussed still earlier, in his  laboratory at Liverpool. After Peierls' wrote\cite{Moore:2021a}{\it ``In the late summer of 1940, when publications from Berkeley on the discovery of plutonium were received, the suggestion was made that this could be produced in quantity by means of a slow-neutron reaction, and that it was likely to be suitable for a military weapon. This suggestion was presented to the M.A.U.D. committee by the group at Cambridge where meanwhile Bretscher had begun some experiments on fast neutrons at Chadwick’s request, but I have no first-hand information on how exactly it originated,"}
 Chadwick adds as a marginal hand-written note (transcribed as a footnote in Ref.~\cite{Moore:2021a}: {\it ``First suggestion that I know of was made by Rotblat about June-Early July 1940. Common topic in my laboratory before Cambridge was brought in.''}.

\subsection{Early Work in the USA: $^{235}$U}
In a 1985 colloquium at Los Alamos Harold Agnew, Los Alamos’ third Director, gave a talk\cite{Agnew:1985} that described the beginnings of US nuclear science and his own role in the Manhattan Project. He noted that the earliest advances were occurring under Compton in Chicago, Fermi and Dunning at Columbia, and Lawrence at Berkeley. Early in 1940, Columbia physicist John R. Dunning studied the fissioning of a sample of $^{235}$U that had been separated in a mass spectrometer by Alfred O. Nier at the University of Minnesota. This work showed that $^{235}$U is the isotope responsible for slow neutron fission in uranium, as predicted by Bohr and Wheeler. The thermal fission cross section that was measured, 400-500\,b, is not far from the evaluated value today, 587.3\,$\pm$1.4\,b. The 1940 state of understanding of nuclear fission was summarized in an extensive Review of Modern Physics paper by Louis Turner\cite{Turner:1940}  (Princeton), 
though that paper had only minimal references to neutron cross section measurements.

In 1942, as the USA expanded its efforts to assess the feasibly of developing an atomic bomb, the growth of nuclear research across 
the country's universities was staggering. John Manley, at Chicago's Met Lab, served as a manager for directing and coordinating the work at nine separate institutions: Berkeley (Seaborg, Segre) , the Carnegie Institution in 
Washington (Heydenburg), Cornell (Rossi, Hollowell, Bacher), MIT,  Minnesota (Williams), Purdue, Rice (Richards), Stanford (Bloch), and Wisconsin (McKibben). Los Alamos's NSRC contains Manley's extensive correspondence\cite{Manley:1943} with these researchers.

  As the preparations began for a Manhattan Project, Oppenheimer wrote\cite{Oppenheimer:1942} on Nov. 1, 1942 to Rudolf Peierls in Britain  to summarize his present understanding and raise some differences of opinion on the underlying science and technology. He summarized the recent $^{235}$U fission cross section experiments made at the Carnegie Institution and at Berkeley:  Radioactive yttrium photons on Be making 220 keV neutrons gave 2.9\,b and radioactive Na+D  gave 1.8\,b (these Berkeley values, described below, were reported as 2.8\,b and 1.9\,b respectively) and $^{12}$C+D making  500-900\,keV neutrons gave 2\,b (note this Carnegie value soon after was reported as 1.1\,b, which we now know to be more accurate). The more accurate data from 
  Wisconsin's Van de Graaff were not yet available.

 The unit of barn, $10^{-28}$m$^2$, has its origin during those times. In Los Alamos Manuscript Series LAMS-523, ``Note on the origin of the term `barn'", Baker and Holloway describe how they were inspired by their rural origins to choose this term, and that it was first concluded that for nuclear processes ``a cross section of order $10^{-24}$cm$^2$ was really as big as a barn''. This was coined in December 1942 when they were at Purdue, working under John Manley's direction. Its first written use was in June 1943 in their Los Alamos LAMS-2 report. They came to this choice, having ruled out (for various reasons) other possible names including an ``Oppy", a ``Bethe", a ``Manley", and a ``John".

At the Carnegie Institution of Washington’s Department of Terrestrial Magnetism (DTM) \cite{Dahl:2002}, Heydenburg in late 1942 \cite{Heydenburg:1942} used a Van de Graaff neutron source from the reaction D+D and $^{12}$C + D$\rightarrow$$^{13}$N +n on targets of natural U and those enriched in $^{235}$U. The neutron energies available were limited by the maximum incident energy of the Carnegie electrostatic generator. The measurements suffered from high energy (1.8 and 5.6 MeV) neutrons from the presence of $^{13}$C which caused fission in $^{238}$U, which was present in the uranium sample. The value $\sigma_F$(25) =1.1\,b was measured for 640 keV neutrons, from the C+D reaction. But the data from these C+D and D+D neutron source  experiments was viewed as problematic at the time \cite{Wilson:1947} for the aforementioned reasons, even though their 1.1\,b measured value is now known to agree exactly with our modern ENDF/B-VIII.0 assessment. It was concluded that Li(p,n) source reactions were preferred, as used at Wisconsin and Liverpool (see further below). 

At Berkeley, Segr{\`e} and collaborators used radioactive sources to make neutrons. In 1942, Chamberlain, Kennedy and Segr{\`e}\cite{Chamberlain:1942} used quasi-monoenergetic photoneutrons from the photodisintegration of Be by yttrium radioactive-decay $\gamma$ rays, and of deuterium by $^{24}$Na gamma rays. The powerful radioactive yttrium and sodium sources were made in the Berkeley cyclotron. Segr{\`e} introduced the idea, following Fermi, of a manganese bath to measure the neutron source strength: the neutrons were thermalized and the resulting activation from the capture reaction was measured.  These experiments concluded that $\sigma_F$(25) = 2.8\,b at 160 keV and 1.7\,b for 430 keV neutrons, with uncertainties estimated as 15\%\cite{Wilson:1947}. A natural uranium sample was used so that measured values could only be inferred for incident neutron energies below the 1 MeV $^{238}$U fission threshold.

A major breakthrough was made by the University of Wisconsin nuclear physics group group in 1942, through the development of quasi-monoenergetic source neutrons from the Li(p,n)Be reaction on enriched and natural  uranium samples. The Van de Graaff electrostatic generator was capable of accelerating single charged species up to approximately 4 MeV, producing monoenergetic neutrons up to 2 MeV. Two different approaches were developed to determine the neutron flux\cite{Wilson:1947} one of which measured recoil protons, comparing against the n-p collision cross section as a primary standard, the other used the manganese bath technique. In April 1943 this led to $^{235}$U fission results of 1.66\,b by Hanson\cite{Hanson:1943} and 1.22\,b by Benedict and Hanson \cite{Benedict:1943}, at a neutron energy of 0.53 MeV, the latter result being impressively close to our best value today, 1.13\,b. The Hanson measurement was the first ever that I am aware of that made 
a fission cross cross section measurement in ratio to the n-p scattering cross section, a standard approach used today. 
The neutron energy dependence was also investigated for the first time, from 0.2 to 1.8 MeV, indicating a constant value above 600 keV, with an increasing cross section from 600 keV to the lower energies. They also showed that $^{238}$U has a threshold of fission of 1.0 MeV, as predicted by Bohr and Wheeler, and rises to 0.45 barns at 1.8 MeV.

\subsection{Work at Los Alamos: $^{235}$U}

\begin{figure*}[t]
\begin{center}
  \includegraphics[width=\textwidth]{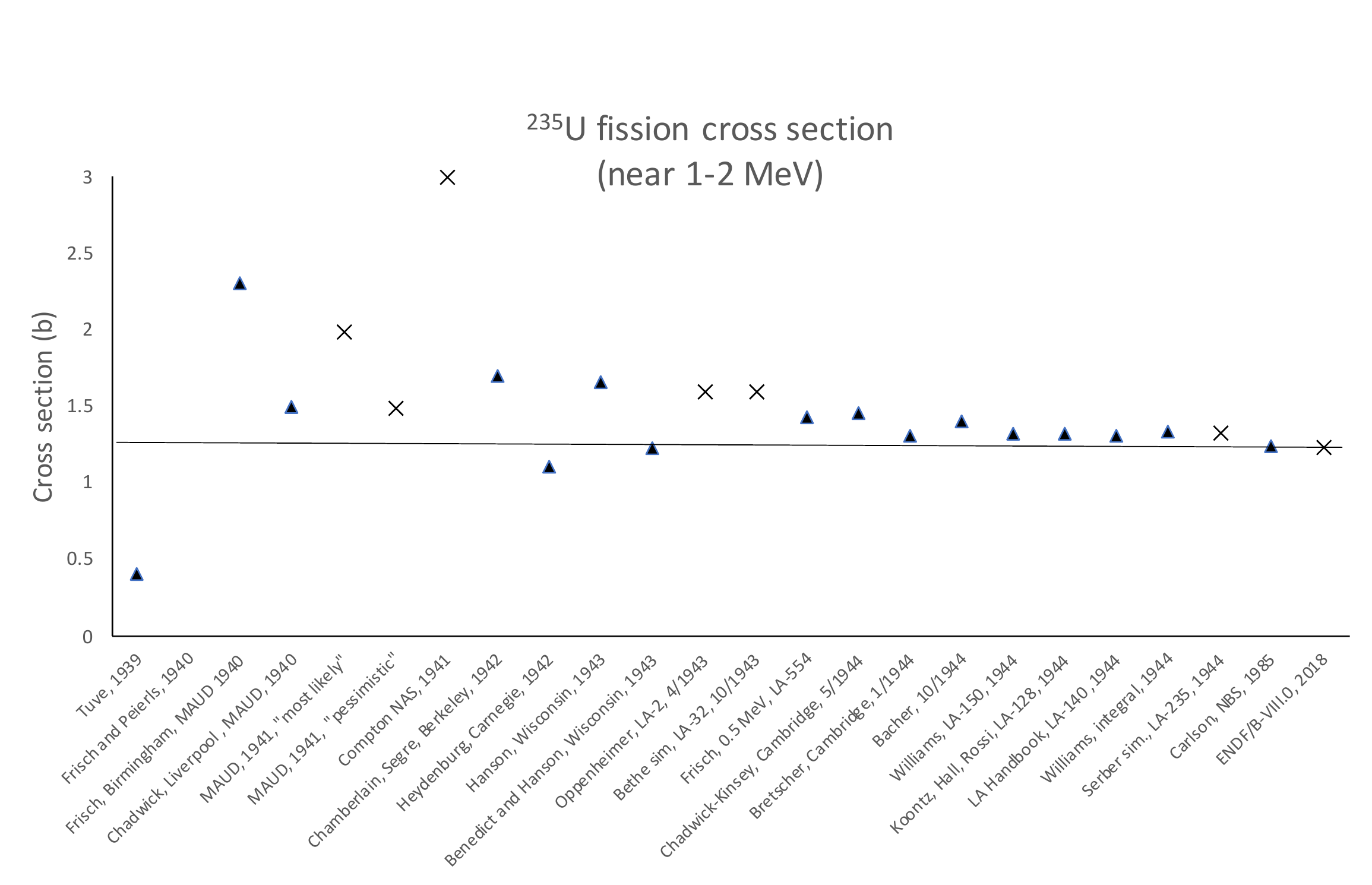}
\caption{The $^{235}$U fission cross section measurements changing with time, for a neutron energy near 1.5 MeV.  Over time the measured values
decreased; the horizontal line shows the best value today. Measurements are shown as solid symbols whereas evaluations or estimates are crosses. Frisch and Peierl's 1940 first guess of 
10\,b can't be seen on this scale. The numerical values are listed in 
Table~\ref{table:u5fiss-1.5}.}
\label{fig:u5fiss.1.5}
\end{center}
\end{figure*}

At the beginning of the Manhattan Project, the state of understanding was discussed in the March 1943 conference and summarized in the LA-2 report 
\cite{Oppenheimer:1943}, with relevant presentations by Oppenheimer, Bethe, and Manley. For the fast neutron energy region, the $^{235}$U fission cross section was thought to be about 1.6\,b, and the $^{239}$Pu cross section was thought to be about twice as large, these assessments being influenced by the recent 1942/1943 uranium measurements from Wisconsin using the Li(p,n) source reaction, \cite{Hanson:1943} and plutonium measurements from the Carnegie Institution. The 
fission neutron average multiplicity
$\bar{\nu}$ for $^{235}$U was taken as 2.2 based on a recent measurement at Chicago by Fermi (Fermi, report CP-257.p.3), and the same value was assumed for $^{239}$Pu, which had not yet been measured. It was guessed, correctly, that the $\bar{\nu}$ for fast neutrons would be just a little larger than for thermal neutrons (which could be more easily measured). It was also recognized that the existing British and US measurements of the fission neutron energy spectrum suffered from serious systematic errors. The summary remarks by Bethe, Manley, and Bacher emphasized the large uncertainties in all these data, and the urgency to more accurately measure the key nuclear constants needed for the project.

Existing accelerator equipment was brought to Los Alamos from various universities across the country. Wisconsin provided two Van de Graaff electrostatic voltage generators (2.4 and 4 million volts), allowing the creation of quasimonoenergetic neutrons from a few tenths of an MeV up to 1.8 MeV using a Li(p,n) reaction and up to 6 MeV with D+D reactions. The ``short-tank'' accelerator had been built mainly by Joseph McKibben, and most of the group there came to Los Alamos. Dick Taschek got his PhD under 
Breit at UW in 1941 and both 
Donald Benedict and Alfred Hanson finished their PhD. theses in 1943 under Ray Herb, the architect of the best Van de Graaffs \cite{Dahl:2002}.
 These Van de Graaff machines, referred to as the “Short Tank” and “Long Tank” proved to be the most useful ones at Los Alamos for precision fission experiments, in William's group. Illinois provided a 600 keV Cockroft Walton generator, creating 2.5-3 MeV neutrons with D+D reactions. Manley had worked with this machine at Illinois, and proved to be Oppenheimer’s “right hand man” in assembling all the accelerator equipment; having moved to Chicago, he put Harold Agnew in charge of moving the Illinois machine to Los Alamos and putting it in the basement of the Z building, for his group. For Robert Wilson’s group, a cyclotron was provided by Harvard that could produce protons up to 7 MeV and deuterons up to 11 MeV. Thermal neutrons were produced following a D+Be reaction with subsequent moderation in graphite, as well as neutrons with energies in the 0.001-100 eV range using time-of-flight methods. Manley describes\cite{Diven:?} the process by which ``I was the one in charge of getting all those damned machines up to Los Alamos”, and the impressive feat of getting them all operational by July 1943.

Los Alamos established the practice of measuring cross sections relative to a H(n,p) “standard”, measuring the recoil protons from a thin film of hydrogenous material in the neutron beam. This was enabled by William’s and Bailey's seminal measurement of the H(n,p) cross section.\cite{Williams:1943,Bailey:1946} The better understanding of the neutron fluence in experiments allowed $^{235}$U fission cross sections to be measured more accurately, versus earlier methods where as we discussed above discrepancies of 30\% or more were common. Once fission cross sections were accurately determined at Los Alamos, they in turn were used as standards to determine other cross sections in relative measurements. 

This method was used by Hall \footnote{Yes, the spy. See A. Carr, ``The Project Y Spies", Los Alamos report LA-UR-28986 (2014).}, Koonz and Rossi\cite{Koontz:1944} in 1944 to more accurately measure the fission cross section in a double chamber that measured fission recoils compared to proton recoils. Hall {\it et al.} used the Li(p,n) reaction with the Van de Graaff high voltage generator in building W for neutron energies up to 1.6 MeV, and a D-D source of neutrons at 2.5 MeV using the  Cockroft-Walton in building Z. They established at 1 MeV a $^{235}$U fission cross section of 1.33$\pm5\%$\,b, which served as a standard cross section at Los Alamos during the Manhattan project. Indeed, this value was reasonably accurate, differing by 10\% with the best ``standard” value today in ENDF/B-VIII.0 at 1 MeV, 1.203b$\pm1.3\%$.

   Later in October 1944, Williams\cite{Williams:1944} reported on similar measurements, from 5 keV to 2 MeV, for a range of actinides that include $^{235,8}$U, $^{239}$Pu, $^{237}$Np,… Compared to the work of Hall {\it et al.}, Williams developed ``long counters” that improved the accuracy below 400 keV and above 3 MeV, with help from Hanson who had come from Wisconsin. Williams used the Van de Graaff with a Li(p,n) source, obtaining data from 5 keV to 2 MeV. Thus, by the Fall of 1944 Williams was able to conclude that ``the general form of $\sigma$(25) as a function of neutron energy is fairly well established … and is known to between 5 and 10 percent from 200 keV to 2 MeV”. This assessment was actually pretty optimistic; we now know that at the important higher energy of 1.5 MeV Williams’ LA-150 $^{235}$U(n,f) data of 1.32\,b was fortunately only 6\% too high, but below 1 MeV his data remained  substantially high (10\%  overestimate at 1 MeV, 25\% at 500 keV, 61\% at 100 keV, 56\% at 30 keV). 
Late in 1944, the Los Alamos Handbook of Nuclear Physics\cite{Bretscher:1944a} (LA-140), 2nd Ed., by Bretscher, Davis, Inglis, Weisskopf
summarized all these data for use by Los Alamos' researchers.

After the war ended, in Nov. 1945 Bailey, Wilson {\it et al.} published \cite{Bailey:1945} updates to the LA-150 data of Williams in the region 25 – 500 keV. This work correctly argued that the previous LA-150 cross sections were too high for the lower neutron energies below 500 keV. These new results brought the fission cross section down, but the Bailey values were still high compared to our best estimates today. Likewise, David Frisch's proportional counter results in 1946 also provided 
lower fission cross section measurements that were lower and more accurate\cite{Frisch:1946}, see Fig.~\ref{fig:u5fiss-fast}.

\begin{table*}
\caption{$^{235}$U fission cross sections near 1.5 MeV ``fast" neutron energy, from measurements and {\it evaluations (italicized).}
The exact energy or ranges of energies are given in parentheses. The ``LA-" references denote work done at Los Alamos.}
\centering
\begin{tabular}{lll}
\hline
Authors & Date & $\sigma_F$  \\
\hline
Tuve, Carnegie Inst. DC \cite{Bohr:1939}  &  1939 & 0.4\,b (0.6 MeV) \\ \hline
{\it Frisch \& Peierls} \cite{Frisch:1940} & 1940 & {\it 10\,b (a bad guess)} \\ \hline
MAUD, Chadwick, Frisch  & 1940 & 2.1\,b (0.35 MeV) \\
Liverpool \cite{MAUD:1941}& & 1.5\,b (0.8 MeV) \\ \hline
MAUD, Frisch &  1940 & 2.3\,b (0.2-1 MeV) \\ 
Birmingham \cite{MAUD:1941}  &  1941 &  \\ \hline
Betscher, Cambridge \cite{Bretscher:1940D37}  &  1940 & $<$ 2.8\,b (1.8 MeV) \\ \hline
{\it MAUD ``most likely''} \cite{MAUD:1941} & 1941 & {\it 2\,b} \\ \hline
{\it MAUD ``pessimistic''} \cite{MAUD:1941}  & 1941 & {\it 1.5\,b} \\ \hline
{\it Compton, NAS} \cite{Reed:2007} & 1941 & {\it 3\,b} \\ \hline
Chamberlain, Kennedy, & 1942 & 2.8\,b\footnote{Note Wilson LA-1009 has 2.5\,b at 160 keV; uncertainties estimated as 15\%}  (220 keV) \\
Segr{\`e} \cite{Chamberlain:1942}, Berkeley & & 1.7\,b (430 keV) \\ \hline
Heydenburg, Meyer\cite{Heydenburg:1942} & 1942 & 1.1\,b (0.64 MeV \\
Carnegie Inst. & & and lower) \\ \hline
Hanson, \cite{Hanson:1943} & 3/43 & 1.66\,b \\
Wisconsin &  &  (0.5-1.8 MeV) \\ \hline
Benedict, Hanson\cite{Benedict:1943} & 3/43 &  1.22\,b  \\ 
Wisconsin & & (0.53 MeV) \\ \hline
{\it Oppenheimer\cite{Oppenheimer:1943}, LA-2} & 3/43 & {\it 1.6\,b} \\ \hline
{\it Bethe,\cite{Bethe:1943} LA-32} & 10/43 & {\it 1.6\,b} \\ \hline
Bretscher\cite{Bretscher:1944} & 1/44 & 1.31\,b (2.05 MeV) \\ 
Cambridge &  & \\ \hline
Chadwick, Kinsey\cite{Chadwick:1944} & 05/44 & 1.45\,b \\
Cambridge &  & (1.5-2 MeV) \\ \hline
Williams,\cite{Williams:1944} LA-150 (1944) & 1944 & 1.32\,b \\  \hline
Koontz, Hall\cite{Koontz:1944}, LA-128 & 1944 &1.31\,b $\pm$5\% \\ \hline
D. Frisch\cite{Frisch:1946} & 1946 & 2.1\,b (35 keV) \\ 
LA-554, LA-1009 & & 1.43\,b (0.5 keV) \\ \hline
Wilson\cite{Wilson:1947} LA-1009 & 1947 & 1.33\,b$\pm$5\% (1 MeV) \\ \hline
Carlson\cite{Carlson:19?} &  1985 & 1.24\,b$\pm$2\% \\
NBS &   &   \\ \hline
{\it ENDF/B-VIII.0\cite{Brown:2019}}& 2018 & {\it 1.24\,b}$\pm$1.3\% \\
\hline
\end{tabular}
\label{table:u5fiss-1.5}
\end{table*}

Over time the better cross section measurements led to smaller cross sections in the fast region, as can be seen in Table~\ref{table:u5fiss-1.5} and Fig.~\ref{fig:u5fiss.1.5} for the fission cross section at 1.5 MeV.  This is because the cross section increases substantially as one goes to energies below 0.5 MeV; earlier measurements were more subject to systematic error contamination processes in which the source neutrons were down-scattered by surrounding material. Dr. Fredrik Tovesson of ANL stated to me that ``I have seen this artificially increase the measured fission cross section before, both for fissile targets and non-fissile targets with fissile contaminants. Additionally, many early measurements in the fast region used quasi-monoenergetic neutron sources, while newer measurements have often been performed at time-of-flight (TOF) facilities such as LANSCE/WNR\cite{Lisowski:2006}. When using neutron TOF you are less susceptible to low-energy neutron background, which supports the hypothesis”. In Table~\ref{table:u5fiss-1.5}  a high-accuracy measurement is shown for comparison, by Dr. Allan Carlson,\cite{Carlson:19?} made at the National Bureau of Standards' linac. This measurement is absolute, with the neutron fluence determined from a well-characterized black detector and  the fission rate determined with a well-understood fission chamber.

Carlson, who worked with some of the early Los Alamos researchers,  adds a useful perspective: 
``Carl Bailey who also worked on cross section measurements with Williams said the work was very primitive by today's standards. The cross sections measured were too high because backgrounds were very high and difficult to properly measure. He noted to me that increased background means more fission detector counts thus a higher cross section. They also were not sure the neutron sources they used were truly monoenergetic. They lucked out on that.''\footnote{Carlson also notes: ``Bailey also has a $^{235}$U fission cross section that he made with Williams before the work of Williams. Both are high. Bailey was doing his PhD thesis work at LA and he mentioned to me that when he defended his thesis the panel was composed of Williams, Wilson, Fermi, Bethe and Kennedy. That could be rather intimidating!''}.

\subsection{Early Work in the USA: $^{239}$Pu}

The first measurements of plutonium fission were made for thermalized neutrons at Berkeley in 1941 (Ref.~\cite{Rhodes:19?}, pp. 355-6.), and 
reported in a Berkeley report A-33 by Seaborg, Segre, Kennedy and Lawrence. 
At thermal energies, Chamberlain {\it et al.}\cite{Chamberlain:19?} found that the fission cross section of $^{239}$Pu was 1.87 times greater than $^{235}$U, an amount that was found later by DeWire  at Los Alamos to be an overestimate owing to the neutrons not being completely thermalized (see below, and Table~\ref{table:pu9u5ratio-thermal} -- the correct value is 1.28). A January 1943 letter from Chamberlain, Kennedy, Segre and Wahl to Manley 
(NSRC A84-019-49-9) reported slow neutron
values for this ratio  of 1.238 and 1.294, much improved. 
 Later,
 Lawrence would comment that the fast fission plutonium cross sections is ten times that of (238) uranium (Ref.~\cite{Rhodes:19?}, p. 368.), while a measurement
 by Seaborg and Segr{\`e} in 1941 found a factor of 3.4  (Ref.~\cite{Hoddeson:1993}, p. 23.) While these might seem to be contradictory claims, both could be
 correct depending uxpon the exact neutron energy, owing to the fast-changing $^{238}$U fission cross section as it rises from its threshold; today we assess this ratio to be 10 at 1.4 MeV, but 3.7 at 2 MeV.

By 1942, Oppenheimer and Manley urged the measurement of fast neutron fission of plutonium.   In October, Seaborg wrote to the Carnegie Institution of Washington to say he was sending them a 10\,$\mu$g
sample of plutonium.
Heydenburg and Meyer  wrote\cite{Heydenburg:1943b} to Manley in Los Alamos, on April 9 1943, on “Comparative Fission Cross Section for Element 49 and Normal Uranium”, communicating their late 1942/1943 measurements of the $^{239}$Pu to $^{235}$U fission cross section ratio. They obtained ratios 1.64 at thermal energies, 1.58 at 650 keV, and 1.68 at 3.95 MeV, and reported an absolute $^{239}$Pu(n,f) cross section of 2.18 b at 650 keV. As seen in Tables~\ref{table:pu9u5ratio-fast} and \ref{table:pu9fiss-1.5}  these measurements are impressively accurate when compared to modern 
ENDF/B-VIII.0 values.  Later, Tashek\cite{Taschek:1943} in LA-28 quotes a Heydenburg\cite{Heydenburg:1943a}  (CF-626) pu9/u5 fission ratio of 1.76 at 650 keV, just slightly different.
In the first Los Alamos conference documented in LA-2 Manley\cite{Oppenheimer:1943} discussed the data measured  at two fast energies with results, “about two times 25” at 0.4 and 6 MeV. This appears to be a reference to the aforementioned Heydenburg measurements, though the 
energies and the ratios quoted aren't exactly the same and the ``two times" seems strangely optimistic.

\subsection{Work at Los Alamos: $^{239}$Pu}

\begin{table}
\caption{Plutonium fission ratio $^{239}$Pu/$^{235}$U measurements in the {\it fast } neutron energy region, compared with ENDF/B-VIII.0. The ``LA-" references denote work done at Los Alamos.}
\centering
\begin{tabular}{llll}
\hline
Authors, Energy & Date & Ratio & Today \\
              &          & & ENDF \\
\hline
Heydenberg,\cite{Heydenburg:1943a}  & 1943 & 1.76 & 1.45 \\ 
650 keV & & & \\ \hline
Heydenberg,\cite{Heydenburg:1943b}  & 4/43 & 1.58 & 1.45 \\ 
650 keV & & & \\ \hline
Heydenberg,\cite{Heydenburg:1943b} & 4/43 & 1.68 & 1.56 \\ 
3.95 MeV & & & \\ \hline
Wiegland, Segr{\`e}\cite{Wiegland:1943} & 8/43 & 1.14$\pm0.13$ & 1.14 \\ 
LA-21, 220 keV & & & \\ \hline
Wiegland, Segr{\`e}\cite{Wiegland:1943} & 8/43 & 1.57$\pm0.16$ & 1.55 \\ 
LA-21, ~3.6 MeV& & & \\ \hline
Taschek, Williams\cite{Taschek:1943} & 10/43 & 1.685 & 1.56 \\ 
LA-28, 1.46 MeV & & \\
\hline
\end{tabular}
\label{table:pu9u5ratio-fast}
\end{table}

As we have seen, at the start of Project Y's work the assessment was that the $^{239}$Pu fission cross section would be ``about two times 25” in the fast neutron energy range, leading 
Oppenheimer, Bethe, Manley and so on to use 3\,b in their initial critical mass calculations\cite{Oppenheimer:1943,McMillan:1943}. These same values can be seen in Serber’s Primer book\cite{Serber:1992} p.15, Fig. 1 where the plutonium (“49”) curve shows 3\,b for higher neutron energies. Williams, chairman of one of the 
conference sessions, talked of the importance of 
``absolute measurements of the fission cross sections of 25, 49, 01 [n] , 11 [p], …Someone should gather these materials''\cite{McMillan:1943}, and getting good data on 
plutonium became the priority; these were the first measurements made at Los Alamos.

\begin{figure}[htbp]
\begin{center}
  \includegraphics[width=3.25in]{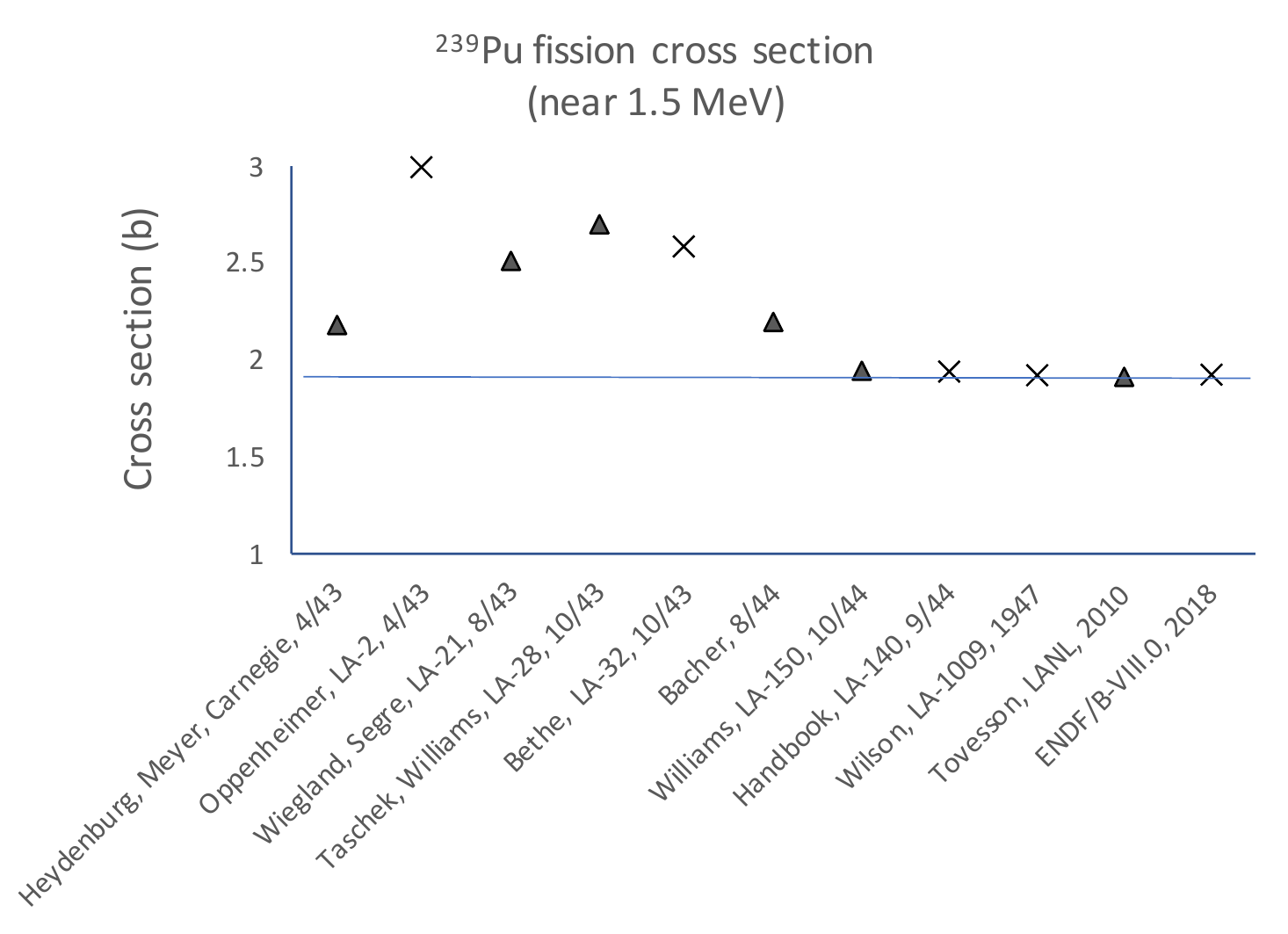}
\caption{The $^{239}$Pu fission cross section measurements changing with time, for a neutron energy near 1.5 MeV.  Over time the measured values
decreased; the horizontal line shows the best value today. Measurements are shown as solid symbols whereas evaluations or estimates are crosses. The numerical values are listed in 
Table~\ref{table:pu9fiss-1.5}. }
\label{fig:pu9fiss.1.5}
\end{center}
\end{figure}

Measurements of plutonium fission posed a challenge for the experimentalists. At first, no samples were available, and then by the summer of 1943, plutonium started arriving but only in tiny $\mu$g quantities from Chicago’s Met Lab. Also, measuring the ionization effects of a fission fragment above the alpha-particle pile-up from alpha decay is challenging. Nevertheless, once measurements began at Los Alamos by Segr{\`e} and collaborators from the summer of 1943 on, the fission ratio data obtained for “49/25”, {\it i.e.} $^{239}$Pu(n,f) in ratio to $^{235}$U(n,f), were remarkably accurate. But when multiplied by the $^{235}$U fission cross sections, which at that time were measured to be too high (see previous section), the resulting plutonium fission cross sections were also too high. The experience during the Manhattan project was, therefore, one in which the plutonium fission cross section measurements decreased during the project, from the initial assessments of 3\,b to just under 2\,b after the Fall of 1944, for fast neutrons around 1.5 MeV, see Fig.~\ref{fig:pu9fiss.1.5}.

The first measurement of plutonium fission cross sections at Los Alamos came from Segr{\`e} and Wiegand, August 31, 1943 (LA-21). Accelerator neutron-sources were not yet used; at that stage they relied on neutron sources from radioactive targets. They had been supplied a 17\,$\mu$g  sample of plutonium together with an enriched $^{235}$U sample containing 35\,$\mu$g  of 235, allowing fission ratio measurements. They found that the fission cross sections of 49 and 25 for 220 keV neutrons have a ratio of 1.14 (using an yttrium-plus-beryllium source). For higher energy radium-plus-beryllium neutrons around 3.6 MeV the fission cross sections of 28, 24 and 49 are respectively 0.32, 0.7 and 1.57 times that of 25. These ratio results are shown in Table~\ref{table:pu9u5ratio-fast} and are seen to be extremely accurate compared to ENDF/B-VIII.0 data today. The earlier 49/25 ratio results by Heydenburg {\it et al}. were  reasonably accurate too.

Two months later, in October 1943 Taschek and Williams measured\cite{Taschek:1943} the 49/25 fission ratio, but this time with the Van de Graaff, and a Li(p,n) source reaction with the 17\,$\mu$g  Pu target. This allowed them to map out the incident energy dependence of the cross section ratio, from 0.1 MeV to 1.5 MeV. Their result at 1.5 MeV incident energy was about 8\% high, and when combined with the uranium $^{235}$U fission value in use at the time (1.6\,b from Wisconsin, also high) they obtained a plutonium fission cross section at 1.5 MeV of 2.7\,b, still very much too high. 

Bethe’s Theoretical Division progress reports\cite{Bethe:1943} show how the new data were quickly adopted. By October 1943, although the $^{235}$U fission cross section was still taken to be 1.6\,b, the value at the beginning of the project that had come from Wisconsin, Bethe had reduced the $^{239}$Pu estimated fast fission cross section to 2.6\,b based on the new Segr{\`e} and the new Taschek fission ratio data, and used these updated values in his critical mass calculations. This plutonium fission cross section of 2.6\,b was considerably too high, but the problem was not in the recent Los Alamos $^{239}$Pu/$^{235}$U fission ratio data but rather in the too-high $^{235}$U  fission of 1.6\,b from Wisconsin.

As the year progressed in 1944, the accurate $^{239}$Pu/$^{235}$U fission cross section measurements combined with the more accurate $^{235}$U fission results led to decreasing estimates of the plutonium fission cross section. Bacher had to report\cite{Bacher:1944} to Oppenheimer and Los Alamos’ Governing Board  that it ``appeared certain that the Pu cross section must be revised downwards by at least 15\%, because of incorrect lifetimes assumed”. He said “The best present values are: $\sigma_F$-25 at 1 MeV$=1.4\,$b;
$\sigma_F$-49 at 1 MeV $ = 2.2\,$b;
$\bar{\nu}$-25$ = 2.4$;
$\bar{\nu}$-49$ = $2.8.''
In fact, as we will see, the plutonium fission value of 2.2\,b would still prove to be an overestimate, being about 13\% higher than best values later found of 1.94\,b, 
see Table~\ref{table:pu9fiss-1.5}.

\begin{table}
\caption{{\it Fast} $^{239}$Pu fission cross sections near 1.5 MeV neutron energy, from measurements and {\it evaluations (italicized).}
The exact energy or ranges of energies are given in parentheses. The ``LA-" references denote work done at Los Alamos. Values in parenthesis indicate that they were ratio measurements to $^{235}$U fission (that were quite accurate, see Table.~\ref{table:pu9u5ratio-fast}) but were multiplied by the  by the $^{235}$U fast fission cross section of that era ({\it e.g.} 1.6\,b in 1943, which we now know was too high).}
\centering
\begin{tabular}{lll}
\hline
Authors & Date & $\sigma_F$ \\
\hline
Heydenburg, Meyer\cite{Heydenburg:1943b} & 1943 & 2.18\,b  \\
Carnegie Inst. & & \\ \hline
{\it Oppenheimer\cite{Oppenheimer:1943}, LA-2} & 3/43 & {\it 3\,b} \\ \hline
Wiegland, Segr{\`e},\cite{Wiegland:1943} & 8/43 & (2.5\,b) \\
LA-21 &  & \\ \hline
Taschek, Williams\cite{Taschek:1943} & 10/43 & (2.7\,b) \\
 LA-28 &  & \\ \hline
{\it Bethe\cite{Bethe:1943}, LA-32} & 10/43 & {\it 2.6\,b} \\ \hline
{\it Bacher}\cite{Bacher:1944} & 8/44 & (2.2\,b)\\ \hline
Williams\cite{Williams:1944}, LA-150 & 10/44 & 1.95\,b \\  \hline
{\it Handbook\cite{Bretscher:1944a} LA-140}& 1944 &{\it 1.95\,b}  \\ \hline
{\it Wilson\cite{Wilson:1947} LA-1009} & 1947 & {\it 1.94\,b} \\ \hline
Tovesson\cite{Tovesson:19?} &  2010 & 1.92\,b$\pm$0.7\% \\ \hline
{\it ENDF/B-VIII.0\cite{Brown:2019}}& 2019 & {\it 1.93\,b}$\pm$1.3\%\\
\hline
\end{tabular}
\label{table:pu9fiss-1.5}
\end{table}

Spring-summer of 1944 was the time when Segr{\`e}’s data on reactor-made plutonium (with a higher $^{240}$Pu content) was showing spontaneous fission rates five times higher than they expected, a shocking discovery that is well-told in the introductory chapter of {\it Critical Assembly} \cite{Hoddeson:1993}.
This forced the gun Pu bomb design (``Thin Man”) to be abandoned owing to risks of pre-initiation and led to the lab pivoting to focus on plutonium implosion. Bacher moved from Physics Division Leader to being put in charge of the new Gadget Division to build the plutonium implosion gadget, and Robert Wilson took over from Bacher as leader of the Research Physics Division. The fundamental nuclear physics data for plutonium fission was all-important, and the intense pressure was captured well in {\it Critical Assembly}, which quoted (p. 196) Wilson\cite{Hoddeson:1993}:
“As Wilson remembers, P-Division realized ``there were mistakes in the measurements,'' some of which were ``very large." The spontaneous fission measurements were the principal worry, but Wilson was also upset by the status of fission cross-section and ~${\bar{\nu}}$ measurements, which were, in his opinion,``a disgrace." Wilson reacted with a tremendous furor. So many important measurements seemed to be ``based on shifting sand .... I mean, how could we call ourselves physicists?" ''

Yet in the late summer of 1944, as for $^{235}$U, Williams’s group made substantial advances\cite{Williams:1944} that allowed them to make excellent plutonium fission cross section measurements on the Van de Graaff, using a quasi monoenergetic Li(p,n) source. Advances in electroplating target production of larger plutonium samples, faster electronics, and use of ``comparison chambers” to measure fission cross section ratios between $^{239}$Pu and $^{235}$U led to the improvements. Their uncertainty estimates were 3\% statistical and 3\% systematic, which was dominated by the mass of the foil. The result they obtained at fast neutron energies, for example ~ 1.95\,b at 1.5 MeV, are in excellent agreement (to 1\%) with our best assessment today, see Table~\ref{table:pu9fiss-1.5} and 
Fig.~\ref{fig:pu9fiss.1.5}.

Later measurements of Williams in LAMS-135 (Ref.~\cite{Hoddeson:1993} p. 342), 1 September 1944, confirmed these results.
As for $^{235}$U,
over time the better plutonium cross section measurements led to smaller cross sections in the fast region, as can be seen in Table~\ref{table:pu9fiss-1.5} and Fig.~\ref{fig:pu9fiss.1.5} for the fission cross section at 1.5 MeV. There, a ``modern'' 2010 measurement is also shown for comparison, by Tovesson, made at Los Alamos' LANSCE facility. This measurement was in ratio to $^{235}$U(n,f) and also normalized to thermal. Other accurate plutonium measurements have been made by Lisowski and by Shcherbakov.

\subsubsection{Thermal fission}

This paper will not describe thermal neutron measurements in detail. But we note that the earliest thermal fission measurements from Berkeley (Chamberlain, Kennedy, Segr{\`e}, Wahl), Carnegie (Heydenburg), and then Los Alamos (Wiegland and Segr{\`e}), gave results that were compromised because the neutrons were not completely thermalized. This was appreciated by J. DeWire, R. Wilson and W. Woodward\cite{DeWire:1944} in June 1944 who discussed the different energy-dependence of $^{235}$U versus $^{239}$Pu fission cross sections in the thermal energy region. DeWire more completely thermalized the neutrons to obtain a fission cross section ratio for $^{239}$Pu(n,f)/$^{235}$U(n,f) of 1.28 at 0.025 eV (2200 m/s), a value that is in perfect agreement with our modern assessments, see Table~\ref{table:pu9u5ratio-thermal}. When combined with their estimated $^{235}$U thermal cross section they obtained a thermal $^{239}$Pu fission cross section of 705\,b\cite{Bretscher:1944}, a value that is only 6\% below our best value today, 752\,b. The thermal $^{235}$U cross section was evaluated to be 542\,b in the 1944 Los Alamos Nuclear Physics Handbook\cite{Bretscher:1944}, 
a value 8\% below our best value today (587 b).

\begin{table}
\caption{{\it Thermal} plutonium fission ratio $^{239}$Pu/$^{235}$U measurements in the  neutron energy region, compared with ENDF/B-VIII.0. The ``LA-" references denote work done at Los Alamos.}
\centering
\begin{tabular}{llll}
\hline
Authors, Energy & Date & Ratio & Today \\
              &          & & ENDF \\
\hline
Heydenberg,\cite{Heydenburg:1943b}  & 4/43 & 1.64 & 1.28 \\ 
Thermal & & & \\ \hline
Chamberlain, Segr{\`e}\cite{Chamberlain:19?} & 1942 & 1.87$\pm$0.14 & 1.28 \\ 
CN-469, thermal& & & \\ \hline
Chamberlain, Segr{\`e} & 1/43 & 1.24,1.29& 1.28 \\ 
letter to Manley& & & \\ \hline
Wiegland, Segr{\`e}\cite{Wiegland:1943} & 8/43 & 1.74$\pm0.2$ & 1.28 \\ 
LA-21, thermal & & & \\ \hline
DeWire\cite{DeWire:1944} & 6/44 & 1.28 & 1.28 \\ 
LA-103, thermal & & \\
\hline
\end{tabular}
\label{table:pu9u5ratio-thermal}
\end{table}



\section{Fission {$\bar{\nu}$} (neutron multiplicity)}

Before Project Y began in Los Alamos, a number of measurements on the average multiplicity of neutrons emitted in thermal $^{235}$U fission had already been made, beginning with Halban’s seminal 1939 measurement in Paris showing that a self-sustaining chain reaction could be produced.  The early 1939-1940 values measured by Halban ($\bar{\nu}$=3.5$\pm$0.7), by Anderson and Fermi ($\bar{\nu}$=2.2), by Zinn and Szilard  ($\bar{\nu}$=2.3), and by Turner ($\bar{\nu}$=3.05), influenced the values adopted in the early calculations of $^{235}$U critical mass by Frisch and Peierls (who assumed $\bar{\nu}$=2.3), and by the MAUD Committee ($\bar{\nu}$=2.5-3.0) and Compton’s NAS committee’s ($\bar{\nu}$=3.0) 
\cite{Reed:2007}.

At Chicago, Fermi’s team had measured the neutrons produced from $^{235}$U per thermal neutron absorbed. This was combined with measurements of the ratio of fission to capture in $^{235}$U (which was not known reliably at thermal) to infer $\bar{\nu}$ \cite{Hawkins:?}  (Par. 1.60); (Fermi, report CP-257).  

The Manhattan Project physicists made some assumptions about $\bar{\nu}$, based on general physics principles, that turned out to be correct. They assumed that the dependence with incident neutron energy ought to be weak, implying that measurements at thermal (where the fission cross section is higher, giving better counting statistics in experiments) provide a good approximation to the $\bar{\nu}$ values needed for fast neutrons.  They also expected $^{239}$Pu’s $\bar{\nu}$ to be at least as large as that of $^{235}$U.

At the first conference at Los Alamos, Oppenheimer reported\cite{Oppenheimer:1943} a value of $\bar{\nu}$=2.2$\pm$0.2 in $^{235}$U for fast neutrons, from Fermi (CP 257, p.3), see Fig.~\ref{fig:oppy-cm}. Nothing was known experimentally for $^{239}$Pu $\bar{\nu}$; as mentioned above, on general physics principles it was assumed that it would be similar to $^{235}$U’s value, and one of the project’s highest priorities was to obtain a measured value for plutonium. Indeed, the very first nuclear physics measurement at Los Alamos was for plutonium $\bar{\nu}$. {\it Critical Assembly} (p.78) describes\cite{Hoddeson:1993} Robert Wilson’s Physics Division view at the time that ``if you’re going to spend a billion dollars” to build an atomic bomb, you have to be sure that $\bar{\nu}$ is large enough to sustain a chain reaction, and that the fraction of fission neutrons that are delayed, versus prompt, needs to be small.

Bacher, the first Physics Division Leader (Fig.~ \ref{fig:BetheBacher}), crafted a research program that pursued complementary and parallel studies, to both mitigate risks and provide a variety of insights. This involved supporting continued thermal measurements at Chicago by Fermi’s team, and new measurements at Los Alamos focused on both thermal and fast neutron energies. Bacher wanted to be able to tie together Fermi’s measurements at thermal with Los Alamos measurements also for thermalized neutrons, and additionally to determine the neutron energy dependence of $\bar{\nu}$ from thermal up to fast neutron energies (Ref.~\cite{Hoddeson:1993} p.184). Diven, Manley and Taschek describe\cite{Diven:?} how hard the fast neutron measurements were, given the small fission cross section at fast versus thermal energies. This is because most neutrons in the experiment go through the target without interacting and provide a large ``noise” background to the measured signal.

In the summer of 1943 at Los Alamos, Williams used\cite{Williams:1943b} the the Van de Graaff  with Li(p,n) source reactions slowed down by paraffin and by water  to make the first plutonium $\bar{\nu}$ measurement.  Not only did the slow neutron source reaction result in a higher fission cross section and better statistics, but it also ensured that the source neutrons were separable from the higher-energy fission neutrons of interest. A tiny ~ 142\,$\mu$g plutonium sample from Chicargo’s Met Lab was prepared into a foil by Art Wahl, and the ratio of prompt $\bar{\nu}$ for plutonium, versus $^{235}$U, was measured to be 1.20$\pm$0.09\cite{Williams:1943c} and plutonium $\bar{\nu}$ itself determined to be 2.64$\pm$0.2 (Ref.~\cite{Hoddeson:1993} p.79). The fact that plutonium’s $\bar{\nu}$ was larger than uranium’s was very good news, confirming their expectation. Richard's memoirs\cite{Richards:1993} relate this very first experiment at Los Alamos, and stated that the few-$\mu$g  speck of plutonium came from  the Washington University (St. Louis) cyclotron  (perhaps a precursor to the aforementioned Met Lab sample?), and the experiment's success earned the group a 
camping trip into the Pecos Wilderness. 
This result was the last good news on $\bar{\nu}$ for quite a while, as a series of contradictory results for the $^{235}$U $\bar{\nu}$ were subsequently found. 

The first of the disappointing results was in the Fall of 1943 when Fermi reported a new measurement that reduced $^{235}$U $\bar{\nu}$ to 2.0 from his previous value of 2.2. Later, this was found to be a change in the wrong direction (the best prompt value today is 2.41 at thermal and 2.57 at the fast neutron energy of 1.5 MeV). Fermi’s reputation was such that his $\bar{\nu}$=2.0 value was immediately adopted for $^{235}$U and $^{239}$Pu, as seen for example in the updated criticality calculations shown in Bethe’s Theoretical Division monthly report for October 1943, LA-32, see Table~\ref{table:u5critmass}. Oppenheimer wrote to Groves stating that ``even this small change means an increase by 40\% in the amount of material required” (Ref.~\cite{Hoddeson:1993} p.191).  

Robert Wilson captured the crisis of confidence with his statement in the introduction to Nuclear Physics, LA-1009, ``We were constantly plagued by worry about some unpredicted or overlooked mechanism of nuclear physics which might make our program unsound”. Oppenheimer explained to Groves “even the most careful experiments in the field may have unexpected sources of error”. He pointed out the value of pursuing independent approaches so that the $\bar{\nu}$ values obtained could be validated by all promising means.

At Oppenheimer’s request, Segr{\`e} traveled to Chicago to make thermal measurements, taking advantage of the reactor’s high neutron fluence. In early 1944, he found a result for $^{235}$U thermal $\bar{\nu}$ of 2.15, while Fermi obtained a new value of 2.18 (Ref.~\cite{Hoddeson:1993} p.195). These were changes in the right direction, but more accurate values had to await improved experimental methods being developed back at Los Alamos, together with larger amounts of ``Clinton'' plutonium that would come from Oak Ridge's Clinton Engineer Work's reactor.

\begin{figure}[htbp]
\begin{center}
  \includegraphics[width=3.25in]{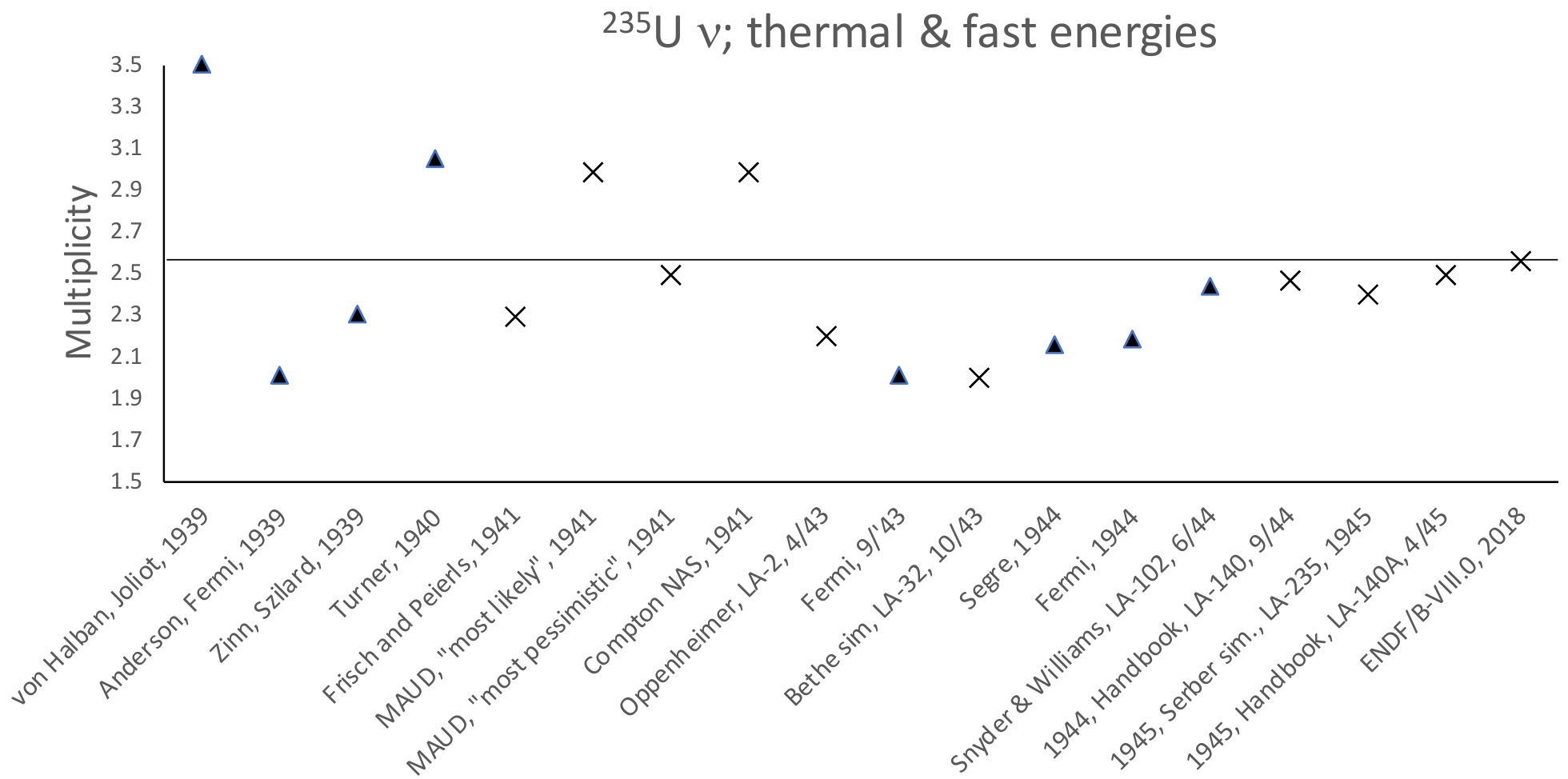}
    \includegraphics[width=3.25in]{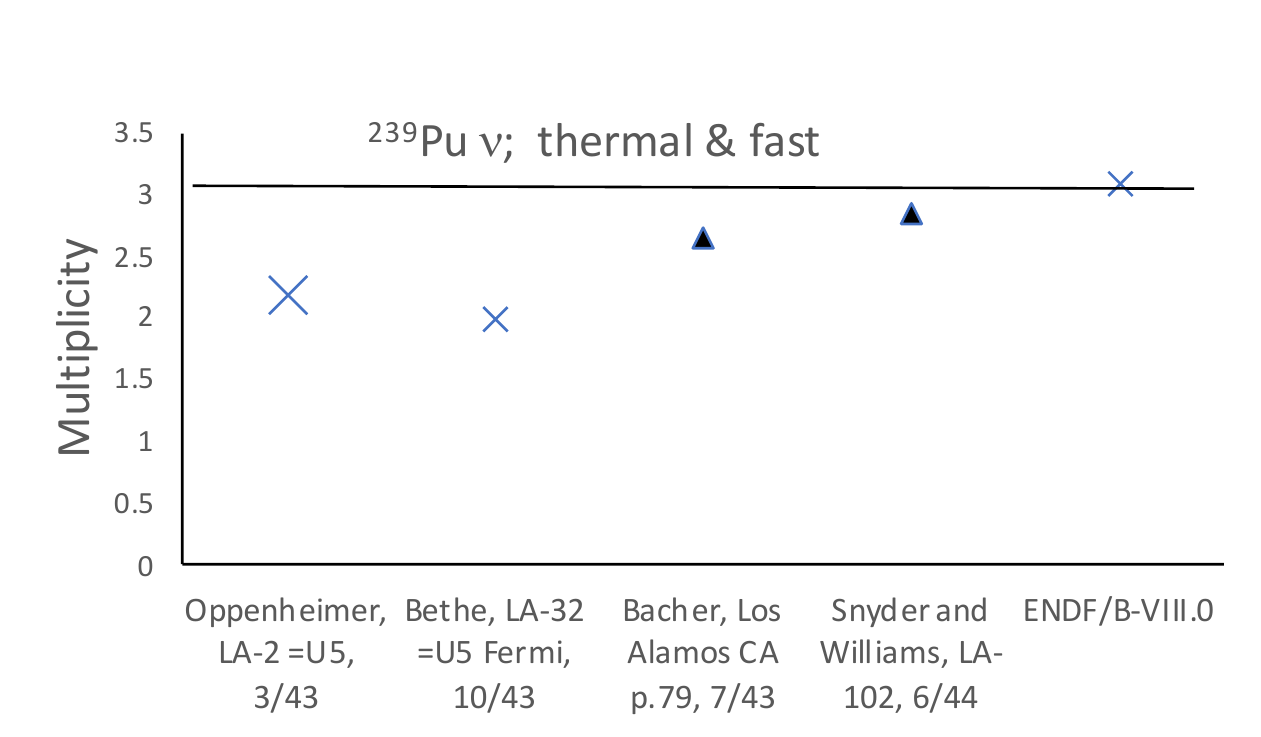}
\caption{The $^{235}$U  prompt $\bar{\nu}$ (upper panel) and  $^{239}$Pu prompt $\bar{\nu}$ (lower panel) measurements changing with time. For simplicity the figure includes 
values measured at different incident energies, from thermal up to fast energies. The horizontal line shows the best value today, for fast neutron energies. Measurements are shown as solid symbols whereas evaluations or estimates are crosses.}
\label{fig:nubar}
\end{center}
\end{figure}

In June 1944, accurate measurements of thermal prompt $\bar{\nu}$ were made by Snyder and Williams\cite{Snyder:1944} using thermalized neutrons from the cyclotron, with an ion chamber for counting fissions. Their result for the $\bar{\nu}$ ratio of $^{239}$Pu/$^{235}$U, was 1.17$\pm$0.02, in excellent agreement with our best value today from the ENDF/B-VIII.0 standards, 1.19. Compared to the first Segr{\`e} measurement in 1943 that had been done with only 142\,$\mu$g, this experiment used 562 mg of Pu. Results were also obtained for an ``integral” broad energy Ra-Be neutron source, with a known source strength, resulting in absolute measurements\cite{Bretscher:1944a} of thermal $\bar{\nu}$-25=2.44, and thermal $\bar{\nu}$-49=2.86, with estimated uncertainties of 0.05, {\it i.e.} about 2\% (LA-140). These agree to 1\% with our best ENDF/B-VIII.0 values today, of 2.41 and 2.87 respectively at thermal. The increasingly-accurate measurements 
of the  $\bar{\nu}$ ratio for $^{239}$Pu/$^{235}$U during the Manhattan Project is also usefully described by Hutchinson in this issue \cite{Hutchinson:2020a}.

Robert Wilson\cite{Wilson:1944}  reported similar excellent results on July 4, 1944 (no time off for celebrations?). They found a similar result for the prompt $\bar{\nu}$ $^{239}$Pu/$^{235}$U ratio of 1.18$\pm$0.01 (compared with 1.19 today). They were also able to show that the prompt neutrons did not include any “slightly delayed” neutrons that were delayed by $>$5.E-9 seconds; an important result for consideration of a fast chain reacting system.

By late in 1944, the (small) energy dependence of $\bar{\nu}$, from thermal to fast energies, had also been established (Ref.~\cite{Hoddeson:1993} p.342).  William’s group found 0.98$\pm$0.04 for $^{235}$U, and 1.01$\pm$0.04 for $^{239}$Pu, for the ratio  $\bar{\nu}$(250 keV)/$\bar{\nu}$(thermal)\cite{Williams:1944b}. These experiments, and similar ones in Manley’s and Wilson’s groups\cite{Manley:1944} confirmed the initial expectation that $\bar{\nu}$ would be similar for fast and thermal energies, 
and agree with our modern assessments: ENDF/B-VIII.0 today shows a ratio of 1.02 and 1.01 for $^{235}$U and $^{239}$Pu, respectively, for 250 keV v. thermal energies\footnote{The incident neutron energy dependence becomes larger as one moves up to higher energies, reaching 1.07  for   $\bar{\nu}$(1.5 MeV)/$\bar{\nu}$(thermal) for both  Pu and U.}.

Combining Snyder and Williams’ $\bar{\nu}$ data with Wilson’s data, the September 1944 LA-140 Nuclear Physics Handbook assessed\cite{Bretscher:1944a} thermal evaluated values of $\bar{\nu}$-25=2.47 and $\bar{\nu}$-49=2.91, which agree with our modern values to 2\% and 1\% respectively. By 1945, the LA-140A Nuclear Physics Handbook had updated the fast $^{235}$U $\bar{\nu}$ to 2.50.  The values of $\bar{\nu}$ data determined throughout the 1940s are shown in Fig.~\ref{fig:nubar}.

\section{Prompt Fission Neutron Spectra, PFNS}
The topic of the energy dependence of prompt neutrons emitted in fission has the distinction of being documented in the first manuscript LAMS-1, an early 1943 paper by W.E. Bennett and H.T. Richards.\cite{Bennett:1943} The prompt fission neutron spectrum (PFNS) was needed during the Manhattan Project for a variety of reasons. It set a substantial part of the neutron energy distribution in fast metal regions of critical systems, defining the average neutron energy for which cross sections and $\bar{\nu}$ needed to be known (around 1--2 MeV) and influencing calculated criticality.  For example, Chadwick\cite{Chadwick:2018} quantified 
the calculational uncertainty on the criticality k-eff  owing to the uncertainty of the average energy of the PFNS; for Godiva, a fast highly enriched uranium $^{235}$U assembly, an uncertainty of 300 keV in the average PFNS energy (roughly the accuracy of understanding in 1943) corresponds to about 1\% relative uncertainty in calculated k-eff, which is about 2 kg out of ~46 kg in the critical mass of an idealized $^{235}$U sphere.  Compared to the other nuclear data 
uncertainties in Table~\ref{table:u5critmass} circa 1944-1945 this is a relatively small effect. But there is another phenomenon in dynamical systems where 
the PFNS is important:
it determines the neutron velocities and therefore the time between subsequent chain reaction fissions, see Lestone's paper\cite{Lestone:2020} in this issue.

Today, neutronics simulations access ENDF data on the PFNS spectra that are also dependent upon the incident neutron energy; the “Chi matrix”. The PFNS are different for $^{235}$U, $^{238}$U, and $^{239}$Pu, with plutonium being “hotter” than uranium. However, in the early days of nuclear science it was correctly appreciated that the PFNS dependence on incident energy, and actinide target type, would be small. This proved to be the case, and so in this section (as was also the case at Los Alamos 1943-1945) I will mostly not concern myself with incident-energy differences, and will show PFNS  plots that combine data from various experiments, sometimes with different incident energies. It was common to intentionally thermalize accelerator-source neutrons to benefit from higher statistics and lower backgrounds owing to the large fission cross section at low energies.

\begin{figure*}[htbp]
\begin{center}
\includegraphics[width=6.25in]{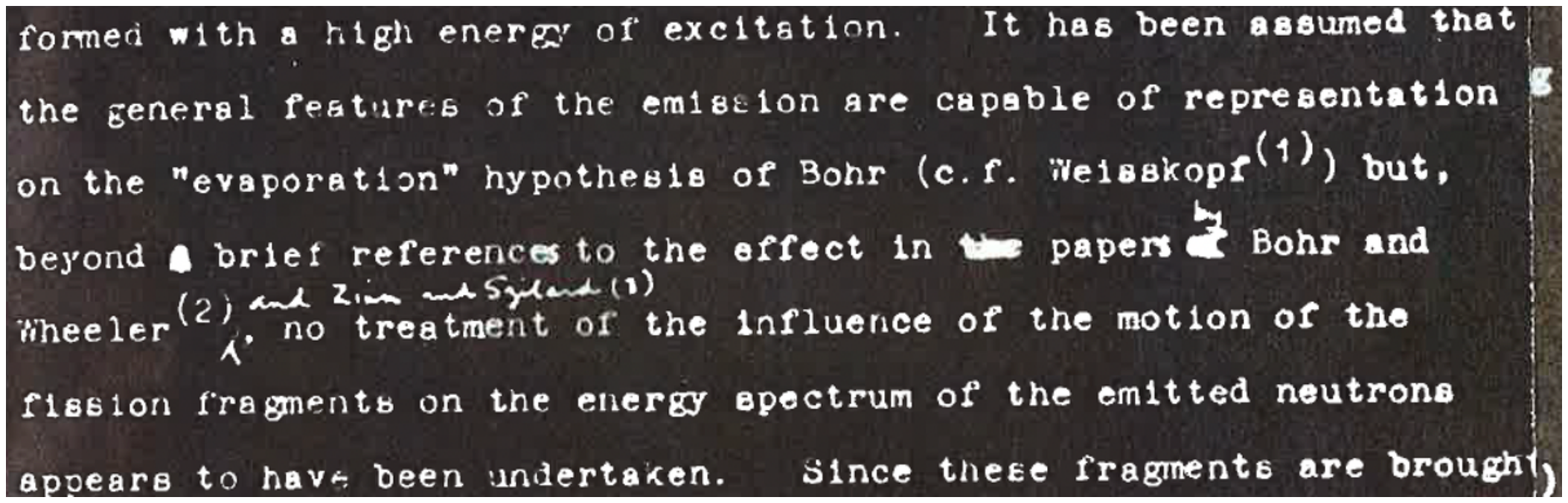}
\vspace{-7.5cm}
\caption{An image extract from Norman Feather's report BM-148, Emission of Neutrons from Moving Fission Fragments, February 1942.}
\label{fig:feather}
\end{center}
\end{figure*}

Early 1939 data on the PFNS from thermal fission neutrons had been published by W.H. Zinn and L. Szilard (PR 56, 619 (1939)) and H. von Halban, F. Joliot, and L. Kowarski (Nature 143, 939 (1939)). These very early measurements were suggestive but could not characterize the spectrum in any detail. The earliest theoretical treatment of the fission process, beyond brief references to the phenomena in Bohr and Wheeler and Zinn and Szilard, was by Norman Feather in Cambridge, February 1942 \cite{Feather:1942}. Feather correctly understood the emission mechanism to be the isotropic evaporation of neutrons from excited fission fragments in their center-of-mass frame, followed by the kinematical boosting of these neutrons based on the fragment’s motion, and he developed a mathematical model to describe them, see Fig.~\ref{fig:feather}.

In the early 1940s, programs to measure the PFNS were established in Britain and  the USA. In Britain, the PFNS was measured at Liverpool by Rotblat, Pickavance, Rowlands, Hall and Chadwick (B report 86), with a photographic emulsion plate method. At Rice, W.E Bennett and H.T. Richards (CF Report 325; H.T. Richards PR59, 796 (1941)) made PFNS measurements by observing recoils in the cloud chamber. Subsequently, Bennett and Richards would move to Minnesota with Williams to measure the PFNS with photographic plates using the Van de Graaff; they thought that this emulsion technique would allow neutrons to be measured over a wide range of energies, keeping scattering material to a minimum.\footnote{Richards moved to Los Alamos for Project Y and subsequently was on the
faculty at 
the University of Wisconsin until 1988. He had 49 graduate students, several continuing in the field of neutron-nuclear physics.} At Los Alamos, many measurements would be made to determine the neutron spectrum through proton recoil studies in ionization chambers, for thermal and fast incident neutrons, using approaches that followed Bloch and Staub’s work\cite{McMillan:1943} (report CF-525)  at the Stanford cyclotron (which proved to be  very accurate, see below). At Chicago, measurements were made with the pile thermal neutrons, with the cyclotron source, and with neutrons from spontaneous decay sources \cite{Hoddeson:1993}. 

As was often the case, it was Bethe who first quantified the likely average energy of the neutrons, and understood their basic spectrum shape from 
theoretical insights. In LA-2 he 
assessed that they should have an average energy of about 1.7 MeV, see Fig.~\ref{fig:bethe-pfns}, influenced by the 1943 Stanford measurements (see below) . This is close to the best value today of 2.0 MeV for thermal 
neutrons on $^{235}$U (2.04 MeV for fast 1.5 MeV incident neutrons). Bethe correctly appreciated that nuclear theory insights could  provide useful 
quantitative guidance, perhaps as good or better than many experiments of that time
period owing to their substantial background scattering problems (see below). This is no longer the case, and nowadays our best 
theoretical treatments tend to be calibrated to high-precision measured data. In the last decade,
flagship experiments\cite{Kelly:2018,Kelly:2020,Kelly:2021} at Los Alamos, by LANL, LLNL and CEA scientists have focused on precise PFNS measurements.

It is interesting to try to follow Bethe's train of thought in the text in Fig.~\ref{fig:bethe-pfns}, given that this represents the first quantitative 
theoretical perspective on the PFNS's average neutron energy, $E_{\rm av}$. First he considers the moving fragments' velocity and notes that this corresponds to a moving velocity of 0.8 MeV/nucleon (which will be used to do a kinematical boost into the lab frame). Our assessments today are that the average kinetic energy for the fragments is 169.1 MeV at thermal, 169.4 MeV at fast energies, and so when divided by the number of nucleons (236 for $^{235}$U+n) we obtain 0.72 MeV/nucleon, close to Bethe's 0.8 MeV/nucleon. Next he states that the average lab-frame energy  of the PFNS neutrons will be this 0.8 MeV plus the mean (center of mass, CM) energy for evaporation, which is $E_{\rm av}$=3/2$kT_{\rm CM}$, $T_{\rm CM}$ being the temperature for evaporation (See Eq.~(3) in 
Ref.~ \cite{Kodeli:2009}). Simply adding the two quantities is correct for isotropic neutron emission in each fragment's center-of-mass system. Next we must consider whether Bethe's text implies that he is (a) using a $T_{\rm CM}$=0.6 MeV from some consideration, see below, to derive an average CM fission fragment kinetic energy of 0.9 MeV, to derive $E_{\rm av}$=1.7 MeV PFNS average energy; or the opposite, (b) assuming 1.7 MeV from ``present measurements" of PFNS to work backwards and infer $T_{\rm CM}$=0.6 MeV and imply it is reasonable. I think it is the latter since Bethe was aware of Bloch's summary in LA-4 that showed Stanford's recent PFNS measurement (reported on March 11, 1943 letter to Manley, NSRC-A84-19-Box49-7) with an average energy of 1.7 MeV (although there was much uncertainty at the time, and early PFNS measurements from Liverpool and Rice  reported  PFNS  average energies that were erroneously high). We can still ask, what might Bethe thought of the value $T_{\rm CM}$=0.6 MeV from his nuclear physics insights? Where might that have come from? 

A Maxwellian evaporation spectrum has a functional form $P(\epsilon)=\sqrt(\epsilon) \rm{exp}(-\epsilon/kT_{\rm CM})$,\cite{Kawano:2013} $\epsilon$ being the neutron CM-frame energy, and has an average CM energy $\epsilon_{\rm av}=3/2kT_{\rm CM}$. A simple estimate of temperature can be obtained from the average post-neutron emission excitation energy $U$ of the decaying fragment and the level density parameter $a$, $U=aT_{\rm CM}^2$ (Bethe\cite{Bethe:1937}, 1937, p.81) with the pre-neutron excitation energy $U'=U+S$, S being the separation energy. If one disregards the complexities of the double-humped fission fragment distribution, with different spectra from the heavy and light peaks, one could estimate an average fission fragment level density parameter $a$ =A/8=(236/2)/8=14.75 MeV$^{-1}$. At the time Bethe thought 
${\overline{\nu}}$ was about 2  (1 per fragment on average);  one might guess a neutron separation energy of ~$S$=5 MeV for neutron-rich fragments;  after neutron emissions occurred and residual systems were bound to further emission, their remaining excitation energies would be somewhere between 0 and 5 MeV and heavily weighted towards the upper end owing to the shape of the evaporation spectrum, about 4.7 MeV on average; so one would estimate an average post-neutron emission excitation 
energy per fragment of $U=\epsilon_{\rm av}$+ 4.7 MeV, which is then equal to $aT_{\rm CM}^2$, so $4.7 +3/2.T_{\rm CM}=14.75.T_{\rm CM}^2$, giving $T_{\rm CM}=0.62$ MeV,  $\epsilon_{\rm av}$=0.93 MeV, close to Bethe's $T_{\rm CM}=$0.6 MeV, $\epsilon_{\rm av}$=0.9 MeV in LA-2, Fig.~\ref{fig:bethe-pfns}.

 So Bethe's approach had a PFNS average energy of $E_{\rm av}$= $0.8+3/2\times0.6$=1.7 MeV, whereas my ``recreation'' of what he might have done gives $0.72 + 3/2\times 0.62$=1.65 MeV. For comparison, a modern estimate of these fission fragment values by Kawano {\it et al.}\cite{Kawano:2013}, based on a statistical model informed by empirical data, gives temperatures in the 0.75-0.85 MeV range -- somewhat higher than
the 0.62 MeV obtained from the above simple argument.\footnote{Further comparisons with Kawano's simulations: He obtains
average pre-neutron emission excitation values of U=13.7 MeV (light) and 8.8 (heavy), so 11.3 MeV on average, versus my U'=10.7 MeV;
average CM temperature of 0.85 MeV (light) and 0.78 MeV (heavy) so 0.815 MeV on average, versus my 0.62 MeV, and average CM neutron evaporation 
energy of 1.22 MeV versus my 0.93 MeV} Such modern modeling results would be expected to give higher temperatures, since Bethe's $E_{\rm av}$=1.7 MeV is 
a little small compared to our best understanding today, 2.0 MeV for thermal incident neutrons.

\begin{figure}[htbp]
\begin{center}
\includegraphics[width=3.25in]{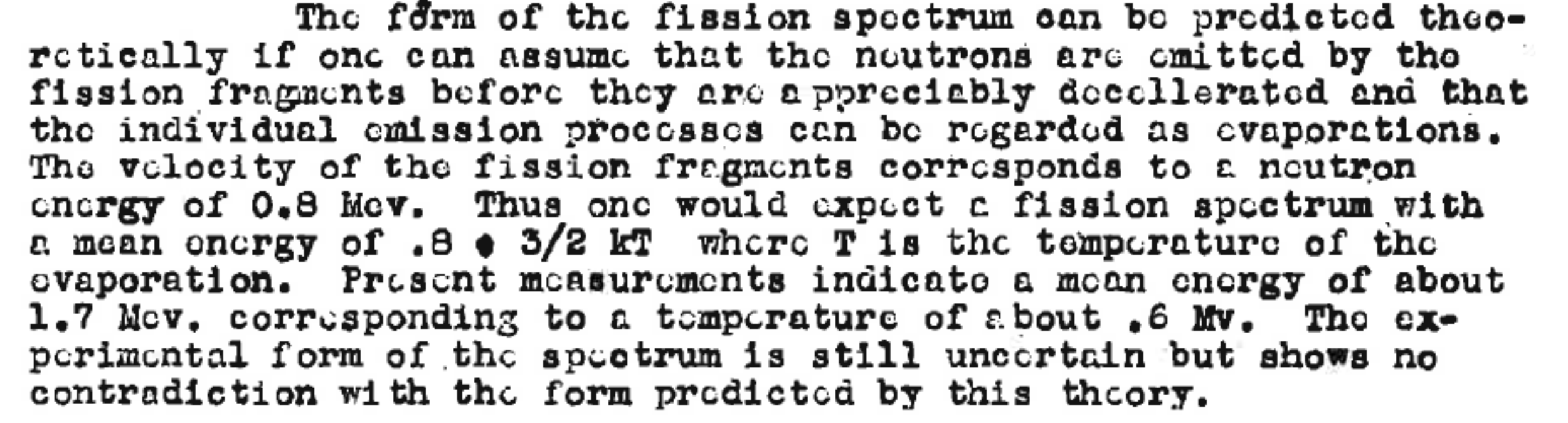}
\caption{An image extract  from Bethe's report in LA-2, April 1943, predicting the average energy of PFNS neutrons to be 1.7 MeV (versus the best evaluations today of 2-2.1 MeV.)}
\label{fig:bethe-pfns}
\end{center}
\end{figure}

In 1942, Christy and Manley\cite{Christy:1942} made a clever and essential integral measurement of the average energy of the PFNS neutrons, at Chicago, by measuring the energy of the neutrons absorbed in water following thermal $^{235}$U fission, an approach that was used at Liverpool too \cite{Bennett:1943}. This was particularly important because of the experimental challenges that were being faced by the many groups that were trying to measure the spectrum directly. Their result, 2.2$\pm$0.2 MeV, is very close to the correct result 2.0 MeV in ENDF/B-VIII.0, and furthermore they correctly assessed that if anything it would be an over-estimate of the true value. It led Serber to state\cite{Serber:1992} the PFNS  has an average energy of 2 MeV (Primer, p.17; p.70), probably a rough average of Christy's measurement (2.2 MeV) and the 
Stanford measurement (1.7 MeV) quoted by Bethe and Manley, and  pointed experimentalists to appreciate possible systematic errors in the PFNS measurements that they were making.

\begin{figure}[htbp]
\begin{center}
\includegraphics[width=3.25in]{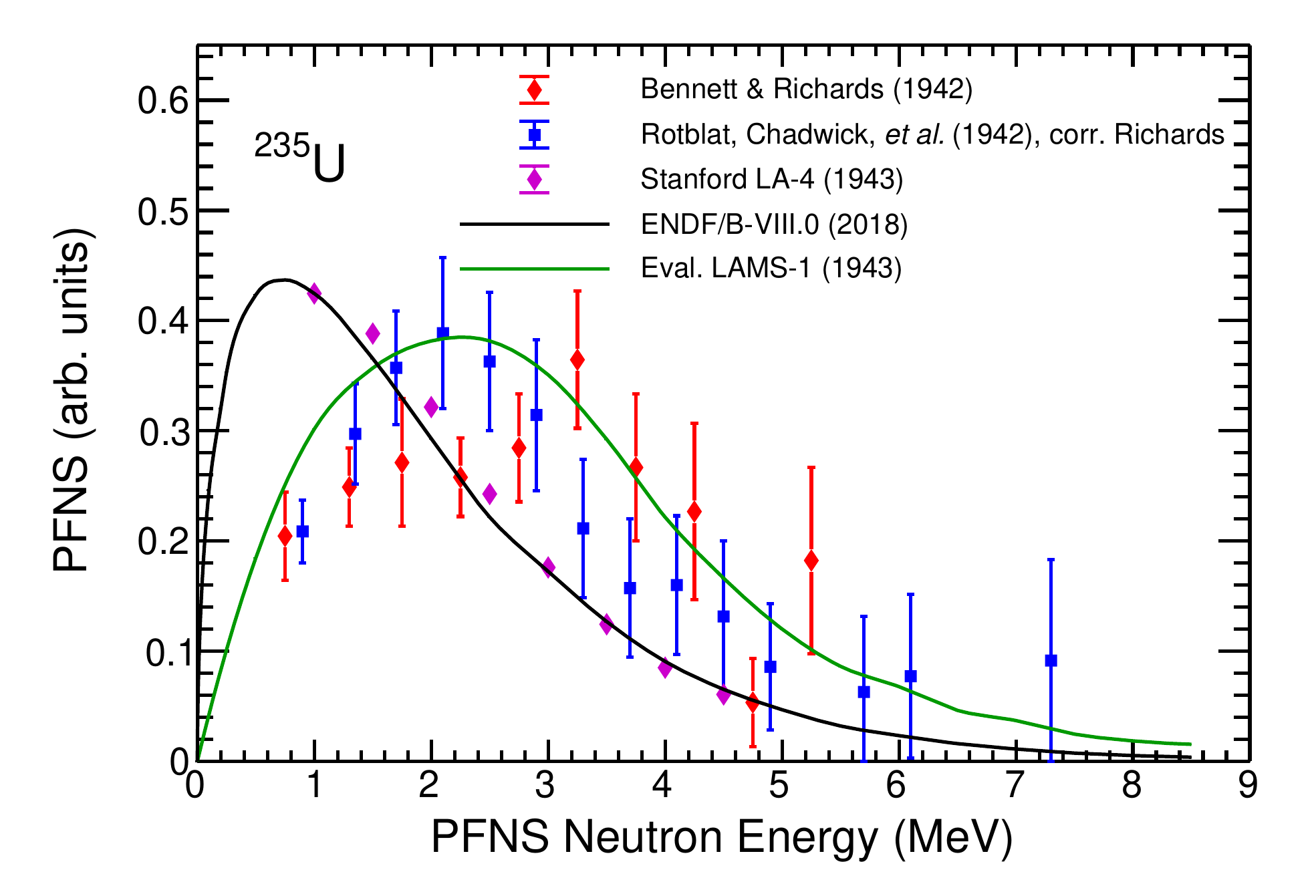}
\caption{The $^{235}$U prompt fission neutron spectrum for thermal neutrons at the beginning of Project Y, as presented in LAMS-1. The black solid line is our 
modern value of the thermal spectrum from ENDF/B-VIII.0, which comes from an IAEA evaluation\cite{Capote:2015a,Capote:2015b,Capote:2015c}, and shows the accuracy of the 1943 Stanford measurement.}
\label{fig:u5-pfns-thermal}
\end{center}
\end{figure}

At the March 1943 conference in Los Alamos, Bethe in LA-2 described the motivation for understanding the PFNS, and the experimentalists present agreed on the unsatisfactory nature of the spectrum data taken so far at Rice and at Liverpool, which suffered from large systematic errors from multiple scattering processes. Bethe said that the measurements were “unsatisfactory largely because too much fission material has been used. If enriched material were available smaller masses could be used and inelastic scattering of the fission neutrons avoided”. However, Bethe did provide optimism in a theoretical understanding of the shape of the spectrum, which “can be predicted theoretically.” A few weeks later, in LA-4, at the April 27 workshop Bloch (p.3) described the Stanford PFNS measured for thermal neutrons on $^{235}$U, from a cyclotron source, which had an average emission energy of  1.7\,$\pm$20\% MeV, and peaked at 1.1\,$\pm$20\% MeV (F. Bloch and H. Staub, CF-525, and see NSRC A84-019-49-7). He noted that these energies were considerably lower than those from Rice and Liverpool, and was concerned that they suffered from inelastic scatterings that distorted the spectrum to lower energies, although we now know the Stanford measurements to be remarkably accurate (ENDF/B-VIII.0 has a PFNS average energy of 2.0 MeV for thermal neutrons on $^{235}$U), see Fig.~\ref{fig:u5-pfns-thermal}. Manley was more positive about the Stanford results, saying ``Data of ion chamber pulse size distributions from Stanford, 
which looks reasonable theoretically, show neutrons tailing off from 1 MeV'' (in LA-2); subsequent experimental methods at Los Alamos would also use ion-chamber pulse distributions and obtain relatively accurate results, as discussed below.

Bennett and Richards (1943) in LAMS-1 noted a likely problem with the Liverpool Rotblat-Chadwick measurement owing to multiple scattering in the experiment; LAMS-1 attempts to correct for this to give the data in their figure II. The curve drawn has a mean energy of 2.9 MeV – “50\% too large if the measurements of mean energy by absorption in a water tank are reliable.” This refers to the aforementioned Chicago integral measurement of 2.2 MeV by Christy and Manley, that was correctly trusted. Figure~\ref{fig:u5-pfns-thermal} shows the PFNS data described in LAMS-1 together with their “best estimate” (green line) of the time - an average of two of the data sets (but unfortunately not the Stanford data, which at the time they thought could be suffering from systematic inelastic scattering errors).   Today’s ENDF/B-VIII.0 spectrum is shown for comparison, and is seen to be substantially softer with an average neutron energy of 2.0 MeV. The Bloch and Staub Stanford 1943 PFNS data are also shown and seen to be remarkably accurate.

\begin{figure}[htbp]
\begin{center}
\includegraphics[width=3.25in]{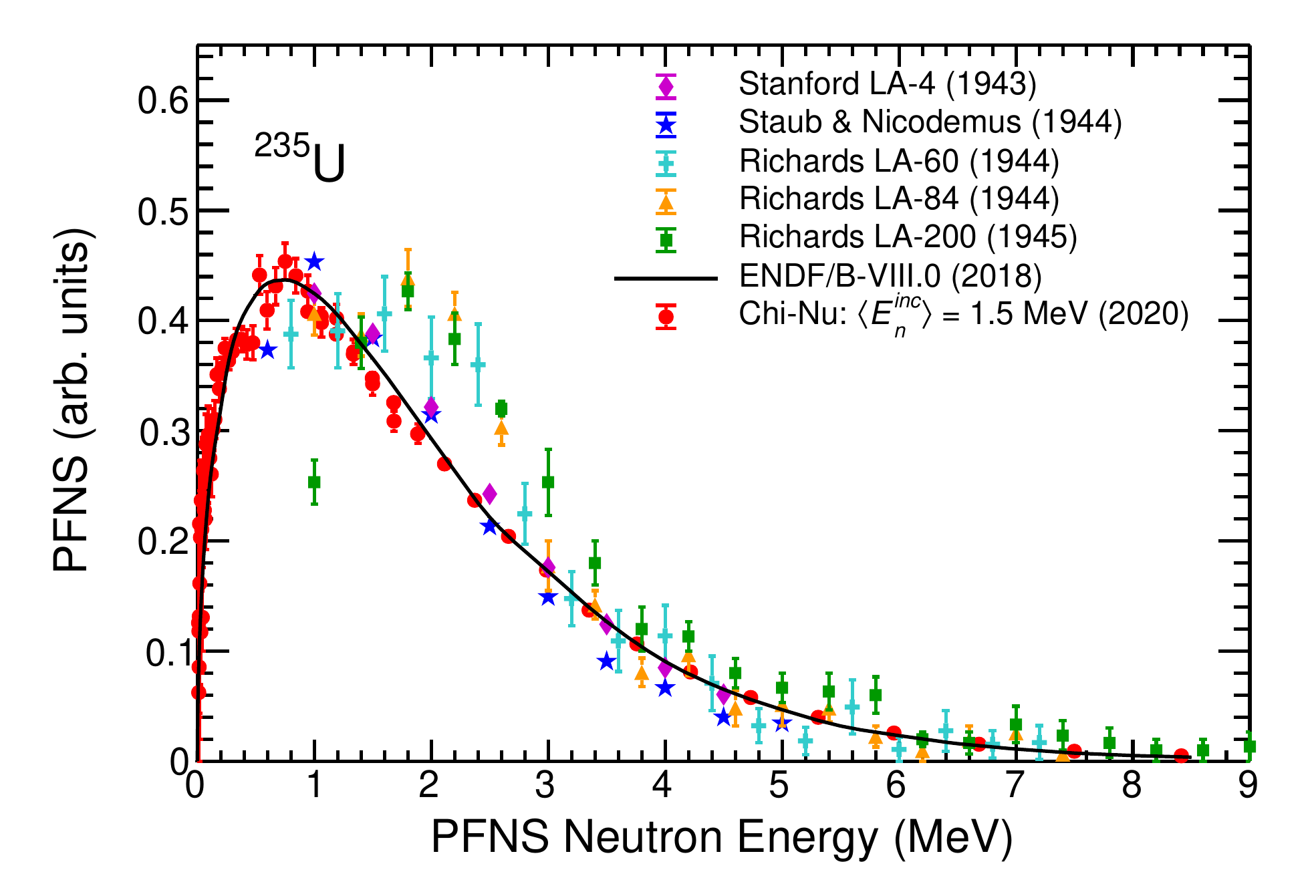}
\caption{The $^{235}$U prompt fission neutron spectrum for neutrons, showing measurements made during Project Y. The incident neutron  energy varies for these different experiments and while in principle the spectrum changes with energy, the changes are much smaller than the differences seen between these data sets. The black solid line is our 
modern value of the spectrum from ENDF/B-VIII.0, and the red points are recent measurements made at Los Alamos for an average incident neutron energy 
of 1.5 MeV.}
\label{fig:u5-pfns-end}
\end{center}
\end{figure}

During the course of the Manhattan project, the spectrum would be measured many times at Los Alamos, at Stanford, and at Minnesota. Richards’ Minnesota data were documented in LA-60\cite{Richards:1944} where an average energy of 1.85 MeV was found – fairly accurate compared to our best result today of 2.0 MeV, see
Fig.~\ref{fig:u5-pfns-end}.  But after that, with an improved enriched $^{235}$U source, instead of finding more accurate PFNS results, the Minnesota data seem to become less accurate over time for reasons I don’t understand -- Fig.~\ref{fig:u5-pfns-end} shows LA-84 (May, 1944) data for thermal fission on $^{235}$U and  LA-200 (Jan 15, 1945) data for fast 300-650 keV neutrons on $^{235}$U, and it is evident that  these orange and green data points were not very accurate. Richards’ measured change in average energy from thermal to fast incident energies (2.3 to 2.6 MeV) is much greater than our best assessments today (2.00 to 2.01 MeV in ENDF/B-VIII.0). Also, plutonium data in LA-84 differed substantially from uranium, a result not seen in subsequent measurements at Los Alamos by Staub and Nicodemus. Richards’ LA-200 paper provides a cautionary example of how one can face “one step forward, two steps back” in science. On the (erroneously) hot 2.6 MeV-average PFNS  he obtained for $^{235}$U, he suggested that it was in fact correct by showing comparisons of an integral of the $^{238}$U(n,f)/$^{235}$U(n,f) ratio obtained with his spectrum, compared to a new direct integral measurement of this fission by Wilson. The good agreement he obtained was presumably from canceling errors.

\begin{figure}[htbp]
\begin{center}
  \includegraphics[width=3.25in]{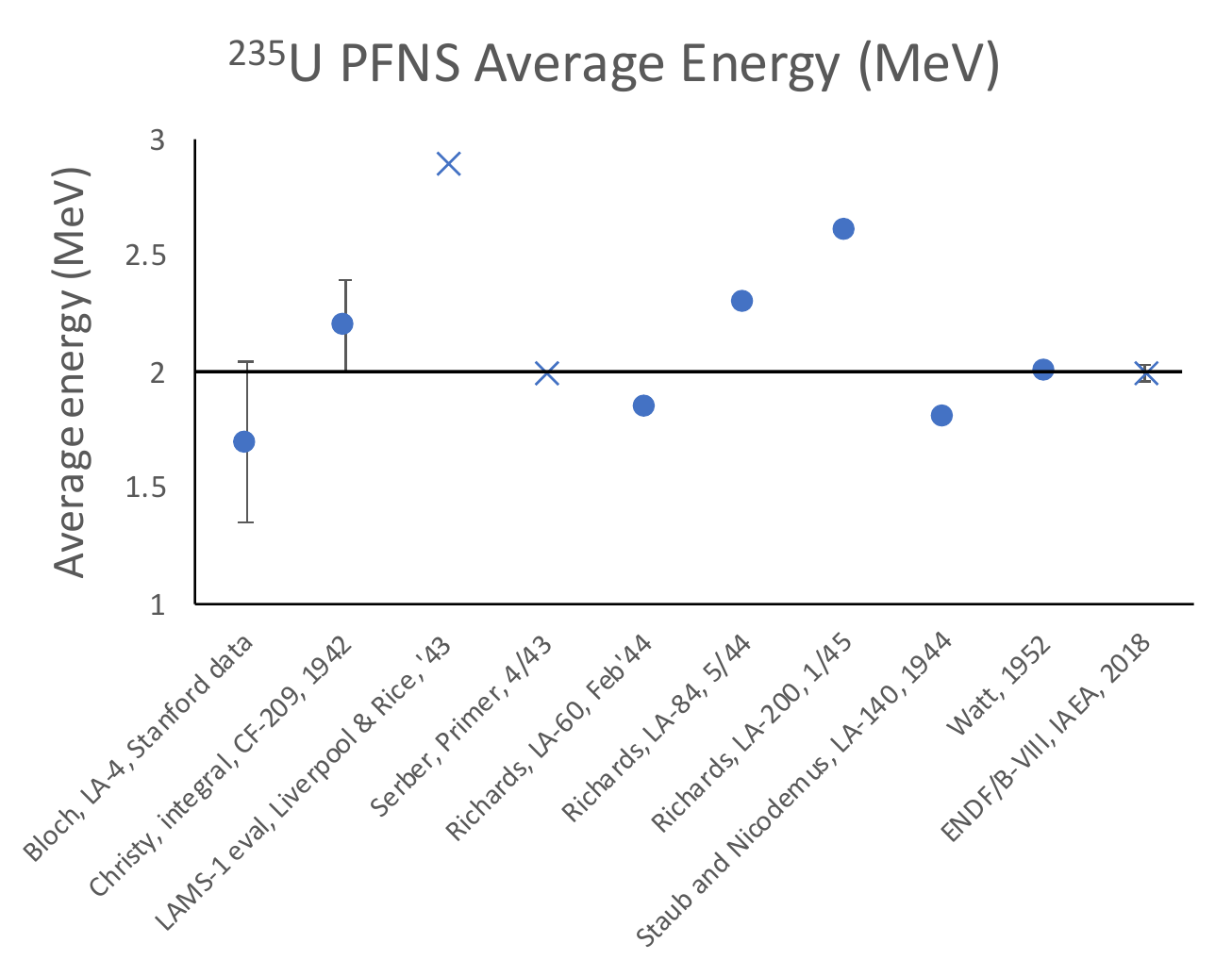}
   \includegraphics[width=3.25in]{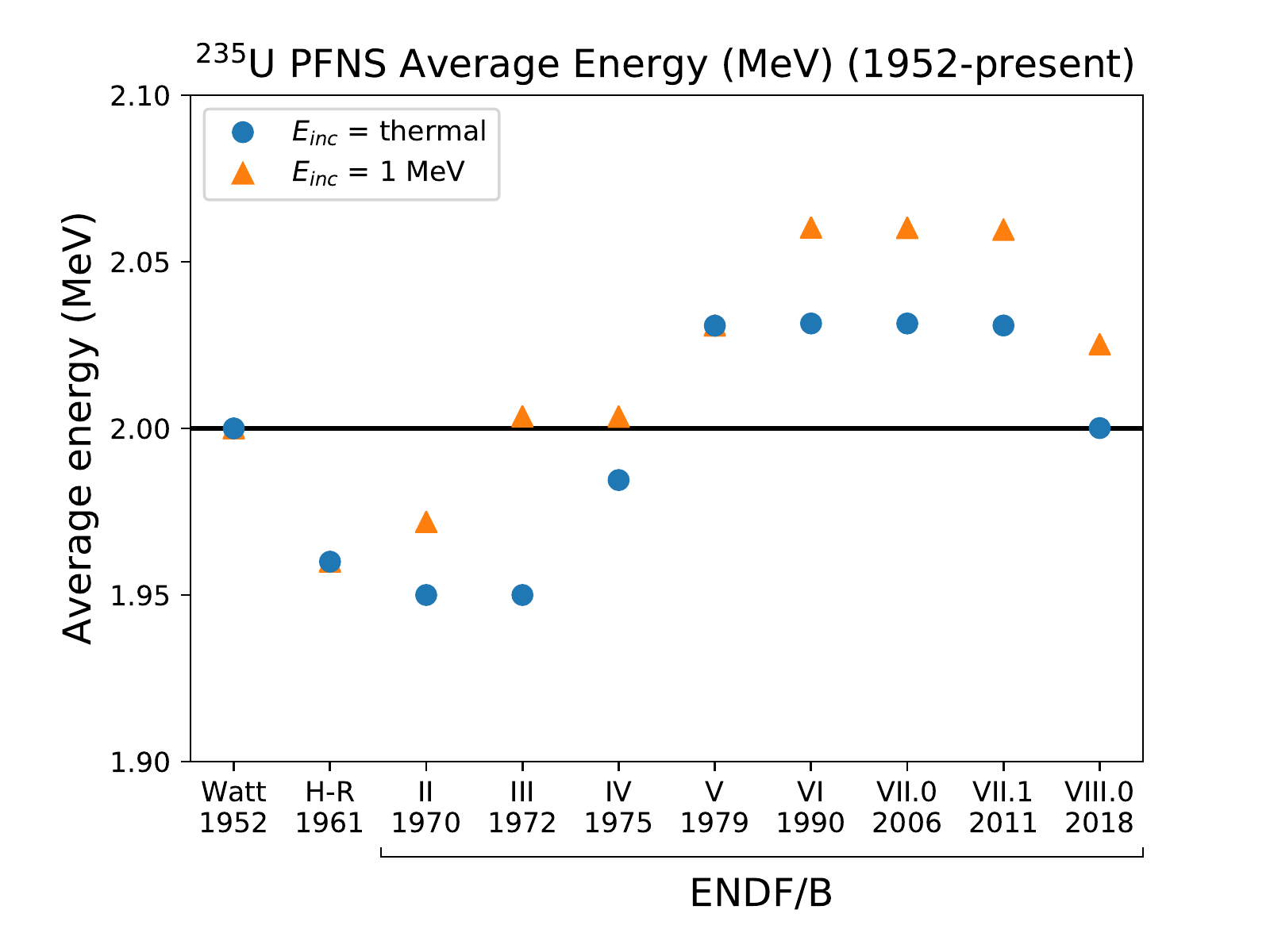}
\caption{The $^{235}$U PFNS fission spectrum's average neutron energy measurements changing with time. (a) Upper panel: Data from 1942 -- 1952. Although not strictly correct, for simplicity the figure includes 
values measured at different incident energies, from thermal to fast energies.  Measurements are shown as solid symbols whereas evaluations or estimates are crosses; (b) Lower panel: Data evaluations from 1952 -- present,  for thermal incident neutrons and for 1 MeV incident energy. H-R 
denotes the 1961 Hansen and Roach evaluation\cite{Hansen:1961}; values from various ENDF database  versions are shown  from 1970-present (Credit: N. Gibson). The horizontal lines show the best thermal value today.}
\label{fig:u5pfns-eav}
\end{center}
\end{figure}

The most accurate $^{235}$U and $^{239}$Pu PFNS data were obtained by Nicodemus and Staub at Los Alamos in 1944 using moderated neutrons from the Li(p,n) reactions at the Van de Graaff generator, and were published after the war\cite{Nicodemus:1953}. These $^{235}$U PFNS data are shown in Fig~\ref{fig:u5-pfns-end} and are seen to agree well with the earlier Stanford data and our best ENDF/B-VIII.0 evaluation today (black curve, which is based on Capote {\it et al.}'s IAEA evaluation\cite{Capote:2015a,Capote:2015b,Capote:2015c}). For comparison, some recent 2020 high-accuracy $^{235}$U “Chi-Nu” data from Los Alamos’ LANSCE facility\cite{Kelly:2018,Kelly:2020,Kelly:2021}  are shown as red points. In April 1944 William’s group also saw small differences between the $^{235}$U and $^{239}$Pu PFNS (Ref.~\cite{Hoddeson:1993}, p.196) between 1 and 2.5 MeV outgoing neutron energy. They were aware of substantial differences between their ionization chamber data and those from Minnesota’s photographic plate method. The Los Alamos results would prove to be the most accurate.

The average energy of these PFNS measurements on $^{235}$U is shown in Fig.~\ref{fig:u5pfns-eav} (upper panel). 
After the Manhattan project, in 1952 Watt\cite{Watt:1952}  at Los Alamos published his seminal paper on how to represent the PFNS  with a straightforward semi-empirical analytic expression $N(E)=c.\rm {exp}(-E/a){\rm sinh}(\sqrt{(bE))}$, where $c$ is a normalization constant and $a, b$ are fit parameters. The average energy of this spectrum is $3a/2 + a^2 b/4$ MeV \cite{Kodeli:2009}. 
 His analysis included the use of 1952 data from Bonner (Los Alamos) and from Hill's (Argonne) reactor experiment, obtaining $a=1, b= 2$ with an average energy of 2.00 MeV for the $^{235}$U thermal PFNS. This is exactly the same as our best ENDF/B-VIII.0 estimate today,  see Fig.~\ref{fig:u5pfns-eav}, which comes from an IAEA standards committee evaluation\cite{Capote:2015a,Capote:2015b,Capote:2015c}, although between 1952 and 2018, the evaluated value drifted down to
 1.95 MeV on the low side and 
  up to 2.03 MeV on the high side ({\it e.g.} ENDF/B-V to VII.1)\cite{Gibson:2021} (Fig.~\ref{fig:u5pfns-eav} lower panel) .

\begin{figure}[htbp]
\begin{center}
\includegraphics[width=3.25in]{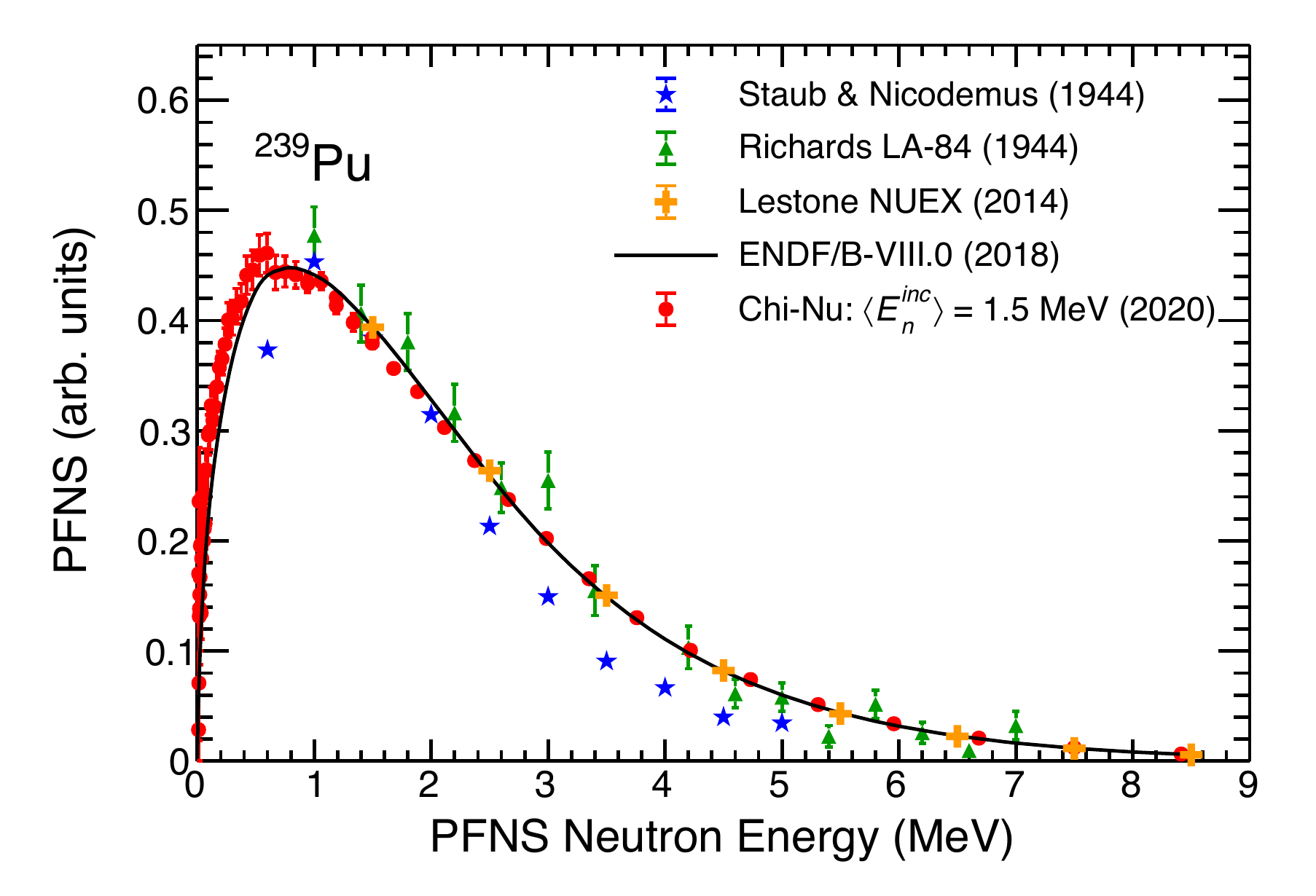}
\caption{The $^{239}$Pu prompt fission neutron spectrum for neutrons, showing measurements made during Project Y. The incident neutron  energy varies for these different experiments and while in principle the spectrum changes with energy, the changes are much smaller than the differences seen between these data sets. The black solid line is our 
modern value of the spectrum from ENDF/B-VIII.0, and the red points are recent measurements made at Los Alamos for an average incident neutron energy 
of 1.5 MeV.}
\label{fig:pu9-pfns-end}
\end{center}
\end{figure}

Plutonium spectrum data are shown in Fig~\ref{fig:pu9-pfns-end}. Both Staub and Nicodemus’ data, and those from Richards (LA-84) are reasonably accurate when compared to the best evaluated data today, ENDF-B-VIII.0 (black line). For comparison, also shown is a recent Los Alamos measurement from the LANSCE/Chi-Nu experiment\cite{Kelly:2020,Kelly:2021} (red symbols), and Lestone and Shore's\cite{Chadwick:2011,Lestone:2014} NUEX experiment  from test data (orange points), 
which is further discussed in the Conclusions.

In summary,  tremendous progress on the PFNS  had been made by the end of the Manhattan project. We now know that the local (Los Alamos) 1944 measurements at the Van de Graaff by Staub and Nicodemus, with Williams’s group, were the most accurate, but at the time it wasn’t so clear. For example, in March 1945, Serber (LA-235) \cite{Serber:1945} in T-Division was using a spectrum based on Richard’s 1944 LA-200 Minnesota data, which we now know was far too hot.

  \section{Transport and scattering}






\begin{figure}[htbp]
\begin{center}
\includegraphics[width=3.25in]{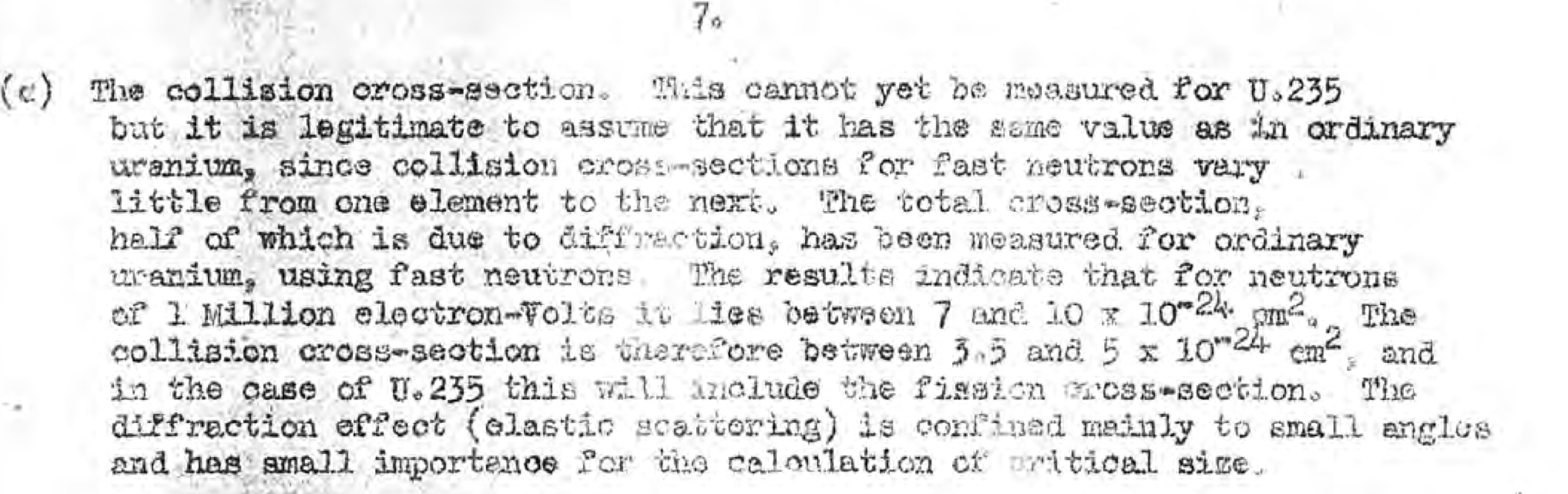}
\caption{An image from the MAUD\cite{MAUD:1941} report on the transport cross section, which they called the ``collision cross section".}
\label{fig:MAUD-transport}
\end{center}
\end{figure}
 
\begin{figure}[htbp]
\begin{center}
  \includegraphics[width=3.in]{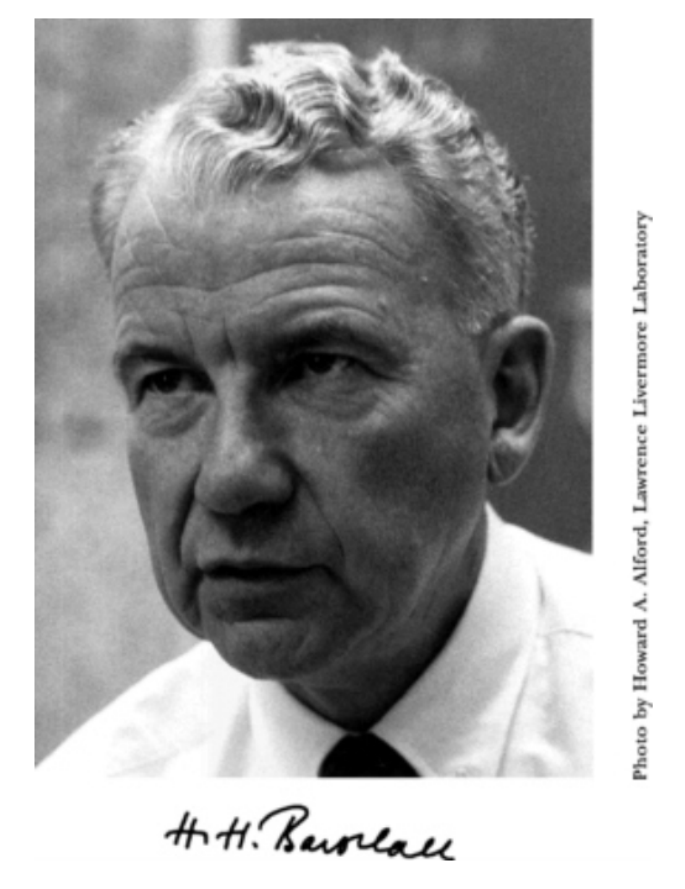}
\caption{Heinz Barschall. Credit: LLNL.}
\label{fig:barschall}
\end{center}
\end{figure}

As discussed in Section II, the transport cross section $\sigma_T$ is a key quantity needed for computing the critical mass, Eq.~\ref{eq:m-crit}.
The 1941 MAUD report\cite{MAUD:1941}  included numerous insightful discussions on the magnitude of nuclear cross sections; one of them is shown in Fig.~\ref{fig:MAUD-transport} for the transport cross section, referred to there as the ``collision cross section''. The ``most likely'' value they assessed was 5\,b for fast neutrons on uranium, with a possible range 3.5-5\,b, agreeing extremely well with our ENDF/B-VIII.0
value today of 4.85\,b at 1.5 MeV for $^{235}$U in ENDF/B-VIII.0. At the beginning of the Los Alamos Project Y, 
Oppenheimer was using 4\,b, but as we see below, Los Alamos soon moved to use values from 4.7-5\,b (See Tables~\ref{table:u5critmass}, \ref{table:pu9critmass}).

 At Los Alamos, Manley's group, 
with Heinz Barschall (Fig.~\ref{fig:barschall}) and other strong experimentalists 
was set up to measure these cross sections. After the war Barschall moved to the University of Wisconsin and maintained close collaborations with 
neutron physics researchers at Los Alamos and Livermore laboratories. With his students, he played an important role in the development of the 
optical model of neutron scattering. His contributions also included advances in nuclear medicine and
long-term service as editor of the Physical Review C.

The transport cross section has various definitions in the literature. In the 1940s, they understood it to be a quantity that included all reactions, that is, the neutron total cross section, but subtracted from this should be neutrons that are scattered into the forward direction (where the collision didn't essentially change the transport process). This led to assessments that used a transport cross section:

$$
 \sigma_T=\sigma_{\rm non} + 2\pi \int \sigma_{\rm el}(\theta)(1-\cos(\theta))\sin (\theta) \rm{d}\theta
$$
 \begin{equation}
 =\sigma_{\rm non} + (1- <{\mu}>)\sigma_{\rm el} ,
 \label{eq:transport}
\end{equation}

where  $\sigma_{\rm non}$ is the total nonelastic cross section (which includes the fission cross
section),  $\sigma_{\rm el}$ is the total elastic cross section,  $\sigma_{\rm el}(\theta)$ is the angle-differential elastic cross section and $\mu$ is the average of the cosine scattering angle. 
With ENDF/B-VIII.0 today, at 1.5 MeV we find for $^{235}$U, $\sigma_{\rm non}$=3.248\,b,  $\sigma_{\rm el}$=3.557\,b, $\mu$=0.549, so that $\sigma_T$=4.852\,b. A cross check using the PARTISN multi-group transport code is useful, since this explicitly computes a transport cross section. Tom Saller obtains 4.83\,b in a 618-group around 1.5 MeV using PARTISN, which is consistent with the point value from ENDF/B-VIII.0. 

Because the transport cross section required a determination of the angular dependence of scattering, $\sigma_{\rm el}(\theta)$, Manley's group established a program to systematically measure such differential elastic and inelastic scattering angular distributions for a wide range of materials: actinides and potential tamper and structural elements. As well as directly measuring the scattered neutron,  Barschall made a breakthrough in which he determined elastic scattering probabilities at different angles by measuring the recoiling kinetic energy of the struck target nucleus; simple kinematics allowed him to infer what neutron angle would have produced such a recoil energy\cite{Wilson:1947}.

\begin{figure}[htbp]
\begin{center}
\includegraphics[width=3.25in]{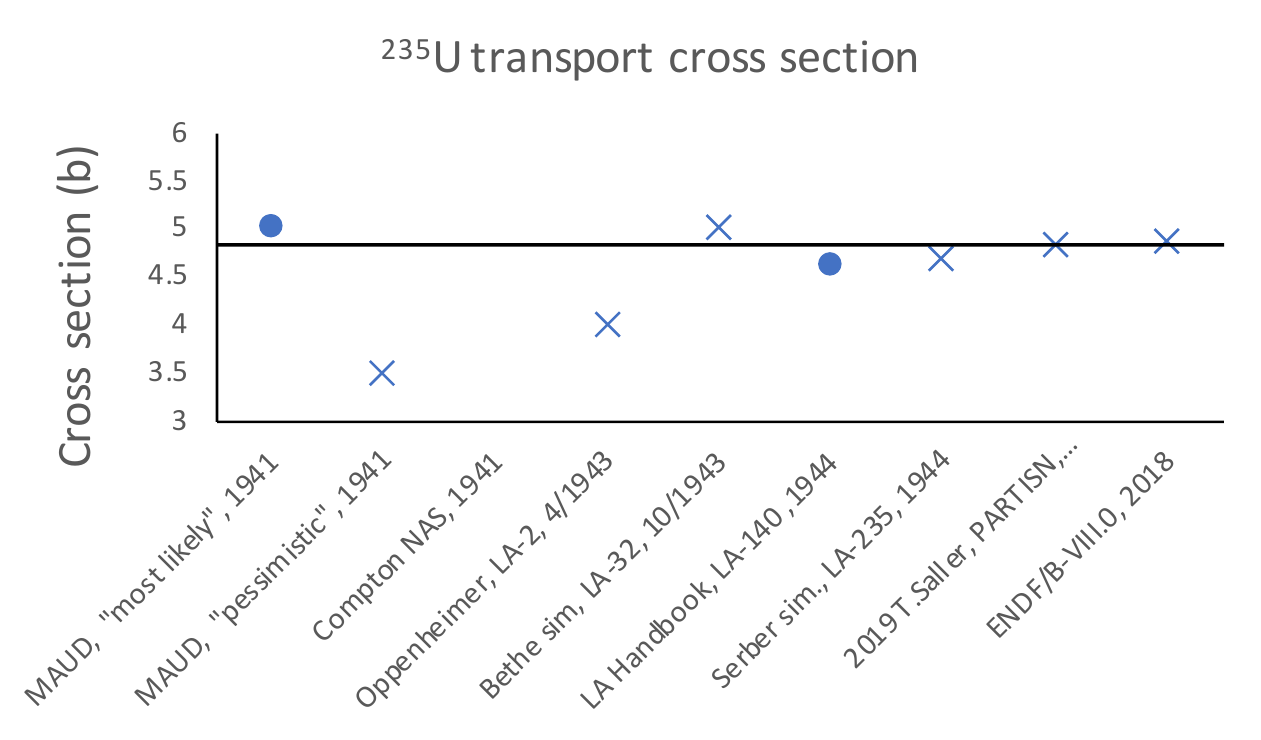}
\caption{The $^{235}$U transport cross section assessments over time. Solid symbols denote measurements, whereas crosses denote values from 
evaluations. The horizontal line shows the best value today. Measurements are shown as solid symbols whereas evaluations or estimates are crosses. }
\label{fig:u5transport}
\end{center}
\end{figure}

Figure~\ref{fig:u5transport} shows the evolution of assessments of the uranium transport cross section over time. Unlike the fission data, this was a case where the earliest assessments from 1941 on proved to be remarkably accurate.

\section{Criticality Uncertainty Reduction}

\begin{figure*}[htbp]
\begin{center}
\includegraphics[width=6.25in]{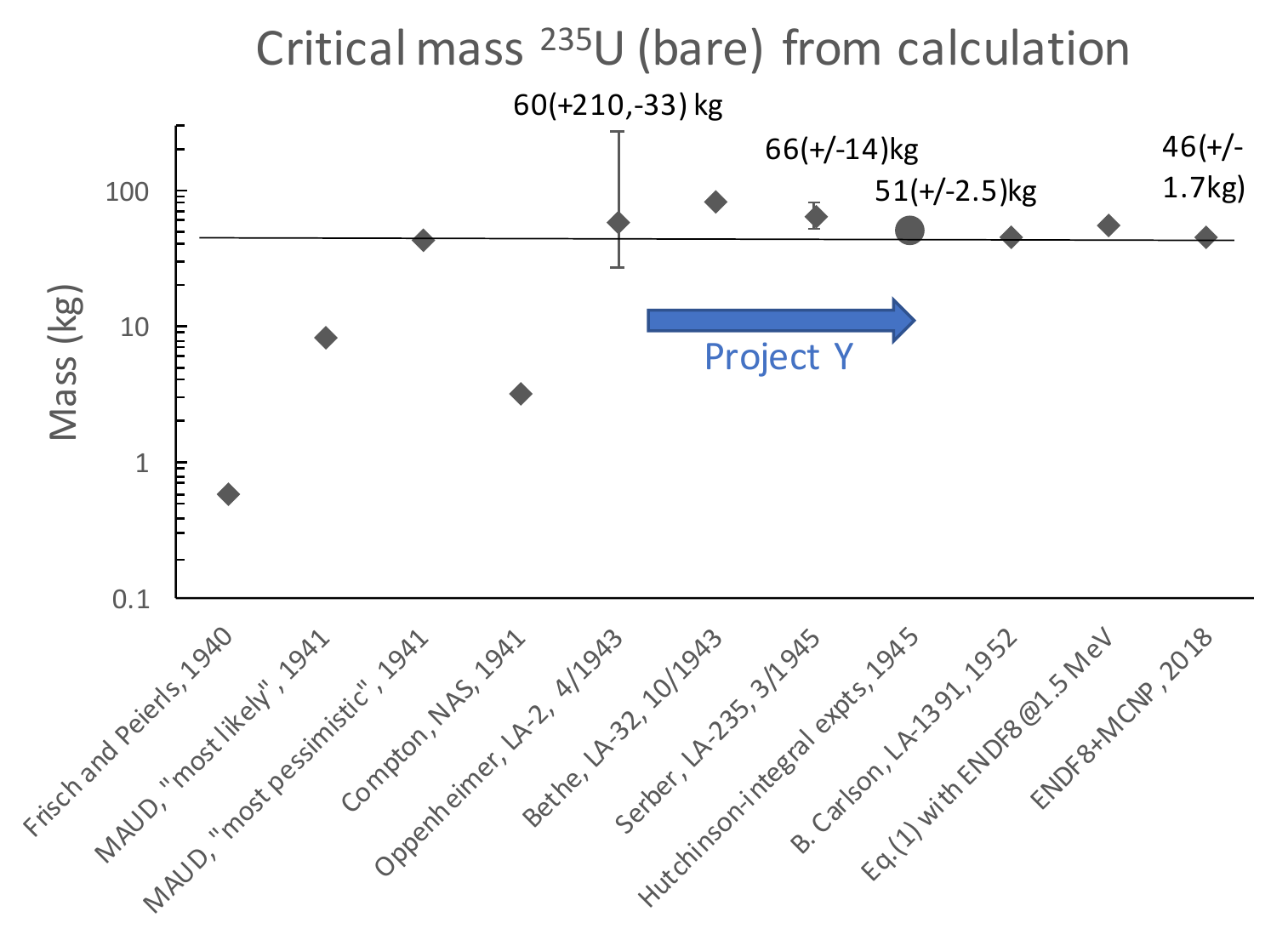}
\caption{The calculated $^{235}$U bare critical mass with uncertainties - based on differential cross section data - at the beginning of Project Y, 60(+210, -33)\,kg and at the end of Project Y,  66($\pm$14)\,kg as well as from the first 
integral experiment measurements at Los Alamos in 1945 as described by Hutchinson {\it et al.}\cite{Hutchinson:2020a}, 50.8$\pm$2.5\,kg (solid circle point).
The horizontal line represents our best understanding today from MCNP6$^{\tiny\textregistered}$ simulations using ENDF/B-VIII.0, 46.4$\pm$1.7\,kg, which were in practice calibrated to the 
Godiva bare HEU critical mass experiment in the 1950s.}
\label{fig:u5critmassUQ}
\end{center}
\end{figure*}

I close by tracing the calculated criticality uncertainty reduction that took place through Project Y's campaign to measure nuclear constants. 
In the 1940s it was not a routine practice to  provide  uncertainty assessments on measured nuclear cross sections.\footnote{
Even decades later, this could be the case. Don Barr was a
remarkable nuclear radiochemist who came to the Laboratory in the 1950s and made numerous excellent fission measurements. Don had a cheerful demeanor when discussing his measurements in comparison to other laboratory's differing results. He was confident that his results  were right, and he was right to be confident.  Once I asked him to assess the uncertainty on a measurement he had made (since today we live in a 
world of ``uncertainty quantification'', UQ). Instead of answering me, he was offended: how could I distrust his work?!}
 In the measurements described in this paper, specific uncertainty estimates were often missing.  The LA reports did, however, frequently discuss experimental problems that would lead to systematic errors, and the Project Y  scientists
worked hard to develop improved methods to mitigate these challenges. In none of the LA documents quoted were numerical uncertainties on calculated critical masses given, except for the aforementioned article by Tolman\cite{Tolman:1943} who, in March 1943,  assessed $\approx$ 50\,\% critical mass uncertainties  [$^{235}$U: 30$\pm$15\,kg;$^{239}$Pu: 10\,kg (5--20\,kg)].

At the beginning of the project, uncertainties in the fission cross section exceeded 25--50\% and by 1945 these were reduced to 5--10\% 
(today they are 1--2\%); uncertainties in the fission neutron multiplicity were initially 15\% or more and were reduced to a few percent (today they are a fraction of a \%); the neutron spectrum PFNS was measured fairly accurately by the end of the project with an average energy measured to about 10\% (today it is known to
less than 2\%), and the transport cross section, initially known to 20\%, was determined to 5\%  (and today this is known to
a few \%).

Based on these uncertainties in the fundamental data, the bare critical mass uncertainty can be determined, either using Eq.~(1) or using our modern 
computational transport codes such as MCNP6$^{\tiny\textregistered}$. Fig.~\ref{fig:u5critmassUQ} shows some calculated uncertainties for a bare $^{235}$U critical sphere using Eq.~(1), with the last value shown (from 2018) instead based on MCNP6$^{\tiny\textregistered}$. The very large error bars shown at the beginning of the project in 1943  were reduced to about 20\% by the spring of 1945 (66 kg $\pm$14 kg). As was noted, it was fortuitous that Oppenheimer's initial 1943 estimate of 60 kg for the mass was very similar 
to what Serber was calculating in the spring of 1945, using the much more accurate nuclear data measured at Los Alamos. At the beginning of Project Y, the fission cross section assessment was overestimated, whereas ${\bar{\nu}}$ was  underestimated, and these errors canceled. 

The best value today from a
ENDF/B-VIII.0 and MCNP6$^{\tiny\textregistered}$  Monte Carlo transport calculation, 46.4$\pm$1.7 kg, is shown as a horizontal line in Figure~\ref{fig:u5critmassUQ}. The figure  also shows an accurately-calculated value of 45 kg obtained in 1952  by Bengt Carlson\cite{Carlson:1952} using the Serber--Wilson neutron diffusion method  \cite{Sood:2021}  with three neutron energy groups (an approach that was developed by the US and the British during the 
Manhattan Project, 1944-1945).  Note that Carlson and Wilson both had connections with the Montreal/Chalk River Manhattan Project laboratories as described in this issue by Andrews, Andrews and Mason\cite{Andrews:2021}; Carlson came to Los Alamos in 1945, where he spent the rest of his career. Carlson would go on to invent the SN neutronics method that has been used so widely at Los Alamos and across the world \cite{Morel:1999}.

The final 1945 calculated uncertainty of 20\% in the critical mass is still quite high, and the calculated value using Eq.~(\ref{eq:m-crit}), 66 kg $\pm$14 kg, was actually 43\% above the value we know today, 46.4\,kg. That is why Los Alamos performed integral critical assembly experiments of criticality, which became possible in 1945 once substantial kg-quantities of HEU began to arrive, see Hutchinson {\it et al.}'s paper in this  Issue\cite{Hutchinson:2020a}. Hutchinson {\it et al.} shows how extrapolations from integral subcritical experiments on ``25" in 1945 gave estimates of the $^{235}$U bare critical mass of 50.8$\pm$2.5\,kg, a value only 9\% above our best understanding today, see the solid circle data point in Figure~\ref{fig:u5critmassUQ}.
Such integral measurements, which were mostly focused on tamped fast assemblies, allowed the critical masses to be determined more accurately. Likewise, today the MCNP6$^{\tiny\textregistered}$  calculation-based uncertainty of $\pm$1.7\,kg remains substantial (corresponding to 1\% uncertainty in calculated k-eff that comes from our  ENDF/B-VIII.0 nuclear data uncertainties\cite{Chadwick:2018}\footnote{This comes from a MCNP6$^{\tiny\textregistered}$ calculation by Jennifer Alwin. It is consistent with the approximate formula k-eff=$(M/M_C)^{0.3}$ \cite{Hutchinson:2021d}.}); our true knowledge of this critical mass is actually about an order of magnitude more accurate because 
the integral Godiva bare HEU assembly was measured to high accuracy (0.1\% in k-eff), a few years after the war (1951).

\section{Conclusions}

Research at Los Alamos 1943-1945 dramatically advanced our understanding of nuclear science.  
Even though many experimental measurement techniques had been established earlier, they were vastly improved and extended at Los Alamos. Methods were developed that allowed fission cross sections to be measured to unprecedented accuracy, over a wide neutron energy range. Compared to the period after the war, the electronics that they had at the time was primitive, but many of Project Y's best experimental physicists had become experts in electronics\cite{Wilson:1947}.
After the war, many of the scientists 
returned to universities across the USA, and some  moved to laboratories at  Berkeley and Oak Ridge, Argonne (1946), Brookhaven (1947), Idaho (1949), and Livermore (1952).  Their subsequent nuclear physics research at these universities and laboratories was built upon the methods in experiment and theory developed during Project Y.

The initial large systematic errors on the fission cross section  were reduced to below 5-10\% by the end of Project Y -- many of the figures in this paper have show the measured data trending with time towards our best evaluated values today. But rather remarkably, in the case of the prompt fission neutron spectrum's (PFNS) average energy,  Serber's 1943 beginning first guess  of 2.0 MeV (with an uncertainty I estimated as 300 keV) is exactly where we have ended, 2.00(1) MeV.  {\it  ``What we call the beginning is often the end, and to make an end is to make a beginning.''}\footnote{Ursula von der Leyen also used this T.S. Eliot line on the occasion of the final Brexit trade deal.}  The  large uncertainty reduction comes  from much-improved experimental and analysis methods forming the basis for the recent 2018 IAEA/ENDF/B-VIII.0 evaluation\cite{Brown:2019,Capote:2015a,Capote:2015b,Capote:2015c} for  thermal neutrons on $^{235}$U.

This paper showed how calculated critical masses from Project Y, using improved cross sections, became much more accurate with  uncertainties reduced to about 20\,\% by 1945 (66\,kg $\pm$14 kg for a bare $^{235}$U sphere). Nevertheless, they were not as accurate as those obtained from the integral criticality experiments which became possible later in 1945 once larger amounts of 
 enriched uranium and plutonium became available (giving 50.8$\pm$2.5\,kg  for $^{235}$U, a value only 9\% above our best understanding today). One might ask whether the differential 
 cross section work was ultimately valuable, given the more accurate integral measurements? It was, because these cross sections then became available for use use in 
 accurate simulation codes \cite{Sood:2021}. Their use extended well beyond addressing the original critical mass question. The rapidly-improving neutronics simulation capability opened up the ability to rapidly explore a wealth of physics questions, to advance the technologies in both defense and civilian nuclear power arenas. 

The Manhattan Project benefited from bringing together a remarkable cohort of first-rate scientists, who brought their best graduate students with them. 
This had an enduring impact on the scientific culture at the Laboratory, creating a self-confidence\footnote{
Allan Carlson also relates an occasion on which he heard Dick Taschek (a Project Y nuclear physicist who later became Los Alamos' P-Division leader) giving the first talk at the Neutron Standards and Flux Normalization meeting at Argonne in 1970; ``He subtly referred to the values of the $^{235}$U fission cross section by a newcomer who probably did not understand the problems involved with making very accurate measurements. I am sure he was thinking of the difference between the measurements of [Los Alamos'] Diven and of Poenitz [from ANL]. He mentioned the many years of effort on fission cross section measurements at Los Alamos, so they got it right.''
The irony in this anecdote, as Carlson relates it,  is that actually the newcomer Poenitz had values closer than Diven to the present best evaluation!}
 in what can be accomplished.
 
Los Alamos National Laboratory has sustained a long-standing reputation for ``all things nuclear'' (to this day it is also 
especially known for material science and computing). An example relates to 
an innovative determination of the prompt fission neutron spectrum from plutonium by Lestone {\it et al.}\cite{Chadwick:2011,Lestone:2014} that was shown in 
Fig~\ref{fig:pu9-pfns-end}. This analysis used experimental data from the nuclear test diagnostic, NUEX, which benefitted from having certain advantages compared to lab experiments (notably, higher neutron fluences with better signal to noise), but had some other complementary systematic uncertainties 
({\it e.g.} for determining the spectra below 1 MeV outgoing neutron energy). When Los Alamos published these data in 2011\cite{Chadwick:2011} and 2014\cite{Lestone:2014}, their claimed accuracy was immediately accepted by the international community, even though the Laboratory was unable to openly publish all the details that went into the analysis. The NUEX result 
was quite important in a subsequent 2016 international assessment led by Capote with a collaboration organized through the  IAEA.\cite{Capote:2015a}
This was a consequence of the international regard for Los Alamos' reputation in nuclear science, and it was only five years later that these 
data could be independently corroborated by  high-precision lab experiments by the CEA, Los Alamos,  and Livermore (The LANL-LLNL ``Chi-Nu'' data\cite{Kelly:2018,Kelly:2020,Kelly:2021}  
also shown in Fig~\ref{fig:pu9-pfns-end} -- compare the red and orange points).


But any pride in our Laboratories' technical accomplishments must, in the end, play second fiddle to the importance of sustaining a 
scientific questioning culture as articulated by our first Director, J. Robert Oppenheimer: {\it ``There must be no barriers to freedom of inquiry. There is no place for dogma in science. The scientist is free and must be free to ask any question, to doubt any assertion, to seek for any evidence, to correct any errors}.''\cite{Barnett:1949}

\vspace{0.25in}

Acknowledgments: It is a pleasure to acknowledge useful discussions with  Keegan Kelly, Jen Alwin, Allan Carlson,  Patrick Talou, Nathan Gibson, Toshihiko Kawano, Jesson Hutchinson, Roberto Capote, Fredrik Tovesson, John Lestone, Denise Neudecker, Jerry Wilhelmy, Robert Haight, Jonathan Katz, Bill Archer, Joe Schmidty, Keith Jankowski, Eric Bauge, Mark Cornock, Michael Bernardin, Bob Block, Craig Carmer, Saleem Alhumaidi, Boris Pritychenko, Tom Kunkle, Alan Carr and Richard Moore. I appreciate the scanning of historic paper documents archived at Oak Ridge by OSTI staff at Oak Ridge National Laboratory. This work was supported by the US Department of Energy through the Los Alamos National Laboratory. Los Alamos National Laboratory is operated by Triad National Security, LLC, for the National Nuclear Security Administration of the US Department of Energy under Contract No. 89233218CNA000001. This document is 
released as Los Alamos document LA-UR-20-30028, DRAFT.


\vspace{0.25in}
\noindent\rule{0.35\textwidth}{.4pt}


%

\bibliographystyle{ans_js}   
  \small\bibliography{bibliography.bib}  

\begin{thebibliography}{100}
\newcommand{\enquote}[1]{``#1''}
\providecommand{\url}[1]{\texttt{#1}}
\providecommand{\urlprefix}{URL }
\expandafter\ifx\csname urlstyle\endcsname\relax
  \providecommand{\doi}[1]{doi:\discretionary{}{}{}#1}\else
  \providecommand{\doi}{doi:\discretionary{}{}{}\begingroup
  \urlstyle{rm}\Url}\fi

\bibitem{Bohr:1939}
\textsc{B.~Bohr} and \textsc{J.~A. Wheeler}, \enquote{The Mechanism of Nuclear
  Fission,} \emph{Physical Review}, \textbf{55}, 426 (1939).

\bibitem{Chadwick:1969}
\textsc{J.~Chadwick}, \emph{Oral History. Interview by Charles Weiner,
  Cambridge, UK, Sunday April 20, 1969. American Institute of Physics} (1969);
  {https://www.aip.org/history-programs/niels-bohr-library/oral-histories/}.

\bibitem{Close:2019}
\textsc{F.~Close}, \emph{Trinity. The Treachery and Pursuit of the Most
  Dangerous Spy in History}, Allen Lane, Penguin Random House, Milton Keynes,
  UK (2019).

\bibitem{Peierls:1939}
\textsc{R.~Peierls}, \enquote{Critical Conditions in Neutron Multiplication,}
  \emph{Mathematical Proceedings of the Cambridge Philosophical Society},
  \textbf{35}, \emph{4}, 610 (1939).

\bibitem{Peierls:1990?}
\textsc{R.~Peierls}, \emph{Atomic Histories}, Springer Verlag, NY (1997).

\bibitem{Frisch:1940}
\textsc{O.~Frisch} and \textsc{R.~Peierls}, \emph{On the construction of a
  superbomb.}, The memorandum came in two parts, a “policy”
  introduction/covering note and a slightly longer technical paper. The first
  part was published in Ronald Clark, Tizard (Methuen 1965), pp. 215-7. The
  second part was published as appendix 1 to Margaret Gowing, Britain and
  atomic energy 1939-45 (Macmillan 1964), pp. 389-93. Note though that there
  are errors in the formula and/or yield estimate (which should be 4E20 ergs)
  in the following books which also reproduce it: the appendix of Serber's
  Primer; in Peierls' Atomic Histories, and in Ferenc Morton Szasz, British
  scientists and the Manhattan Project: the Los Alamos years (Macmillan 1992),
  pp. 141-7. (1940).

\bibitem{Bernstein:2011}
\textsc{J.~Bernstein}, \enquote{A memorandum that changed the world,}
  \emph{American J. Physics}, \textbf{79}, 440 (2011).

\bibitem{MAUD:1941}
\textsc{G.~P. Thomson} and \textsc{J.~Chadwick}, \emph{Report by MAUD Committee
  on the use of uranium for a bomb}, Reprinted in appendix 2 to Margaret
  Gowing, Britain and atomic energy 1939-45 (Macmillan 1964), pp. 394-426. An
  almost complete version that made its way to the US National Archives,
  missing only appendix 5, is online here: http://ipfmlibrary.org/maud.pdf.
  (1941).

\bibitem{Moore:2020c}
\textsc{R.~Moore}, \enquote{Trinity and the British Mission movie,}
  \emph{YouTube https://www.youtube.com/watch?v=7DyWmovR5EY} (2021).

\bibitem{Gowing:196?}
\textsc{M.~Gowing}, \emph{Britain and Atomic Energy 1939--1945}, St. Martin's
  Press, New York (1964).

\bibitem{NAS:1941}
\textsc{A.~Compton}, \enquote{Report to the President of the National Academy
  of Sciences by the Academy Committee on Uranium,}
  https://www.documentcloud.org/documents/3913458-Report-of-the-Uranium-Committee-Report-to-the.html,
  National Academy of Sciences, Washington DC (November 6, 1941).

\bibitem{Reed:2007}
\textsc{B.~C. Reed}, \enquote{Arthur Compton’s 1941 Report on explosive
  fission of U-235: A look at the physics,} \emph{American J. Physics},
  \textbf{75}, 1065 (2007).

\bibitem{Rhodes:19?}
\textsc{R.~Rhodes}, \emph{The Making of the Atomic Bomb}, Touchstone, New York
  (1986).

\bibitem{MED:1946}
\textsc{War-Department}, \enquote{Background Information on Development of
  Atomic Energy under Manhattan Project,}  Manhattan District History Book 1,
  General, Vol. 4, Chapter 8 Press Releases - Part I, Oak Ridge, OSTI (December
  31, 1946).

\bibitem{Chadwick:2020}
\textsc{M.~B. Chadwick}, \enquote{Manhattan Project Cross Sections \&
  Criticality In the Light of Today’s Understanding,}  LA-UR-20-24607;
  presentation June 30, 2020, Los Alamos National Laboratory (2020).

\bibitem{Norman:2015}
\textsc{E.~B. Norman}, \textsc{K.~J. Thomas}, and \textsc{K.~Telhami},
  \enquote{Seaborg's Plutonium?} \emph{arXiv:1412.7590 (2015); A. Extance,
  ``Manhattan Project Plutonium, Lost to Obscurity, Recovered by Scientists'',
  Scientific American, Jan 15 (2015)} (2015).

\bibitem{Bernstein:1996}
\textsc{J.~Bernstein}, \emph{Hitler's Uranium Club}, AIP Press, Woodbury, NY
  (1996).

\bibitem{Brown:?}
\textsc{P.~Brown}, \emph{The Neutron and the Bomb}, Oxford University Press
  (1997).

\bibitem{Bretscher:1940D37}
\textsc{E.~Bretscher}, \enquote{Report on work carried out in Cambridge,
  September-December 1940, and Report II,}  BRER D.37, Churchill Archives
  Centre, Cambridge, Tube Alloys, Dec 19 (1940).

\bibitem{Capote:2020}
\textsc{R.~Capote~{\it et al.}}, \enquote{Unrecognized Sources of Uncertainties
  (USU) in Experimental Nuclear Data,} \emph{Nuclear Data Sheets},
  \textbf{163}, 191 (2020).

\bibitem{Poenitz:1970}
\textsc{W.~P. Poenitz}, \enquote{Interpretation and intercomparison of standard
  cross sections,} \emph{Proc. EANDC Symp. on Neutron Standards and Flux
  Normalization, CONF-701002}, 338, IL, Chicago, USA (1970).

\bibitem{Hutchinson:2020a}
\textsc{J.~Hutchinson}, \textsc{J.~Alwin}, \textsc{A.~McSpaden},
  \textsc{W.~Myers}, \textsc{M.~Rising}, and \textsc{R.~Sanchez},
  \enquote{Criticality Experiments with Fast 25 and 49 Metal and Hydride
  Systems During the Manhattan Project,} \emph{This issue} (2021).

\bibitem{Hutchinson:2020b}
\textsc{R.~Kimpland}, \textsc{T.~Grove}, \textsc{P.~Jaegers},
  \textsc{R.~Malenfant}, and \textsc{W.~Myers}, \enquote{``Tickling'' the
  Dragon,} \emph{This issue, and LA-UR-20-30291-Draft} (2021).

\bibitem{Hutchinson:2020c}
\textsc{R.~Kimpland}, \textsc{T.~Grove}, \textsc{P.~Jaegers},
  \textsc{R.~Malenfant}, and \textsc{W.~Myers}, \enquote{Water Boiler Reactors
  at Los Alamos,} \emph{This issue, and LA-UR-20-30349-Draft} (2021).

\bibitem{Hutchinson:2021d}
\textsc{J.~Hutchinson}, \textsc{M.~Alwin, Rising}, and \textsc{R.~Sanchez},
  \enquote{Criticality for Trinity,} \emph{Los Alamos National Laboratory
  journal WRL} (2021).

\bibitem{Wilson:1947}
\textsc{R.~Wilson~{\it et al.}}, \enquote{Nuclear Physics,}  LA-1009, Los
  Alamos National Laboratory (1947).

\bibitem{Hoddeson:1993}
\textsc{L.~Hoddeson}, \textsc{P.~W. Henriksen}, \textsc{R.~A. Meade}, and
  \textsc{C.~Westfall}, \emph{Critical Assembly: A Technical History of Los
  Alamos During the Oppenheimer Years, 1943-1945}, Cambridge University Press,
  Cambridge, UK (1993).

\bibitem{Richards:1993}
\textsc{H.~T. Richards}, \emph{Through Los Alamos, 1945: Memoirs of a Nuclear
  Physicist}, The Arlington Place Press, Madison, WI (1993).

\bibitem{Brown:2019}
\textsc{D.~Brown}, \textsc{M.~Chadwick}, \textsc{R.~Capote},
  \textsc{A.~Kahler}, \textsc{A.~Trkov}, \textsc{M.~Herman},
  \textsc{A.~Sonzogni}, \textsc{Y.~Danon}, \textsc{A.~Carlson},
  \textsc{M.~Dunn}, \textsc{D.~Smith}, \textsc{G.~Hale}, \textsc{G.~Arbanas},
  \textsc{R.~Arcilla}, \textsc{C.~Bates}, \textsc{B.~Beck}, \textsc{B.~Becker},
  \textsc{F.~Brown}, \textsc{R.~Casperson}, \textsc{J.~Conlin},
  \textsc{D.~Cullen}, \textsc{M.-A. Descalle}, \textsc{R.~Firestone},
  \textsc{T.~Gaines}, \textsc{K.~Guber}, \textsc{A.~Hawari},
  \textsc{J.~Holmes}, \textsc{T.~Johnson}, \textsc{T.~Kawano},
  \textsc{B.~Kiedrowski}, \textsc{A.~Koning}, \textsc{S.~Kopecky},
  \textsc{L.~Leal}, \textsc{J.~Lestone}, \textsc{C.~Lubitz}, \textsc{J.~M.
  Dami{\'a}n}, \textsc{C.~Mattoon}, \textsc{E.~McCutchan},
  \textsc{S.~Mughabghab}, \textsc{P.~Navratil}, \textsc{D.~Neudecker},
  \textsc{G.~Nobre}, \textsc{G.~Noguere}, \textsc{M.~Paris}, \textsc{M.~Pigni},
  \textsc{A.~Plompen}, \textsc{B.~Pritychenko}, \textsc{V.~Pronyaev},
  \textsc{D.~Roubtsov}, \textsc{D.~Rochman}, \textsc{P.~Romano},
  \textsc{P.~Schillebeeckx}, \textsc{S.~Simakov}, \textsc{M.~Sin},
  \textsc{I.~Sirakov}, \textsc{B.~Sleaford}, \textsc{V.~Sobes},
  \textsc{E.~Soukhovitskii}, \textsc{I.~Stetcu}, \textsc{P.~Talou},
  \textsc{I.~Thompson}, \textsc{S.~van~der Marck}, \textsc{L.~Welser-Sherrill},
  \textsc{D.~Wiarda}, \textsc{M.~White}, \textsc{J.~Wormald},
  \textsc{R.~Wright}, \textsc{M.~Zerkle}, \textsc{G.~\v{Z}erovnik}, and
  \textsc{Y.~Zhu}, \enquote{{ENDF/B-VIII.0}: The {8$^{th}$} Major Release of
  the Nuclear Reaction Data Library with {CIELO}-project Cross Sections, New
  Standards and Thermal Scattering Data,} \emph{Nuclear Data Sheets},
  \textbf{148}, 1  (2018); {https://doi.org/10.1016/j.nds.2018.02.001}.,
  \urlprefix\url{https://www.sciencedirect.com/science/article/pii/S0090375218300206},
  special Issue on Nuclear Reaction Data.

\bibitem{Chadwick:2019}
\textsc{M.~Chadwick}, \textsc{P.~Moller}, and \textsc{J.~E. Lynn},
  \emph{Plutonium Nuclear Science}, In the Plutonium Handbook, Vol. 1 pp.
  17-92, Eds. D.L. Clark, D.A Geeson, R.J. Hanrahan Jr., American Nuclear
  Society Publishing, IL (2019).

\bibitem{Oppenheimer:1942}
\textsc{J.~R. Oppenheimer}, \enquote{Letter to R. Peierls, Nov. 1., (1992),}
  NSRC A-84-19-64-19, Los Alamos National Laboratory (1943).

\bibitem{Martz:2020}
\textsc{J.~Martz}, \textsc{F.~Freibert}, \textsc{D.~Clark}, and
  \textsc{S.~Crockett}, \enquote{The Taming of Plutonium: Pu Metallurgy and the
  Manhattan Project,} \emph{This issue} (2021).

\bibitem{Hanson:2020}
\textsc{S.~Hanson} and \textsc{W.~Oldham}, \enquote{Radiochemical methods from
  Trinity to Today,} \emph{This issue} (2021).

\bibitem{Oppenheimer:1943}
\textsc{J.~R. Oppenheimer}, \enquote{Los Alamos Conference, April 15, 1943,}
  LA-2, Los Alamos National Laboratory (1943).

\bibitem{McMillan:1943}
\textsc{E.~McMillan}, \enquote{Los Alamos Conference, April 15, 1943: Follow on
  discussions,}  LA-4, Los Alamos National Laboratory (1943).

\bibitem{Serber:1992}
\textsc{R.~Serber}, \emph{The Los Alamos Primer}, University of California
  Press (1992).

\bibitem{Tolman:1943}
\textsc{R.~L. Tolman}, \enquote{Memorandum on Los Alamos Project as of March
  1943,}  NSRC A-84-19-64-19, Los Alamos National Laboratory (1943).

\bibitem{Chadwick:2020a}
\textsc{T.~A. Chadwick} and \textsc{M.~Chadwick}, \enquote{Who Invented the
  Trinity Nuclear Test's Christy Gadget?: Patents and Evidence from the
  Archives,} \emph{This issue; Los Alamos report LA-UR-20-27638, version 2}
  (2021).

\bibitem{Chadwick:2010}
\textsc{H.~D. Selby}, \textsc{M.~R. Mac~Innes}, \textsc{D.~W. Barr},
  \textsc{A.~L. Keksis}, \textsc{R.~A. Meade}, \textsc{C.~J. Burns},
  \textsc{M.~B. Chadwick}, and \textsc{T.~C. Wallstrom}, \enquote{Fission
  product data measured at Los Alamos for fission spectrum and thermal neutrons
  on 239Pu, 235U, 238U,} \emph{Nuclear Data Sheets}, \textbf{111}, \emph{12},
  2891 (2010).

\bibitem{Archer:2020a}
\textsc{B.~Archer}, \enquote{The Computing Facility at Los Alamos During the
  Manhattan Project,} \emph{This issue} (2021).

\bibitem{Archer:2020b}
\textsc{B.~Archer}, \enquote{Los Alamos Hydrodynamics Simulations during World
  War II,} \emph{This issue} (2021).

\bibitem{Lewis:2020}
\textsc{N.~Lewis}, \enquote{Trinity by the Numbers: The Computing Effort that
  Made Trinity Possible,} \emph{This issue and Los Alamos report
  LA-UR-20-30196} (2021).

\bibitem{Bethe:1943}
\textsc{H.~Bethe},  LA--32, Los Alamos National Laboratory (1943).

\bibitem{Serber:1945}
\textsc{R.~Rarita} and \textsc{R.~Serber}, \enquote{Critical Masses and
  Multiplication Rates,}  LA--235, Los Alamos National Laboratory (1945).

\bibitem{Bethe:1943b}
\textsc{H.~A. Bethe} and \textsc{R.~Christy}, \enquote{The Los Alamos Handbook
  of Nuclear Physics,}  LA--11, Los Alamos (1943).

\bibitem{Hansen:1961}
\textsc{G.~Hansen} and \textsc{W.~Roach}, \enquote{SIX AND SIXTEEN GROUP CROSS
  SECTIONS FOR FAST AND INTERMEDIATE CRITICAL ASSEMBLIES,}  LAMS-2543, Los
  Alamos National Laboratory (1961).

\bibitem{Goldsmith:1947}
\textsc{H.~H. Goldsmith}, \textsc{H.~W. Ibser}, and \textsc{B.~T. Feld},
  \enquote{Neutron Cross Section of the Elements,} \emph{Review Modern
  Physics}, \textbf{19}, \emph{4}, 259 (1947).

\bibitem{NSR1952HUZZ}
\textsc{D.~J. {Hughes}}, \textsc{T.~W. {Bonner}}, \textsc{H.~{Goldstein}},
  \textsc{W.~W. {Havens}}, \textsc{W.~W. {Havens}}, \textsc{L.~{Kaplan}},
  \textsc{C.~O. {Muehlhause}}, \textsc{A.~H. {Snell}}, \textsc{J.~R. {Stehn}},
  \textsc{T.~M. {Snyder}}, \textsc{R.~F. {Taschek}}, \textsc{A.~{Wattenberg}},
  and \textsc{C.~W. {Zabel}}, \enquote{Neutron cross sections; a compilation of
  the AEC Neutron Cross Section Advisory Group. May 15, 1952,}  BNL-170 (1952).

\bibitem{NSR1955HUZY}
\textsc{D.~J. {Hughes}} and \textsc{J.~A. {Harvey}}, \enquote{Neutron Cross
  Sections. July 1, 1955,}  BNL-325 (1955).

\bibitem{Holden:2001}
\textsc{N.~Holden}, \enquote{A short history of nuclear data and its
  evaluation,}  BNL-52675; 51st Meeting of the USDOE CSEWG: A CSEWG
  Retrospective, 35th Anniversary of CSEWG, Brookhaven National Laboratory
  (1945).

\bibitem{Plompen:2020}
\textsc{A.~J.~M. Plompen}, \textsc{O.~Cabellos}, \textsc{C.~De~Saint~Jean},
  \textsc{M.~Fleming}, \textsc{M.~Algora}, \textsc{M.~Angelone},
  \textsc{P.~Archier}, \textsc{E.~Bauge}, \textsc{O.~Bersillon}
  \textsc{et~al.}, \enquote{The joint evaluated fission and fusion nuclear data
  library, JEFF-3.3,} \emph{The European Physical Journal A}, \textbf{56}, 108
  (2020).

\bibitem{Chadwick:2006}
\textsc{M.~Chadwick}, \textsc{P.~Oblo{\v z}insk{\' y}}, \textsc{M.~Herman},
  \textsc{N.~Greene}, \textsc{R.~McKnight}, \textsc{D.~Smith},
  \textsc{P.~Young}, \textsc{R.~MacFarlane}, \textsc{G.~Hale},
  \textsc{S.~Frankle}, \textsc{A.~Kahler}, \textsc{T.~Kawano},
  \textsc{R.~Little}, \textsc{D.~Madland}, \textsc{P.~Moller},
  \textsc{R.~Mosteller}, \textsc{P.~Page}, \textsc{P.~Talou},
  \textsc{H.~Trellue}, \textsc{M.~White}, \textsc{W.~Wilson},
  \textsc{R.~Arcilla}, \textsc{C.~Dunford}, \textsc{S.~Mughabghab},
  \textsc{B.~Pritychenko}, \textsc{D.~Rochman}, \textsc{A.~Sonzogni},
  \textsc{C.~Lubitz}, \textsc{T.~Trumbull}, \textsc{J.~Weinman},
  \textsc{D.~Brown}, \textsc{D.~Cullen}, \textsc{D.~Heinrichs},
  \textsc{D.~McNabb}, \textsc{H.~Derrien}, \textsc{M.~Dunn},
  \textsc{N.~Larson}, \textsc{L.~Leal}, \textsc{A.~Carlson}, \textsc{R.~Block},
  \textsc{J.~Briggs}, \textsc{E.~Cheng}, \textsc{H.~Huria}, \textsc{M.~Zerkle},
  \textsc{K.~Kozier}, \textsc{A.~Courcelle}, \textsc{V.~Pronyaev}, and
  \textsc{S.~van~der Marck}, \enquote{{ENDF/B-VII.0}: Next Generation Evaluated
  Nuclear Data Library for Nuclear Science and Technology,} \emph{Nuclear Data
  Sheets}, \textbf{107}, \emph{12}, 2931 (2006); {10.1016/j.nds.2006.11.001}.,
  \urlprefix\url{http://www.sciencedirect.com/science/journal/00903752}.

\bibitem{Chadwick:2011}
\textsc{M.~Chadwick}, \textsc{M.~Herman}, \textsc{P.~Oblo{\v z}insk{\' y}},
  \textsc{M.~Dunn}, \textsc{Y.~Danon}, \textsc{A.~Kahler}, \textsc{D.~Smith},
  \textsc{B.~Pritychenko}, \textsc{G.~Arbanas}, \textsc{R.~Arcilla},
  \textsc{R.~Brewer}, \textsc{D.~Brown}, \textsc{R.~Capote},
  \textsc{A.~Carlson}, \textsc{Y.~Cho}, \textsc{H.~Derrien}, \textsc{K.~Guber},
  \textsc{G.~Hale}, \textsc{S.~Hoblit}, \textsc{S.~Holloway},
  \textsc{T.~Johnson}, \textsc{T.~Kawano}, \textsc{B.~Kiedrowski},
  \textsc{H.~Kim}, \textsc{S.~Kunieda}, \textsc{N.~Larson}, \textsc{L.~Leal},
  \textsc{J.~Lestone}, \textsc{R.~Little}, \textsc{E.~McCutchan},
  \textsc{R.~MacFarlane}, \textsc{M.~MacInnes}, \textsc{C.~Mattoon},
  \textsc{R.~McKnight}, \textsc{S.~Mughabghab}, \textsc{G.~Nobre},
  \textsc{G.~Palmiotti}, \textsc{A.~Palumbo}, \textsc{M.~Pigni},
  \textsc{V.~Pronyaev}, \textsc{R.~Sayer}, \textsc{A.~Sonzogni},
  \textsc{N.~Summers}, \textsc{P.~Talou}, \textsc{I.~Thompson},
  \textsc{A.~Trkov}, \textsc{R.~Vogt}, \textsc{S.~van~der Marck},
  \textsc{A.~Wallner}, \textsc{M.~White}, \textsc{D.~Wiarda}, and
  \textsc{P.~Young}, \enquote{{ENDF/B-VII.1} Nuclear Data for Science and
  Technology: Cross Sections, Covariances, Fission Product Yields and Decay
  Data,} \emph{Nuclear Data Sheets}, \textbf{112}, \emph{12}, 2887  (2011);
  {10.1016/j.nds.2011.11.002}.,
  \urlprefix\url{http://www.sciencedirect.com/science/article/pii/S009037521100113X},
  special Issue on ENDF/B-VII.1 Library.

\bibitem{Chamberlain:1942}
\textsc{O.~Chamberlain}, \textsc{J.~W. Kennedy}, and \textsc{E.~Segr{\`e}},
  \enquote{Fission Cross Section of $^{235}$U for Intermediate Velocity
  Neutrons,}  CF-403 (available from OSTI, ORNL); see Kennedy and Segre letter
  to Manley in NSRC, A84-19-Box 49-9, Nov 2, 1942; see also Wilson LA-1009 Sec.
  3.4.3 and Critical Assembly p.51., University of California, Berkeley (Oct
  24, 1942).

\bibitem{Heydenburg:1942}
\textsc{N.~P. Heydenburg} and \textsc{R.~Meyer}, \enquote{Fission Cross
  Sections of $^{235}$U and $^{238}$U for Fast Neutrons,}  CF-636 (available
  from OSTI, ORNL), Carnegie Institution of Washington (1943).

\bibitem{Hanson:1943}
\textsc{A.~O. Hanson}, \enquote{Measurement of Fission Cross Sections by the
  Use of a Coincidence Proportional Counter,}  CF-618 (available from OSTI,
  ORNL), University of Wisconsin (April, 1943).

\bibitem{Benedict:1943}
\textsc{D.~L. Benedict} and \textsc{A.~O. Hanson}, \enquote{A Measurement of
  the 25 Fission Cross Section for 0.53 MeV Neutrons by the Manganese Solution
  Method,}  CF-638 (available from OSTI, ORNL), University of Wisconsin (April,
  1943).

\bibitem{Williams:1943}
\textsc{J.~H. Williams}, \enquote{The Scattering Cross Section of Hydrogen and
  the Cross Section of Carbon for Neutrons of Energies 0.35 to 6.0 MeV,}
  CF-599 (available from OSTI, ORNL), Los Alamos National Laboratory (April,
  1943).

\bibitem{Bretscher:1944}
\textsc{E.~Bretscher}, \textsc{A.~P. French}, and \textsc{E.~B.~M. Martin},
  \enquote{Determination of the U(235) and U(238) fission cross section for
  neutron energies of 2 to 4 MeV,}  AB 4/405; reported in Koontz LA-128 Fig.
  19, LA-1009 Fig.~7., UK National Archives (Jan 1, 1944).

\bibitem{Chadwick:1944}
\textsc{B.~B. Kinsey} and \textsc{S.~G. Cohen}, \enquote{Measurement of the
  fission cross-section of uranium isotopes,}  AB 4/480; reported as
  Chadwick-Kinsey data in Koontz LA-128 Fig. 18, UK National Archives (May 1,
  1944).

\bibitem{Wiegland:1943}
\textsc{C.~Wiegland} and \textsc{E.~Segr{\`e}}, \enquote{Fission Cross Section
  of 94-239 for 220 keV Neutrons and Ra Plus Be Neutrons,}  LA--21, Los Alamos
  National Laboratory (1943).

\bibitem{Koontz:1944}
\textsc{P.~G. Koontz}, \textsc{T.~A. Hall}, and \textsc{B.~Rossi},
  \enquote{Absolute Values of the Fission Cross Sections of 25 and 28 from 0.28
  to 2.50 MeV,}  LA--128, Los Alamos National Laboratory (August 23, 1944).

\bibitem{Williams:1944}
\textsc{J.~H. Williams}, \enquote{Cross Sections for Fission of 25, 49, 28, 11,
  37, 00, 02, Boron and Lithium,}  LA-150, Los Alamos National Laboratory
  (1944).

\bibitem{Snyder:1944}
\textsc{T.~M. Snyder} and \textsc{J.~H. Williams}, \enquote{Number of Neutrons
  per Fission for 25 and 49,}  LA-102, Los Alamos National Laboratory (1944).

\bibitem{Wilson:1944}
\textsc{R.~Wilson}, \enquote{Neutrons per Fission from 49 Compared with 25 --
  Slightly Delayed Neutrons from 49 to 25,}  LA-00104, Los Alamos National
  Laboratory (1944).

\bibitem{Wattenburg:1947}
\textsc{A.~Wattenburg}, \enquote{Photo-Neutron Sources,}  National Research
  Council Division of Mathematical and Sciences Committee on Nuclear Science
  report; NP-1100, Argonne National Laboratory (1947).

\bibitem{Peierls:1945}
\textsc{R.~Moore}, \enquote{Peierls's 1945 Paper `Outline of the Development of
  the British Tube Alloy Project',} \emph{This issue} (2021).

\bibitem{Bretscher:1940D37b}
\textsc{E.~Bretscher}, \enquote{Attempt to determine the fission cross section
  of uranium for neutrons of several hundred kv. energy,}  Tube Alloys, BRER
  D.37, Churchill Archives Centre, Cambridge (1940).

\bibitem{Moore:2021a}
\textsc{R.~Moore}, \enquote{Peierls's 1945 Paper ``Outline of the Development
  of the British Tube Alloy Project'',} \emph{Nuclear Technology, this issue}
  (2021).

\bibitem{Agnew:1985}
\textsc{H.~Agnew}, \enquote{Colloquium: Manhattan Project Veteran Harold Agnew
  Presents, `The Way It Was No: Regrets',}  Youtube, Los Alamos National
  Laboratory (1985).

\bibitem{Turner:1940}
\textsc{L.~A. Turner}, \enquote{Nuclear Fission,} \emph{Review Modern Physics},
  \textbf{12}, \emph{12}, 1 (1940).

\bibitem{Manley:1943}
\textsc{J.~H. Manley},  NSRC A84-019-49-9, Los Alamos National Laboratory
  (1983).

\bibitem{Dahl:2002}
\textsc{P.~F. Dahl}, \emph{From Nuclear Transmutation to Nuclear Fission,
  1932-1939}, Institute of Physics, Bristol (2002).

\bibitem{Diven:?}
\textsc{B.~C. Diven}, \textsc{J.~H. Manley}, and \textsc{R.~F. Taschek},
  \enquote{Nuclear Data,}  Los Alamos Science, Los Alamos National Laboratory
  (1983).

\bibitem{Bailey:1946}
\textsc{C.~L. Bailey}, \textsc{W.~E. Bennett}, \textsc{T.~Bergstralh},
  \textsc{R.~G. Nuckolls}, \textsc{H.~T. Richards}, and \textsc{J.~H.
  Williams}, \enquote{The Neutron-Proton and Neutron-Carbon Scattering Cross
  Sections for Fast Neutrons,} \emph{Physical Review}, \textbf{70}, \emph{9},
  583 (1946).

\bibitem{Bretscher:1944a}
\textsc{E.~Bretscher}, \textsc{Davis}, \textsc{Inglis}, and
  \textsc{V.~Weisskopf},  LA-140, Los Alamos National Laboratory (August 23,
  1944).

\bibitem{Bailey:1945}
\textsc{C.~Bailey}, \enquote{Fission cross section of uranium 235 from 20 to
  500 KeV,}  LA-00447, Los Alamos National Laboratory (1945).

\bibitem{Frisch:1946}
\textsc{D.~Frisch}, \enquote{Measurement of Neutron Fluxes in the Region 30 to
  500 keV with Proportional Counters,}  LA--00554, Los Alamos National
  Laboratory (April 22, 1946).

\bibitem{Carlson:19?}
\textsc{A.~D. Carlson}, \emph{Proc. International Conf. on Nucl. Data for
  Sci.and Tech., Antwerp, Belgium, 1982}, (and IAEA-TECDOC-335,162, (1985)),
  338 (1982).

\bibitem{Lisowski:2006}
\textsc{P.~W. Lisowski} and \textsc{K.~F. Schoenberg}, \enquote{The Los Alamos
  Neutron Science Center,} \emph{Nucl. Instrum. Methods}, \textbf{A562},
  \emph{2}, 910 (2006).

\bibitem{Chamberlain:19?}
\textsc{O.~Chamberlain}, \textsc{J.~W. Kennedy}, \textsc{E.~Segr{\`e}}, and
  \textsc{A.~Wahl},  quoted in Weigland, LA-21 as CN-469 but perhaps this is a
  typo and should be CF-469; the author has not yet been able to track this
  down, University of California, Berkeley (1942 - a guess).

\bibitem{Heydenburg:1943b}
\textsc{N.~P. Heydenburg} and \textsc{R.~C. Meyer}, \enquote{Comparative
  Fission Cross Section for Element 49 and Normal Uranium,}  Los Alamos NSRC
  A-84-019, 49-9, April 9 1943 letter to Manley at Los Alamos, Carnegie
  Institution of Washington (1943).

\bibitem{Taschek:1943}
\textsc{R.~F. Taschek} and \textsc{J.~Williams},  LA-28, Los Alamos National
  Laboratory (1943).

\bibitem{Heydenburg:1943a}
\textsc{N.~P. Heydenburg} and \textsc{R.~C. Meyer},  CF-626 (OSTI, ORNL),
  quoted in Taschek, LA-28, Carnegie Institution of Washington (1943).

\bibitem{Bacher:1944}
\textsc{R.~Bacher}, \enquote{R. Bacher to J.R. Oppenheimer and Los Alamos’
  Governing Board,}  NSRC Box Folder 46 (and Critical Assembly p.196), Los
  Alamos National Laboratory (Aug 20, 1944).

\bibitem{Tovesson:19?}
\textsc{F.~Tovesson} and \textsc{T.~Hill}, \enquote{Cross Sections for
  $^{239}$Pu(n, f) and $^{241}$Pu(n, f) in the Range En = 0.01 eV to 200 MeV,}
  \emph{Nucear Science and Engineering}, \textbf{165}, \emph{2}, 224 (2010).

\bibitem{DeWire:1944}
\textsc{J.~DeWire}, \enquote{The Ratio of the Fission Cross Sections of 49 and
  25 for Thermal Neutrons and the Ratio [1+$\alpha$(49)]/[1+$\alpha$(25)],}
  LA--103, Los Alamos National Laboratory (June 30, 1944).

\bibitem{Hawkins:?}
\textsc{D.~Hawkins}, \enquote{Manhattan District History Project Y. The Los
  Alamos Project,}  LA-02532-MS; Book 8 Vol. 2 Technical, Los Alamos National
  Laboratory (April 29, 1947).

\bibitem{Williams:1943b}
\textsc{J.~Williams},  P-Div Progress report LAMS-4 July 1943; LAMS-7, Los
  Alamos National Laboratory (Aug 15, 1943).

\bibitem{Williams:1943c}
\textsc{J.~H. Williams}, \enquote{Measurement of $\nu_{49}/\nu_{25}$,}  LA-25,
  Los Alamos National Laboratory (1943).

\bibitem{Williams:1944b}
\textsc{J.~H. Williams},  R-Div Progress Report 4, LAMS-151, Los Alamos
  National Laboratory (Dec 1, 1944).

\bibitem{Manley:1944}
\textsc{J.~Manley},  LAMS-95, Progress Report Number 20 of the Experimental
  Physics Division, Los Alamos National Laboratory (May 1, 1944).

\bibitem{Bennett:1943}
\textsc{W.~E. Bennett} and \textsc{H.~T. Richards}, \enquote{A Discussion of
  the Fission Neutron Spectrum,}  LAMS-1, Los Alamos National Laboratory
  (1943).

\bibitem{Chadwick:2018}
\textsc{M.~B. {Chadwick}}, \textsc{R.~{Capote}}, \textsc{A.~{Trkov}},
  \textsc{M.~W. {Herman}}, \textsc{D.~A. {Brown}}, \textsc{G.~M. {Hale}},
  \textsc{A.~C. {Kahler}}, \textsc{P.~{Talou}}, \textsc{A.~J. {Plompen}},
  \textsc{P.~{Schillebeeckx}}, \textsc{M.~T. {Pigni}}, \textsc{L.~{Leal}},
  \textsc{Y.~{Danon}}, \textsc{A.~D. {Carlson}}, \textsc{P.~{Romain}},
  \textsc{B.~{Morillon}}, \textsc{E.~{Bauge}}, \textsc{F.~J. {Hambsch}},
  \textsc{S.~{Kopecky}}, \textsc{G.~{Giorginis}}, \textsc{T.~{Kawano}},
  \textsc{J.~{Lestone}}, \textsc{D.~{Neudecker}}, \textsc{M.~{Rising}},
  \textsc{M.~{Paris}}, \textsc{G.~P.~A. {Nobre}}, \textsc{R.~{Arcilla}},
  \textsc{O.~{Cabellos}}, \textsc{I.~{Hill}}, \textsc{E.~{Dupont}},
  \textsc{A.~J. {Koning}}, \textsc{D.~{Cano-Ott}}, \textsc{E.~{Mendoza}},
  \textsc{J.~{Balibrea}}, \textsc{C.~{Paradela}}, \textsc{I.~{Duran}},
  \textsc{J.~{Qian}}, \textsc{Z.~{Ge}}, \textsc{T.~{Liu}},
  \textsc{L.~{Hanlin}}, \textsc{X.~{Ruan}}, \textsc{W.~{Haicheng}},
  \textsc{M.~{Sin}}, \textsc{G.~{Noguere}}, \textsc{D.~{Bernard}},
  \textsc{R.~{Jacqmin}}, \textsc{O.~{Bouland}}, \textsc{C.~{De Saint Jean}},
  \textsc{V.~G. {Pronyaev}}, \textsc{A.~V. {Ignatyuk}}, \textsc{K.~{Yokoyama}},
  \textsc{M.~{Ishikawa}}, \textsc{T.~{Fukahori}}, \textsc{N.~{Iwamoto}},
  \textsc{O.~{Iwamoto}}, \textsc{S.~{Kunieda}}, \textsc{C.~R. {Lubitz}},
  \textsc{M.~{Salvatores}}, \textsc{G.~{Palmiotti}}, \textsc{I.~{Kodeli}},
  \textsc{B.~{Kiedrowski}}, \textsc{D.~{Roubtsov}}, \textsc{I.~{Thompson}},
  \textsc{S.~{Quaglioni}}, \textsc{H.~I. {Kim}}, \textsc{Y.~O. {Lee}},
  \textsc{U.~{Fischer}}, \textsc{S.~{Simakov}}, \textsc{M.~{Dunn}},
  \textsc{K.~{Guber}}, \textsc{J.~I. {Marquez Damian}}, \textsc{F.~{Cantargi}},
  \textsc{I.~{Sirakov}}, \textsc{N.~{Otuka}}, \textsc{A.~{Daskalakis}},
  \textsc{B.~J. {McDermott}}, and \textsc{S.~C. {van der Marck}},
  \enquote{CIELO Collaboration Summary Results: International Evaluations of
  Neutron Reactions on Uranium, Plutonium, Iron, Oxygen and Hydrogen,}
  \emph{Nucl.Data Sheets}, \textbf{148}, 189 (2018).

\bibitem{Lestone:2020}
\textsc{J.~P. Lestone}, \emph{This issue} (2021).

\bibitem{Feather:1942}
\textsc{N.~Feather}, \enquote{Emission of Neutrons from Moving Fission
  Fragments',}  BM-143, British Mission (February, 17, 1942).

\bibitem{Kelly:2018}
\textsc{M.~Devlin}, \textsc{K.~J. Kelly}, and \textsc{J.~A. Gomez},
  \enquote{The Prompt Fission Neutron Spectrum of $^{235}$U(n, f) below 2.5 MeV
  for Incident Neutrons from 0.7 to 20 MeV,} \emph{Nuclear Data Sheets},
  \textbf{148}, 322 (2018).

\bibitem{Kelly:2020}
\textsc{K.~J. Kelly~{\it et al.}}, \enquote{Measurement of the $^{239}$Pu(n, f)
  prompt fission neutron spectrum from 10 keV to 10 MeV induced by neutrons of
  energy 1–20 MeV,} \emph{Physical Review C}, \textbf{102}, \emph{3}, 322
  (2018).

\bibitem{Kelly:2021}
\textsc{K.~J. Kelly}, \textsc{M.~P.}, \textsc{J.~Taieb}, \textsc{M.~Devlin},
  \textsc{D.~Neudecker}, \textsc{R.~C. Haight}, \textsc{G.~Be{\'e}lier},
  \textsc{B.~Laurent}, \textsc{E.~Bauge}, \textsc{M.~B. Chadwick}
  \textsc{et~al.}, \enquote{{Comparison of Results from Recent NNSA and CEA
  Measurements of the $^{239}$Pu($n$,$f$) Prompt Fission Neutron Spectrum},}
  \emph{Nucl.~Data Sheets}, LA--UR--20--30,506 (submitted).

\bibitem{Kodeli:2009}
\textsc{I.~Kodeli}, \textsc{A.~Trkov}, \textsc{R.~Capote}, \textsc{Y.~Nagaya},
  and \textsc{V.~Maslov}, \enquote{Evaluation and use of the Prompt Fission
  Neutron Spectrum and Spectra Covariance Matrices in Criticality and
  Shielding,} \emph{Nucl. Instr. Methods}, \textbf{A610}, 540 (2009).

\bibitem{Kawano:2013}
\textsc{T.~Kawano}, \textsc{P.~Talou}, \textsc{I.~Stetcu}, and \textsc{M.~B.
  Chadwick}, \enquote{Statistical and evaporation models for the neutron
  emission energy spectrum in the center-of-mass system from fission
  fragments,} \emph{Nuclear Physics}, \textbf{A913}, 51 (2013).

\bibitem{Bethe:1937}
\textsc{H.~Bethe}, \enquote{Nuclear Physics. B. Nuclear Dynamics, Theoretical,}
  \emph{Physical Review}, \textbf{9}, \emph{2}, 69 (1937).

\bibitem{Christy:1942}
\textsc{R.~Christy} and \textsc{J.~Manley}, \enquote{The Energy of Fission
  Neutrons,}  CF-209 (available from OSTI, ORNL), Chicago University (July 31,
  1942).

\bibitem{Capote:2015a}
\textsc{R.~Capote~{\it et al.}}, \enquote{Prompt Fission Neutron Spectra of
  Actinides,} \emph{Nuclear Data Sheets}, \textbf{131}, 1 (2015).

\bibitem{Capote:2015b}
\textsc{A.~Trkov}, \textsc{R.~Capote}, and \textsc{V.~G. Pronyaev},
  \enquote{Current Issues in Nuclear Data Evaluation Methodology: $^{235}$U
  Prompt Fission Neutron Spectra and Multiplicity for Thermal Neutrons,}
  \emph{Nuclear Data Sheets}, \textbf{123}, 8 (2015).

\bibitem{Capote:2015c}
\textsc{A.~Trkov} and \textsc{R.~Capote}, \enquote{Evaluation of the Prompt
  Fission Neutron Spectrum of Thermal-neutron Induced Fission in $^{235}$U,}
  \emph{Physics Procedia}, \textbf{64}, 48 (2015).

\bibitem{Richards:1944}
\textsc{H.~T. Richards}, \enquote{The Fission Spectrum of 25,}  LA--60, Los
  Alamos National Laboratory (February 11, 1944).

\bibitem{Nicodemus:1953}
\textsc{D.~B. Nicodemus} and \textsc{H.~H. Straub}, \enquote{Fission Neutron
  Spectrum of $^{235}$U,} \emph{Physical Review}, \textbf{89}, 1288 (1953).

\bibitem{Watt:1952}
\textsc{B.~E. Watt}, \enquote{Energy Spectrum of Neutrons from Thermal Fission
  of $^{235}$U,} \emph{Physical Review}, \textbf{87}, 1037 (1946).

\bibitem{Gibson:2021}
\textsc{N.~Gibson}, \enquote{Fission Data in Evaluations 1952-Present,} ,
  \emph{LA-UR-2021} (2021).

\bibitem{Lestone:2014}
\textsc{J.~Lestone} and \textsc{E.~Shores}, \enquote{Uranium and Plutonium
  Average Prompt-Fission Neutron Spectra (PFNS) from the Analysis of NTS NUEX
  Data,} \emph{Nuclear Data Sheets}, \textbf{119}, 213 (2013).

\bibitem{Carlson:1952}
\textsc{B.~Carlson}, \enquote{Multi-Velocity Serber-Wilson Neutron Diffusion
  Calculations,}  LA--1391, Los Alamos National Laboratory (1952).

\bibitem{Sood:2021}
\textsc{A.~Sood}, \textsc{A.~Forster}, \textsc{R.~C. Little}, and \textsc{B.~J.
  Archer}, \enquote{Neutronics Calculation Advances at Los Alamos: Manhattan
  Project to Monte Carlo,} \emph{Nuclear Technology, this issue} (2021).

\bibitem{Andrews:2021}
\textsc{S.~A. Andrews}, \textsc{M.~T. Andrews}, and \textsc{T.~E. Mason},
  \enquote{Canadian Contributions to the Manhattan Project and Early Nuclear
  Research,} \emph{Nuclear Technology, this issue} (2021, submitted).

\bibitem{Morel:1999}
\textsc{J.~E. Morel}, \enquote{Deterministic Transport Methods and Codes at Los
  Alamos,}  LA-UR-99-3566, Los Alamos National Laboratory (1999).

\bibitem{Barnett:1949}
\textsc{L.~Barnett}, \enquote{J.R. Oppenheimer Quotation by L. Barnett in `J.
  Robert Oppenheimer',} \emph{Life, International Edition}, \textbf{7},
  \emph{9} (October 24, 1949).

\end{thebibliography}

\end{document}